\documentclass[fleqn,usenatbib]{mnras}
\usepackage{newtxtext,newtxmath}
\usepackage[usenames, dvipsnames]{color}

\usepackage[T1]{fontenc}
\DeclareRobustCommand{\VAN}[3]{#2}
\let\VANthebibliography\thebibliography
\def\thebibliography{\DeclareRobustCommand{\VAN}[3]{##3}\VANthebibliography}

\usepackage{graphicx}	
\usepackage{amsmath}	
\usepackage[dvipsnames,svgnames,table,xcdraw]{xcolor}
\usepackage{physics}
\usepackage{xspace}
\usepackage{ulem}

\usepackage{macros_ap}
\usepackage{newtxtext,newtxmath}

\usepackage{multirow,tabularx}

\usepackage[export]{adjustbox}
\usepackage{calc}
\newsavebox{\twofigures}

\title[The \simu simulations]{\simu simulations -- Anisotropic thermal conduction, black hole physics, and the robustness of massive galaxy cluster scaling relations}

\author[A. Pellissier et al.]{
Alisson Pellissier,$^{1,2}$\thanks{E-mail: apellissier@cea.fr}
Oliver Hahn,$^{3,4}$ Chiara Ferrari$^{2}$
\\
$^{1}$AIM, CEA, CNRS, Universit\'e Paris-Saclay, Universit\'e Paris Diderot, Sorbonne Paris Cit\'e, F-91191 Gif-sur-Yvette, France\\
$^{2}$Universit\'e C\^ote d'Azur, Observatoire de la C\^ote d'Azur, CNRS, Laboratoire Lagrange, Bd de l'Observatoire, CS 34229, 06304 Nice cedex 4, France\\
$^{3}$Department of Astrophysics, University of Vienna, Türkenschanzstraße 17, 1180 Vienna, Austria\\
$^{4}$Department of Mathematics, University of Vienna, Oskar-Morgenstern-Platz 1, 1090 Vienna, Austria\\
}

\date{Accepted XXX. Received YYY; in original form ZZZ}

\pubyear{2022}

\begin{document}
\label{firstpage}
\pagerange{\pageref{firstpage}--\pageref{lastpage}}
\maketitle

\begin{abstract}

We present the \simu simulations that extend the \Rg suite of massive galaxy clusters at the $M_{\rm vir}\sim10^{15}\Msun$ scale with cosmological magneto-hydrodynamic zoom-in simulations that include anisotropic thermal conduction, modified supermassive black hole (SMBH) feedback, new SMBH seeding and SMBH orbital decay model. These modelling improvements have a dramatic effect on the SMBH growth, star formation and gas depletion in the proto-clusters. We explore the parameter space of the models and report their effect on both star formation and the thermodynamics of the intra-cluster medium (ICM) as observed in X-ray and SZ observations. 
We report that the star formation in proto-clusters is strongly impacted by the choice of the SMBH seeding as well as the orbital decay of SMBHs. Feedback from AGNs is substantially boosted by the SMBH decay, its time evolution and impact range differ noticeably depending on the AGN energy injection scheme used. Compared to a mass-weighted injection whose energy remains confined close to the central SMBHs, a volume-weighted thermal energy deposition allows to heat the ICM out to large radii which severely quenches the star formation in proto-clusters. By flattening out temperature gradients in the ICM, anisotropic thermal conduction can reduce star formation early on but weakens and delays the AGN activity. 
Despite the dissimilarities found in the stellar and gaseous content of our haloes, the cluster scaling relations we report are surprisingly insensitive to the subresolution models used and are in good agreement with recent observational and numerical studies.

\end{abstract}

\begin{keywords}
methods: numerical -- cosmology: large-scale structure of Universe -- galaxies: clusters: intra-cluster medium -- X-rays: galaxies: clusters -- conduction
\end{keywords}


\section{Introduction}

Forming the nodes of the cosmic web, clusters of galaxies are the largest virialised structures in our Universe and their matter content reflects that of the Universe. 
Originating from the highest peaks in the initial cosmic density field \citep{Kaiser1984,Bardeen1986}, their spatial distribution and abundance carry the imprints of the process of structure formation and are heavily sensitive to the underlying cosmology. 
Therefore, they represent veritable crossroads of astrophysics and cosmology as they provide valuable information from the physics driving structure formation to the nature of dark matter and dark energy \citep{Voit2005,Allen2011,Kravtsov2012,Weinberg2013}.

Counting galaxy clusters (GCs) as a function of their mass and cosmic time provides an excellent (late Universe) probe of cosmological parameters \citep[e.g][]{Planck2018,DES2022} including dark energy, the summed neutrino masses \citep[e.g][]{Madhavacheril2017} and modifications of gravity \citep[e.g][]{Wilcox2015}.

However, the predictive power of GCs as cosmological probes is limited principally by our ability to accurately measure their masses using X-ray, Sunyaev-Zeldovich (SZ) or gravitational lensing analyses. Mass estimations require high-quality data and rely on various assumptions that are challenged by the presence of possible biases caused by several factors, such as deviations from hydrostatic equilibrium, triaxiality, or instrumental features. Hence, cluster cosmological surveys depend heavily on well-calibrated scaling relations that relate directly observed quantities - so-called mass proxies - such as the X-ray luminosity, to the underlying cluster mass \citep[see][for a recent review]{Pratt2019}.\\

GCs grow hierarchically over cosmic time as gravity pulls baryonic and dark matter to form collapsed structures. They also grow in mass via major mergers that represent the most energetic phenomena since the Big Bang.
Moreover, the feedback from supernovae (SN) and active galactic nuclei (AGN) in cluster galaxies injects a substantial amount of energy into the intra-cluster medium (ICM).
Such processes continually shape the baryonic components of clusters and can inject up to $10^{64}\,{\rm ergs}$ of gravitational potential energy during one cluster crossing time ($\sim$\,Gyr), primarily dissipated by shocks into heating of the intra-cluster gas to high (X-ray emitting) temperatures \citep{Markevitch2007}, but also through large-scale ICM motions generating cluster-wide turbulence \citep{Hitomi2016,Hitomi2018,Li2020}. A fraction of this energy can also be channeled into non-thermal plasma components such as cosmic rays \citep{Brunetti2014,Bykov2019} and magnetic field amplification \citep{Donnert2018} as revealed by the presence of extended radio emission \citep{Ferrari2008,Feretti2012,vanWeeren2019}.\\
These energetic processes are expected to contribute to the deviation of cluster properties from self-similar predictions, which only account for gravitational evolution in scale-free cluster evolution \citep{Kaiser1986}. 
Galaxy cluster observables are therefore a complex interplay of both cosmology and astrophysics. They require detailed understanding for precisely calibrating cluster scaling relations to fully exploit the power of galaxy clusters as cosmological probes.\\

In this context, numerical simulations provide valuable information, as they can follow the evolution of galaxy clusters with exactly known properties. They can capture the effects of physical processes during cluster formation and predict the resulting observables self-consistently.
However, astrophysical processes related to galaxy formation cannot be resolved in hydrodynamical cosmological simulations due to limited computational resources. Astrophysical processes occurring below the typical resolution are accounted for by so-called sub-grid models. 
Significant recent advances in the development and calibration of efficient sub-grid models has led to cosmological simulations that can reproduce a large number of the observed galaxy properties \citep[see][for a review]{Vogelsberger2020}. However, reproducing the galaxy cluster entropy profiles and the cool-core/non-cool-core dichotomy remains challenging \citep{Rasia2015,Hahn2017, Barnes2017a}.
In cosmological simulations of galaxy clusters, special attention has been dedicated to the effects of feedback from stars and AGN feedback.
These advances have yielded a range of simulations from several independent groups that reproduce various cluster observables, such as the X-ray and SZ scaling relations \citep[e.g. ][]{Barnes2017a,Barnes2017b,LeBrun2017,Truong2018,Cui2018,Henden2019}.

The outcomes of such simulations can rely heavily on the parameter choice of the sub-grid AGN feedback models. \citet{Lebrun2014} showed that the entire baryon gas profile can be varied by tuning the energy accumulation threshold $\Delta T$ of the feedback model of \citet{Booth2009}. In the \Rg simulations of \citet[][]{Hahn2017}, changes in the AGN feedback model had no significant impact on the gas outside the cluster core region. This suggests that the sub-grid modelling of astrophysical processes needs to be improved or that the addition of new physics is necessary to increase the realism of such simulations. Indeed, the addition of thermal conduction in cluster simulations has been shown to significantly affect the properties of the ICM and provides an additional source of gas heating \citep{Voit2011,Voit2015}. Allowing heat transport in the ICM with the implementation of thermal conduction could help to decrease the dependence of numerical simulations on the feedback sub-grid model parameters. While it does not provide enough heating to offset cooling losses in clusters, recent studies have shown that thermal conduction has various impacts on AGN activity and ICM mixing \citep{Yang2016,Kannan2017,Barnes2019,Beckmann2022b}.
In this context, the properties of the intra-cluster gas intimately depend on the physical processes modelled in simulations. Quantifying such dependencies in simulations is therefore fundamental to providing robust GC scaling relations that are focused on the (hot) intra-cluster gas.\\

In this paper, we perform cosmological magneto-hydrodynamic simulations of massive galaxy clusters ($M_{\rm vir}=10^{15.0\pm0.1}\,\Msun$) that include anisotropic conduction, radiative cooling, stellar and AGN feedback to study their effects on cluster scaling relations. In \Sec{sec:methods}, we introduce our \simu sample of zoom-in simulations and the numerical methods and models that we employ for this study. We focus in \Sec{sec:smbh} on our super-massive black hole (SMBH) modelling, presenting a new `tidal friction' model that efficiently controls their orbits and studying different AGN feedback models. We show the impact of anisotropic thermal conduction (ATC) on the cluster stellar and gaseous properties in Sections \ref{sec:atc} and \ref{sec:galandicm}. In \Sec{sec:sr}, we study the impact of AGN feedback models and ATC on the evolution of the simulated clusters along several mass-observable scaling relations relevant to cosmology. Finally, we summarise our results and conclude in \Sec{sec:conclusion}. Additionally, we show in \App{app:atc} the impact of anisotropic thermal conduction on the ICM in idealised configurations.
We also provide in \App{app:Tx} and \ref{app:ci} the bias on cluster scaling relations induced by the choice of the X-ray temperature estimates in the simulations and the impact of core inclusion/exclusion on X-ray observables, respectively.


\section{Methods}
\label{sec:methods}

\subsection{The \simu sample and initial conditions}

This paper presents the \simu project\footnote{where the `C' denotes for the inclusion of anisotropic thermal conduction.}, a suite of high-resolution zoom-in magneto-hydrodynamical (MHD) simulations of nine haloes in the $M_{\rm vir}=10^{15.0\pm0.1}\Msun$ range.
The haloes were selected using the cosmICweb database from the {\sc Rhapsody-New} simulation\footnote{See \href{https://cosmicweb.astro.univie.ac.at}{https://cosmicweb.astro.univie.ac.at} for more details} (Buehlmann et al. 2023, in prep.). We list the properties of these haloes in \Tab{tab:sims}.

\begin{table}
    \caption{Description of the \simu runs presented in this work. In the top part of the table we list the minimum cell size ($\Delta_x$), the initial mass per hydro cell ($m_{\rm gas}$), the dark matter ($m_{\rm dm}$) and minimum stellar particle mass ($m_{\rm *,min}$). The middle part describes the physical models used in the simulations studied in this work. In the bottom part, we list the properties of each \simu haloes: the internal halo ID in the {\sc Rhapsody-New} simulation, the number of substructures ($N_{\rm sub}$), the virial mass ($\Mvir$), radius ($\Rvir$) and concentration ($c_{\rm vir}$) as well as the radius enclosing 500 times the critical density of the Universe  ($\Rfiveh$) and the total mass within ($\Mfiveh$).}
    \label{tab:sims}
    \small
    \begin{tabular*}{0.485\textwidth}{c @{\extracolsep{\fill}} ccc}
\multicolumn{4}{l}{\sc Summary of the \simu simulations}\Bstrut\\
\hline
$\Delta x\,[\kpc]$ & $m_{\rm dm}\,[\Msun]$ & $m_{\rm gas}\,[\Msun]$ & $m_{\rm *,min}\,[\Msun]$\Tstrut\Bstrut\\
\hline
2.82 & $1.54\times10^8$ & $1.68\times10^7$ & $6.58\times10^6$ \Tstrut\Bstrut\\
\end{tabular*}

\vspace{0.3cm}

\begin{tabular*}{0.485\textwidth}{c @{\extracolsep{\fill}} cccc}
\multicolumn{5}{l}{\sc Sub-grid modelling and baryonic processes}\Tstrut\Bstrut\\
\hline
      & Cooling, & AGN energy & AGN energy   & Anisotropic \\
Label & SF,      & deposition & accumulation & thermal     \\
      & SN       & scheme     & threshold    & conduction  \Tstrut\Bstrut\\
\hline
NR    & --  & -- & -- & --\Tstrut\\
VW    & $\checkmark$ & volume-weighted & $10^7\,{\rm K}$ & -- \\
MW    & $\checkmark$ & mass-weighted   & $10^7\,{\rm K}$ & -- \\
MC    & $\checkmark$ & mass-weighted   & $10^7\,{\rm K}$ & $\checkmark$ \\
MW6   & $\checkmark$ & mass-weighted   & $10^6\,{\rm K}$ & -- \\
MW8   & $\checkmark$ & mass-weighted   & $10^8\,{\rm K}$ & -- \Tstrut\Bstrut\\
\end{tabular*}

\vspace{0.3cm}

\begin{tabular*}{0.485\textwidth}{c @{\extracolsep{\fill}} cccccc}
\multicolumn{7}{l}{\sc Halo Properties}\Bstrut\\
\hline
\multirow{2}{*}{ID} & \multirow{2}{*}{$N_{\rm sub}$} & \multirow{2}{*}{$c_{\rm vir}$} & $M_{\rm vir}$ & $R_{\rm vir}$ & $M_{500}$ & $R_{500}$  \\
      &   &  & $[10^{15}\Msun]$ & $[{\rm Mpc}]$ & $[10^{14}\Msun]$ & $[{\rm Mpc}]$\Tstrut\Bstrut\\
\hline
174742934 & 111 & 6.02 & 1.22 & 2.82 & 6.80 & 1.37 \\ 
176970005 & 117 & 6.69 & 1.19 & 2.78 & 7.55 & 1.41 \\ 
174824666 & 124 & 6.35 & 1.16 & 2.77 & 7.56 & 1.42 \\ 
173505201 & 135 & 5.68 & 1.12 & 2.80 & 6.98 & 1.38 \\ 
176144520 & 100 & 6.80 & 1.01 & 2.64 & 6.32 & 1.33 \\ 
176061412 & 105 & 9.52 & 1.00 & 2.63 & 6.62 & 1.35 \\ 
173917492 & 111 & 8.64 & 0.99 & 2.62 & 6.73 & 1.36 \\ 
174743229 &  83 & 6.50 & 0.98 & 2.62 & 6.24 & 1.33 \\ 
173587157 &  84 & 5.45 & 0.86 & 2.51 & 5.19 & 1.25 \Bstrut\\ 
\hline
\end{tabular*}

\end{table}

Sharing similar masses at $z=0$, our haloes have different assembly histories and probe extreme as well as median cluster properties: two haloes have extreme concentrations ($c_{\rm vir}>8$), high and low number of subhaloes ($N_{\rm sub}>120$ and $N_{\rm sub}<85$ respectively). 
On average, our haloes share the same number of substructures and concentrations as the \Rg sample \citep{Wu2015,Hahn2017,Martizzi2016}. 

We use the $\Lambda$CDM cosmology of the {\sc Rhapsody-New} simulation with density parameters $\Omega_{\rm b}=0.049$ for baryons, $\Omega_{\rm m}=0.309$ for total matter and $\Omega_\Lambda=0.691$ for the cosmological constant. The primordial spectral index, the amplitude normalisation and the Hubble constant are $n_s=0.9667$, $\sigma_8=0.8159$ and $H_0=67.74\, {\rm km/s/Mpc}$ respectively \citep{Planck2016}. In this new cosmology, we have a lower baryon fraction of $f_{\rm b}=0.1586$ compared to the \Rg's value of 0.18. The updated value is also much closer to more recent constraints from \citet{Planck2018} of $f_{\rm b}=0.1564$.

We generated the initial conditions using {\sc Music} \citep{Hahn2011} for our nine clusters at $z=49$ from the minimum bounding ellipsoid matrix retrieved from the cosmICweb database using a traceback-radius of $2\Rvir$ centered in a $1\hGpc$ box with an effective resolution of $8192^3$ particles\footnote{We share the same resolution as the \Rg 8K run.}. 
All initial conditions were performed using second-order Lagrangian perturbation theory (LPT) with dark matter and baryon perturbations at $z=49$. Compared to the original \Rg simulations, we do not use the local Lagrangian approximation for the construction of the baryon density field.
Baryons and dark matter did not co-move prior to recombination and sub-percent effects are expected at cluster scales \citep[see e.g.][]{Angulo2013,Hahn2021,Khoraminezhad2021}. 
However, for the simulations used here, we assume that baryons fully trace cold dark matter perturbations.

\subsection{Numerical approach}

For our cluster zoom simulations, we use the Eulerian adaptive mesh refinement {\sc Ramses} code \citep{Teyssier2002} to follow the non-linear evolution of the initial conditions.
Gas dynamics are computed using a second-order unsplit Godunov scheme for the ideal MHD equations \citep{Fromang2006,Teyssier2006}
while collisionless dark matter particles as well as stars and sink particles are evolved using a particle-mesh solver.
Our simulations use the method introduced by \citet{Dubois2016} for solving the anisotropic diffusion of heat using an implicit finite-volume method which is independent of the Courant time step constraint of the MHD scheme.\\
We employ a Lagrangian overdensity-based refinement strategy that splits cells if they reach an overdensity of eight: the refinement of the base grid by $n$ additional levels requires a density of $8^n\bar{\rho}$. Our simulation boxes of $1\hGpc$ on a side, reach a maximum refinement level by maintaining a smallest cell size of physical $\Delta x = 2.8 \kpc$. The dark matter $N$-body particle mass is $1.54\times10^8\Msun$ and initial mass per hydro cell is $1.68\times10^7\Msun$.
The high-resolution Lagrangian ellipsoid patch, from which the $2\Rvir$ sphere centred on each cluster will form, is tagged using a passive scalar colour field that is advected with the gas. Dynamic refinement is restricted to the regions where this colour field is non-zero and no refinement is allowed outside the zoom region. We thus focus most of the computational resources on the forming cluster and its immediate environment.\\

The rest of this section details the various physical ingredients used in our high-resolution zoom-in simulations. See \Sec{sssec:coolmodel} for gas cooling and heating as well as the star formation and stellar feedback, \Sec{sssec:agnmodel} for the subgrid modelling of SMBH formation, evolution and AGN feedback ang finally in \Sec{sssec:Bmodel} we describe the magnetic field evolution with the anisotropic thermal conduction.
The reader can skip to \Sec{sec:smbh} and \Sec{sec:atc} for the scientific results regarding the impact of the the various BH-related sub-grid models and anisotropic thermal conduction respectively on the stellar and gaseous content of a GC, or directly to Sections \ref{sec:galandicm} and \ref{sec:sr} for properties of cluster galaxies and ICM as well as the the evolution of our clusters along various scaling relations respectively.

\subsubsection{Radiative gas cooling, metallicity and stellar evolution}
\label{sssec:coolmodel}

Radiative gas cooling is calculated according to the tabulated rates of \citet{Sutherland1993} for Hydrogen, Helium and metal line cooling.
The total gas metallicity is not evolved separately but treated as a single species. It is advected with the MHD equations as a passive scalar and is sourced by the supernovae feedback model.
We consider an UV background radiation according to the \citet{Haardt1996} model. An instantaneous reionisation takes place at $z=10$ to take into account an earlier reionisation in the particularly overdense proto-cluster regions that we simulate.
The unresolved cold and dense gas that will constitute the inter-stellar medium (ISM) of galaxies is approximated using a temperature floor given by a polytropic equation of state,
\begin{equation}
    T_{\rm floor} = T_*\left(\frac{n_{\rm H}}{n_*}\right)^{\gamma_*-1},
\end{equation}
with $n_{\rm H}$ the Hydrogen number density of the gas, ${n_*}=0.1\pcmcb$ and $T_*=\tento{4}\,{\rm K}$ being respectively the star formation density threshold and the ISM polytropic temperature with $\gamma_*=5/3$ being the ISM polytropic index.
In practice, gas can be heated above the temperature floor, but cannot cool below it.

Star formation occurs when the gas density exceeds ${n_*}$. A  portion of the gas in a cell is converted into a star particle that decouples from the gas. We have a minimum stellar particle mass of $5.6\times 10^6\Msun$. The star particles are randomly drawn from Poisson process \citep{Rasera2006} following a Schmidt law
\begin{equation}
    \dot{\rho_*}= \epsilon_*\,\rho_{\rm gas}\,/\,t_{\rm ff},
\end{equation}
with $\epsilon_*=0.01$ and $t_{\rm ff}=\left(3\pi/32{\rm G}\rho_{\rm gas}\right)^{-1/2}$, the local free-fall time.

Stellar feedback is included using the model of \citet{Dubois2008b} in which each newly formed star that traces a continuous stellar mass distribution following the \citet{Salpeter1955} initial mass function and releases, after $20\,{\rm Myr}$, a fraction $\eta=0.1$ of its mass and metals with a yield of $y=0.1$. Therefore $y\eta = 0.01$ of the time-integrated SFR is returned as metals in the ISM. In addition, each SN feedback event injects a thermal energy of $10^{51}\,{\rm erg}$ into the surrounding ISM.
Compared to the original \Rg simulations, we chose to enable the delayed cooling of the SN heated gas with a dissipation time scale of $20\,{\rm Myr}$. This additional sub-grid model mimics the effect of non-thermal processes, such as turbulence or CRs \citep{RodriguezMontero2022}, which can dissipate energy on longer time scales before being radiated away.
The calibration of the free parameters of the SN feedback listed above is able to reproduce stellar masses consistent with abundance-matching results at masses lower than $10^{12}\Msun$ for resolved haloes with at least 1000 particles (see \Sec{sec:smhmrelations}).

\subsubsection{Black holes and active galactic nuclei}
\label{sssec:agnmodel}

{\sc Ramses} uses collisionless sink particles to model black hole growth and evolution. 
The SMBH formation and evolution follow the model of \citet{Biernacki2017}, itself based on the precedent models of \citet{Dubois2010} and \citet{Teyssier2011}, and build on a sink particle implementation developed within the context of star-forming molecular clouds \citep{Bleuler2014}.\\

\paragraph*{Super-massive black hole seeding.}
The {\sc Phew} clump finder \citep{Bleuler2014}, directly implemented in {\sc Ramses}, determines potential sites for SMBH sink particle formation by identifying relevant peaks in the density field. We will briefly discuss the main steps and free parameters of the sink seeding model that we use. First, all density peaks above a threshold $\rho_{\rm peak}$ are identified as well as their connecting saddle points. To keep only relevant density peaks, we merge all peaks that have a peak-to-saddle ratio lower than 3 to the neighbouring peak with which it shares the highest density saddle point.
This merging process, or noise removal, is halted when a saddle density falls below the $\rho_{\rm saddle}$ threshold. In short, a noise removal is performed on the density field to select only the relevant peaks above a density $\rho_{\rm peak}$ which are later divided by the saddle density threshold $\rho_{\rm saddle}$ into clumps to finally yield the sink formation sites. 
The gas in the spherical region of radius equal to 4 (highest) resolution elements $\Delta x$, defining the sink sphere, is investigated to make sure that the gravitational field is compressive, strong enough to overcome internal gas support and not only accelerated toward the sink sphere centre but that this gas is contracting. As a proximity check, we forbid the gas that is infalling to an already existing sink to create another sink.
While the choice for the initial seed mass is arbitrary, we set it to be the same as our $N$-body dark matter particle mass with $m_{\rm BH,seed} = 10^8\Msun$.

\paragraph*{Gas accretion and black hole dynamics.}
Once SMBHs are formed, they grow in mass at the (un-boosted) Bondi-Hoyle accretion rate \citep{Hoyle1939,Bondi1944,Bondi1952} capped by the Eddington rate :
\begin{equation}
  \dot{M}_{\rm acc}={\rm min}\left(\dot{M}_{\rm Edd}\, , \,\dot{M}_{\rm Bondi} \right),
\end{equation}
 with :
\begin{align}
  &\dot{M}_{\rm Bondi}=4\pi\rho_\infty r_{\rm Bondi}^2v_{\rm Bondi},\\
  &\dot{M}_{\rm Edd}=\frac{4\pi {\rm G}M_{\rm BH}m_p}{\epsilon_r\sigma_T c}=\frac{M_{\rm BH}}{t_S},
\end{align}
where $\sigma_T$ is the Thomson cross-section, $G$ the gravitational constant, $M_{\rm BH}$ and $m_p$, sink and proton mass respectively, $\epsilon_r=0.1$ is the \citet{Shakura1973} radiative efficiency for a SMBH and $t_S\sim45\,{\rm Myr}$ is the Salpeter time. We also have $\rho_\infty=\bar{\rho}/\alpha(x_{\rm sink})$ with $\alpha$ is the dimensionless density profile of the Bondi self-similar solution \citep[see][]{Biernacki2017}, $\bar{\rho}$ the mean density inside the sink sphere, $x_{\rm sink}=r_{\rm sink}/r_{\rm Bondi}$ and the sink radius and velocity defined as follows :
\begin{align}
    r_{\rm Bondi} &= \frac{{\rm G}M_{\rm BH}}{v_{\rm Bondi}^2},\\
    v_{\rm Bondi} &= \sqrt{c_s^2 + v_{\rm rel}^2},
\end{align}
with $v_{\rm rel}$ the relative velocity of the sink to the average gas velocity inside the sink sphere and $c_s$, the local sound speed. While we use MHD, we generically find high plasma beta values in our simulations (ranging from $\beta_e\sim 10^{14}$ in the core to $5\times10^{12}$ in the outskirts due to the lack of magnetic field amplification from $10^{-12}$~G seeds). Therefore, in the Bondi formula, the magneto-sonic speed effectively reduces to the adiabatic sound speed.
In addition to gas accretion, SMBHs can also grow via mergers. In this work, we do not check if two sinks form a bound system but directly merge if they are less than one accretion radius apart, i.e. $4\Delta x$.

The dynamics of a single SMBH cannot be resolved in cosmological simulations. This can lead to spurious oscillations of the SMBH in the potential well of its host halo, due to external perturbations and the finite resolution effects, particularly during merger events.
\citet{Biernacki2017} implemented in {\sc Ramses} a physically motivated model based on Eddington-limited accretion. Their main assumption is that the gas accretion rate onto the accretion disc is set by the Bondi formula ($\dot{M}_{\rm Bondi}$) which corresponds to a large scale accretion flow, while the accretion onto the SMBH is set by the Eddington rate ($\dot{M}_{\rm Edd}$). 
 The difference between the two rates therefore gives the amount of gas not being accreted by the central SMBH. Instead, it should be pushed away from the accretion disc by the Eddington radiation pressure at a rate $\dot{M}_{\rm dec} = \dot{M}_{\rm Bondi} - \dot{M}_{\rm acc}$, which we however do not model explicitly in this work. We also stress that we are not using radiation hydrodynamics in this work. This process of gas accretion and ejection leads to an additional momentum exchange between the gas and the sink particle, hence an additional drag force.
This additional drag force is modelled by requiring a fixed center of mass of the joint gas+sink system during the accretion and a conserved total momentum.
We implemented in {\sc Ramses} a further modification to the model of \cite{Biernacki2017} to move the SMBHs towards the potential minimun (described in \Sec{sec:SMBHdescent}).

\paragraph*{Active galactic nucleus feedback}

The accretion rate of gas onto the SMBH sink particle is always computed from the cells in the sink sphere (of radius $4\Delta x$) using mass-weighting.
Following \cite{Booth2009}, we do not inject the thermal AGN energy at each time-step but store the rest-mass energy of the accreted gas until it would be enough to raise the gas temperature inside the sink sphere by $\Delta T=10^7\,{\rm K}$ (unless specified otherwise, see e.g. studies of \citealt{Teyssier2011} and \citealt{Lebrun2014} using this $\Delta T$ threshold strategy). In other words, we inject this accumulated AGN energy when
\begin{equation}
    E_{\rm AGN} > \frac{3}{2}m_{\rm gas}\kB \Delta T
\end{equation}
in every gas cell of the sink sphere in a mass- or volume-weighted way (see \Sec{sec:agn}) with a maximum allowed temperature of the AGN feedback set to $T_{\rm AGN}=1.5\times\tento{11}\,{\rm K}$. The rate at which this thermal energy is released to the ambient gas is given by :
\begin{equation}
    \dot{E}_{\rm AGN} = \epsilon_c \epsilon_r \dot{M}_{\rm acc} {\rm c}^2,
\end{equation}
where $\epsilon_c=0.15$ is the coupling efficiency \citep{Dubois2012}, i.e. the fraction of radiated energy that couples to the surrounding gas, and is calibrated on the local $M_{\rm BH}-M_*$ relation.

\subsubsection{Magnetic fields and anisotropic thermal conduction}
\label{sssec:Bmodel}

To solve the MHD equations, the {\sc Ramses} code uses the second-order unsplit Godunov method based on the monotonic upstream-centred scheme for conservation laws (MUSCL-Hancock method, \citealt{vanLeer1977,Evans1988}). The constrained transport approach is used to evolve the induction equation
\begin{equation}
  \frac{\partial \bvec{B}}{\partial t} = \bnabla\times\bvec{u}\times\bvec{B},
  \label{eq:induction}
\end{equation}
where $\bvec{u}$ is the gas velocity and $\bvec{B}$ the magnetic field. The scheme satisfies the solenoidal constraint $\bnabla\cdot\bvec{B}=0$ to machine precision \citep{Teyssier2006}. 
The 2D Riemann problem at the cell edges is solved using the approximate Harten-Lax-van Leer-Discontinuities (HLLD) solver from \citet{Miyoshi2005} to compute time averaged electromotive forces.\\
In the ideal MHD limit, the generation of magnetic fields from a previously unmagnetised fluid is impossible. Therefore, the magnetic fields must be seeded in our simulations. For simplicity, we seed a uniform magnetic field along the box $z$ axis with a comoving magnitude of $B_0 = 1.56\times 10^{-12}\,{\rm G}$, which ensures a divergence-free initial field.\\

In the presence of a magnetic field, the conduction of heat in a plasma becomes anisotropic since the motion of charged particles perpendicular to the field lines is restricted. 
We use the implementation of \citet{Dubois2016}  using an implicit finite-volume method for solving the anisotropic diffusion of heat through electrons \citep{Braginskii1965}
\begin{equation}
    \frac{\partial \rho \epsilon_{\rm e}}{\partial t} =-\bnabla\cdot{\bf Q}_{\rm cond} , \label{energyeq}
\end{equation}
with $\epsilon_{\rm e}$ the specific internal energy of electrons.\footnote{We use the original version of the solver which does not adopt the regularised version of \citet{Dashyan2020} that includes a minmod slope limiter to preserve the monotonicity of the flux. Therefore, the solver we use in this work does not exactly conserve energy and does not prevent heat to diffuse against the temperature gradient in certain situations. Although in practice, the addition of the limiter had only a minor impact on the results in terms of e.g. mass outflow rate, and SFR \citep[see][for more details]{Dashyan2020}.}
The conductive heat flux, ${\bf Q}_{\rm cond}$, can saturate once the characteristic length scale of the electron temperature gradient $\ell_{T_{\rm e}}$ is comparable to or less than the mean free path of electron $\lambda_{\rm e}$. Hence, following~\citet{Sarazin1986}, we introduce an effective conductivity which interpolates between the unsaturated (Spitzer conductivity) and saturated regime by
\begin{align}
    {\bf Q}_{\rm cond, sat} &= - f_{\rm sat}\,\kappa_{\rm Sp}\bnabla T_{\rm e},\nonumber\\
        &= - f_{\rm sat}\left[-\kappa_\parallel{\bf b}\left({\bf b}\cdot\bnabla\right)T_{\rm e}\right]- f_{\rm sat}\left(-\kappa_{\rm iso}\bnabla T_{\rm e}\right)\label{eq:conductivity},
\end{align}
with $f_{\rm sat} =\left(1+4.2 \lambda_{\rm e}/\ell_{T_{\rm e}}\right)^{-1}$, ${\bf b}={\bf B}/\lvert{\bf B}\rvert$ the unit vector in the direction of the local magnetic field, $T_{\rm e}$ the electronic temperature, , $\kappa_{\rm iso}$ and $\kappa_\parallel$ the isotropic and parallel conduction coefficient (with respect to the magnetic field lines) respectively with $\kappa_\parallel=\kappa_{\rm Sp}-\kappa_{\rm iso}$. In many astrophysical cases, $\kappa_{\rm iso}/\kappa_\parallel\ll 1$ since the Larmor radius, $\lambda_{\rm L}$, is much smaller than the mean-free-path of electrons, $\lambda_{\rm e}$. For instance, in a hot intra-cluster plasma with $T_{\rm e}=3\keV$, $n_{\rm e}=10^{-2}\pcmcb$ and $B=1\mu{\rm G}$, we have $\lambda_{\rm L}=10^8\cm$ and $\lambda_{\rm e}=10^{21}\cm$. Here, we set a perpendicular conductivity coefficient of 1 per cent to ensure numerical stability \citep{Dubois2016}.\\
The electron energy is tracked separately from that of the ions as described in \citet{Dubois2016} and the rate of energy transfer between the electron and ion temperatures is given by 
\begin{equation}
    Q_{\rm e \leftrightarrow i}= \frac{ T_{\rm i}-T_{\rm e}} {\tau_{\rm eq, ei}} \frac{n_{\rm e} {\rm k_B}} {\gamma-1},
\end{equation}
with the equilibrium timescale
\begin{align}
\tau_{\rm eq, ei}&=\frac{3 m_{\rm e}m_{\rm p}} {8 \sqrt{2 \pi} n_{\rm i} q_{\rm e}^4 \ln \Lambda} \left( \frac{\rm k_B T_{\rm e}} {m_{\rm e}} \right)^{\frac{3}{2}}.
\end{align}
Both ion and electron adiabatic indexes are equal to $\gamma=5/3$.\\

By modelling the anisotropic transport using Braginskii MHD (\Eq{energyeq}), we follow the Spitzer ansatz that assumes a high degree of electron-ion collisionality, which is a good assumption in cluster cores (in which we are particularly interested here), but would need to be corrected in cluster outskirts.
Additionally, we do not take into account the suppression of thermal conduction by the ion mirror instability \citep{Komarov2016} caused by magnetic trapping of electrons by magnetic field strength fluctuations, or the Whistler instability \citep{Levinson1992,Pistinner1998,Roberg-Clark2016,Roberg-Clark2018} where electron-whistler scatterings can significantly alter conduction at very sharp temperature gradients such as in cold fronts \citep{Komarov2018} or at temperature scale lengths below the critical value $\beta_{\rm e} \lambda_{\rm e}$ (\citealp{Drake2021}, with typical values of $\beta_{\rm e}\sim100$ and $\lambda_{\rm e}$ ranging from $0.1\kpc$ in cool-cores to $50\kpc$ in cluster outskirts)\footnote{where $\beta_{\rm e}=8\pi n_{\rm e} {\rm k_B}T_{\rm e}/B^2$ is the electron plasma beta, the ratio of electron thermal to magnetic field pressure.}.
The recent idealised simulations of \citet{Berlok2021} and \citet{Beckmann2022b}, showed that the suppression of heat conductivity by the electron whistler and ion mirror instabilities respectively have only a small impact on the ICM. In this work, we hence model the upper limit of anisotropic thermal conduction within the ICM, which is sufficient for our purposes to study the potential impact on cosmological observables.


\section{The modelling of super-massive black holes}
\label{sec:smbh}

The key ingredients of our SMBH formation and evolution models are: ({\it a}) the conditions for the formation of the SMBH and the SMBH seed mass, ({\it b}) the SMBH dynamics with a possible inclusion of a dynamical friction model, ({\it c}) the SMBH growth by mass accretion at the Bondi-Hoyle-Lyttleton rate limited to the Eddington rate and finally ({\it d}) the induced AGN feedback which affects the surrounding gas which couples back to all the previous model ingredients.
\Ramses uses the so-called sink particle technique \citep{Bate1995} to model SMBH formation and evolution, which is a point mass which can move through the fluid accretion and interact with it by the ejection of mass, energy and momentum.\\
Motivated by the low efficiency of the AGN feedback model in the \Rg simulations, our sub-grid models for the SMBH formation, evolution and AGN feedback need to be revised. In this section we test how the free parameters in the model influence the cluster evolution. Respectively, for ({\it a}) we investigate in \Sec{sec:SMBHseeding} the effect of different SMBH seeding scenarii on the gaseous and stellar content on one of our proto-clusters. Regarding ({\it b}), we will present in \Sec{sec:SMBHdescent} our new `tidal friction' model which allow to control SMBH orbits. Lastly, we study for ({\it d}) different AGN feedback models in \Sec{sec:SMBHseeding} which impact completely differently cluster evolution.
These analyses are all carried out on a fiducial halo (173917492). In the simulations discussed in this section, we do not implement yet the anisotropic thermal conduction.

\subsection{Seeding of SMBHs}
\label{sec:SMBHseeding}
The specifics of SMBH seeding in simulations is an important aspect of controlling the effect of AGN feedback in simulations. Different models for black hole (BH) seeding are used in cosmological simulations such as placing a BH particle in the centre of every massive halo \citep{Schaye2015,Weinberger2017,McCarthy2017} or models that use thresholds of local gas properties such as metallicity, density, temperature and velocity \citep{Dubois2014,Tremmel2017,Habouzit2017,Dubois2021}. 

In this work, we generally adopt the same procedure as in \Rg for black hole seeding, albeit with modified parameters. Following \citet{Biernacki2017}, we use the `minimal' Jeans mass corresponding to the highest refinement level of our simulation to define the initial SMBH sink particle mass $M_{\rm seed}=10^8\Msun$ which also correspond to our dark matter particle mass.
Sink particle formation sites are identified on-the-fly using the {\sc {Phew}} clump finder algorithm which identifies density peaks with a given contrast relative to the next saddle-point \citep[see][for a detailed description]{Bleuler2014} which is directly implemented in the \Ramses code. The {\sc Phew} parameters adopted for the original \Rg simulations favoured the seeding of a sink particle in fewer but larger patches of gas. 
Due to the stochastic nature of star formation and supernova feedback that impact the local gas properties (hence the SMBH seeding), we observed a large variability in the efficiency of AGN feedback in this case. For the new suite of simulations discussed in this paper, we followed a more systematic investigation into the impact of seeding on the proto-cluster region.
In particular we studied the following scenarios where we varied peak density and saddle thresholds but kept all other parameters fixed:
\begin{itemize}
    \item $\rho_{\rm peak}=0.5\,\bar{\rho}$, $\rho_{\rm saddle}=2\,\bar{\rho}$: {\sc Phew} parameters as the original \Rg setup, with $\bar{\rho}=\Omega_m\rho_c$ the mean matter density.
    \item $\rho_{\rm peak}=8\,\bar{\rho}$, $\rho_{\rm saddle}=20\,\bar{\rho}$. With a higher $\rho_{\rm peak}$ value, only the highest density regions in the simulation are probed. In that case, smaller gas patches are selected but spatially more frequent. Thus, it allows to seed more SMBHs in the simulation compared to the original \Rg configuration.
    \item $\rho_{\rm peak}=8\,\bar{\rho}$, $\rho_{\rm saddle}=200\,\bar{\rho}$. We increase the saddle density threshold by a factor of 10. As the result, a much lesser number of peaks are merged which results in an increased number of SMBH seeds.
    \item $\rho_{\rm peak}=8\,\bar{\rho}$, $\rho_{\rm saddle}=15\,\bar{\rho}$, with a lower saddle threshold which induce more peak merging hence a lowered number of SMBH seeds in the simulation.
\end{itemize}

\begin{figure}
    \centering
    \includegraphics{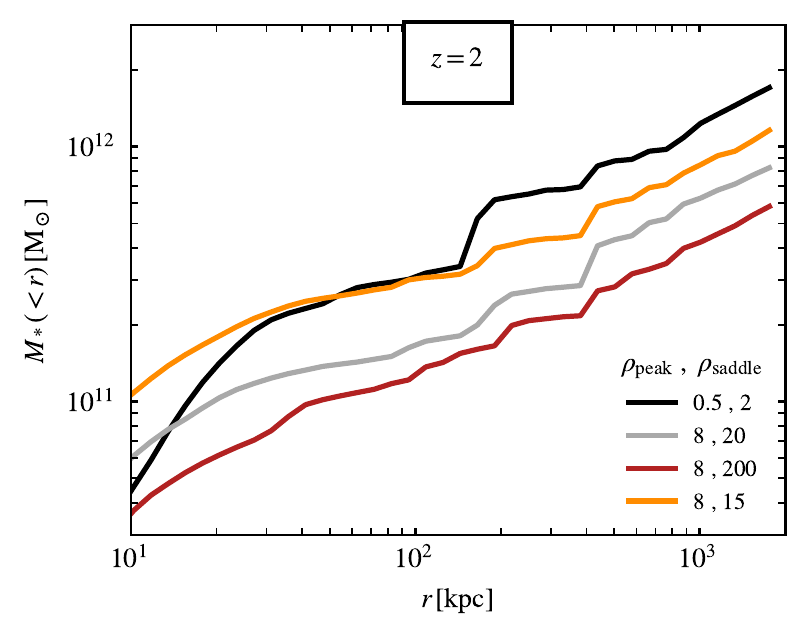}
    \caption{Total stellar mass radial profiles at $z=2$ for the simulations sharing different seeding strategies controlled by the peak and saddle thresholds. The original SMBH seeding of the \Rg simulations is shown in black and the strategy used for our \simu simulations is shown in grey.}
    \label{fig:SMBH_Seeding}
\end{figure}

We show the impact of these choices on the enclosed total stellar mass in the proto-cluster region in \Fig{fig:SMBH_Seeding} at $z=2$. At that time, we find 21, 102, 168, 113 SMBHs of mean masses $3.5\times10^8\,\Msun$, $1.8\times10^8\,\Msun$, $1.6\times10^8\,\Msun$ and $2.1\times10^8\,\Msun$ inside the virial radius respectively in the above-mentioned simulations. 
It demonstrates the tight connection of the $\rho_{\rm peak}$ parameter with the total number of created sinks and the mean mass.
The \Fig{fig:SMBH_Seeding} clearly indicates the resulting effect on the star formation suppression in the proto-cluster environment:  The simulations hosting a higher number of SMBHs (being also spatially more frequent), shows a greater amount of AGN feedback energy injected in haloes. As a result, this more profuse AGN heating will reduce the gas cooling in haloes which decrease the accretion of cold gas onto the central SMBH. The resulting mass accretion rates are seen to be inversely proportional to the number of SMBHs in our simulations. In consequence, the total stellar mass in the proto-cluster is consistently reduced with an increasing number of SMBHs in the simulations. We see that the total stellar mass for the simulation using $\rho_{\rm peak}=8\,\bar{\rho}$, $\rho_{\rm saddle}=200\,\bar{\rho}$ is reduced by a factor of 5 while the number of SMBHs is increased by the same factor approximately. The total stellar mass in the proto-cluster can be directly controlled by the number of SMBHs seeded in the simulations.\\
Galaxy masses at $z=2$ were found to be in agreement with abundance matching results with the use of $\rho_{\rm peak}=8\,\bar{\rho}$, $\rho_{\rm saddle}=20\,\bar{\rho}$ parameters (see in \Sec{sec:smhmrelations}). Therefore, we use these parameters for the simulations of the whole \simu sample.

\subsection{Decaying black hole orbits}
\label{sec:SMBHdescent}
In cosmological simulations, with force resolution larger than tens of pc to kpc, the dynamics of individual SMBHs cannot be resolved.
The lack of resolution manifests itself in spurious oscillations of the SMBH in the potential well of its host halo due to external perturbations, particularly so during mergers. In realistic systems, the SMBH is expected to sit in a much deeper potential well and perturbations are expected to decay much more rapidly. The SMBH should experience a drag force due to its tight gravitational coupling with the surrounding gas and the nuclear star cluster which prevent it from violent perturbations in the host galaxy (e.g. \citealp{Kochanek1987,Portegies2002,Devecchi2009,Davies2011,Stone2017,Neumayer2020}).
As a consequence, the accretion onto off-center SMBHs becomes suppressed which effectively renders SMBH feedback inefficient.

However, the formation of SMBHs from a rapid growth of less massive seeds requires SMBHs to sink efficiently towards the galactic centers and remain trapped there \citep{Ma2021}.
The problem motivated various ad hoc BH centering prescriptions that have been proposed in the literature: e.g SMBHs are directly placed (and fixed) at the local minimum of the potential field \citep{Weinberger2017} or artificially pushed in the direction of the stellar centre \citep{Gabor2013}. Other authors have implemented sub-resolution models for dynamical friction \citep[e.g.,][]{Hirschmann2014,Tremmel2015,Volonteri2016,Pfister2019}.

\subsubsection{Method and Implementation}
In this paper, we follow a novel `sub-grid' approach loosely motivated by the dynamical friction and tidal forces due to the presence of a nuclear star cluster \citep[e.g.][]{Biernacki2017,Ogiya2020} as well as dense gas. These processes keep SMBHs centered even after mergers, any off-centering perturbation will lead to counteracting tides and dynamical friction. To account for these unresolved processes, we adopt a simple SMBH orbital decay model which acts as a dynamical gradient descent in the gravitational potential.

To model the orbital decay as a gravitational potential descent, we employ a slight modification of the Borzilai-Borwein gradient descent algorithm \citep{Barzilai1988}.
Similarly to the classical steepest descent method proposed by \citet{Cauchy1847}, we seek to reduce the sink particle potential at every fine time-step (i.e. the time-step of the maximum level of refinement $l_{\rm max}$)\footnote{In \Ramses, all sink variables are always updated at the highest resolution level, $l_{\rm max}$\citep{Lupi2015}}. The sink particle displacement, $\Delta \bvec{x}$, along the steepest gradient is computed at a time $t$ and its position is updated at the next time iteration $t+\Delta t$.\\
Let $\bvec{x}^{n}$ be the sink position at the time $t$ and $\bvec{x}^{n+1}$ at the next time step $t+\Delta t$. The potential $\phi(\bvec{x}^{n+1})$ that we wish to minimize can be approximated by 
\begin{align}
    \phi(\bvec{x}^{n+1}) &= \phi\left(\bvec{x}^{n} + \Delta\bvec{x}\right), \nonumber\\
    &\approx \phi(\bvec{x}^n) + \Delta \bvec{x}^\intercal\cdot \bnabla\phi(\bvec{x}^n) + \frac{1}{2}\Delta \bvec{x}^\intercal \cdot \btensor{H}_\phi(\bvec{x}^n)\cdot\Delta\bvec{x},\label{eq:taylor_potential}
\end{align}
where $\btensor{H}_\phi(\bvec{x}^n)=\bnabla\bnabla\phi(\bvec{x}^{n})$ is the Hessian matrix of the potential at the position $\bvec{x}^{n}$ in the simulation box. Minimising the second order expansion of the potential~(\ref{eq:taylor_potential}) with respect to the sink particle displacement $\Delta x$, one finds
\begin{equation}
    \bvec{x}^{n+1} = \bvec{x}^{n} - \btensor{H}^{-1}_\phi(\bvec{x}^n)\,\bnabla\phi(\bvec{x}^n) 
\end{equation}
We see that $\bnabla\phi(\bvec{x}^n):=\bvec{f}(\bvec{x}^n)$ is the force exerted on the SMBH sink particle. The inverse of the tidal tensor $\btensor{H}^{-1}_\phi$ multiplying it has dimensions of time squared and thus determines both the `time step' and a possible directional deviation from the local gradient for the descent. As the Hessian is difficult to estimate reliably (and also to invert) since it is less smooth than the force, it is preferable to approximate its value using only a first derivative of the potential. This is achieved by the \citet{Barzilai1988} approximation which sets
\begin{equation}
    \btensor{H}^{-1}_\phi(\bvec{x}^n) \approx  \frac{\left| (\bvec{x}^{n+1} - \bvec{x}^{n})^{\rm T}\,\cdot\,\left[\bvec{f}(\bvec{x}^{n+1})-\bvec{f}(\bvec{x}^n)\right]\right|}{\left\| \bvec{f}(\bvec{x}^{n+1})-\bvec{f}(\bvec{x}^n) \right\|^2} =: \alpha,
\end{equation}

We have implemented this orbital decay model in \Ramses\footnote{The model is implemented in the public \Ramses version, available from \url{https://bitbucket.org/rteyssie/ramses}.} by taking the geometric mean of the simulation time step $\Delta t$ and the descent time step $\sqrt{\alpha}$ so that no descent step occurs if either of them vanishes. The full descent update to the sink positions then reads
\begin{align}
    \bvec{x}^{n+1} = \bvec{x}^{n} - \tilde{\alpha}\,\bvec{f}(\bvec{x}^n)\qquad\textrm{with}\qquad \tilde{\alpha} := f_d\sqrt{\alpha}\,\Delta t, \label{BBupdate}
\end{align}
where $0\le f_d\le 1$ is a dimensionless control parameter. It allows to adjust the effective sink displacement, as we found the SMBH descent to be very effective. 
We further limit the step by requiring that sink particles travel less than $0.2\times \Delta x$ over a time step, which effectively imposes a Courant-Friedrichs-Lewy-type criterion. This sub-grid model only involves positions and forces which therefore ensures the sink displacement to be Galilean invariant\footnote{Note that this is a distinct advantage over a standard friction that would rather use a modification of the velocity update including a drag/friction term of the form $\dot{\vecb{v}} = -\epsilon (\vecb{v}-\vecb{v}_{\rm env}) - \bnabla\phi$ relative to the average environment velocity $\vecb{v}_{\rm env}$, which is not easily estimated, particularly so in the presence of strong outflows driven by the sink itself.}.

\begin{figure}
    \includegraphics{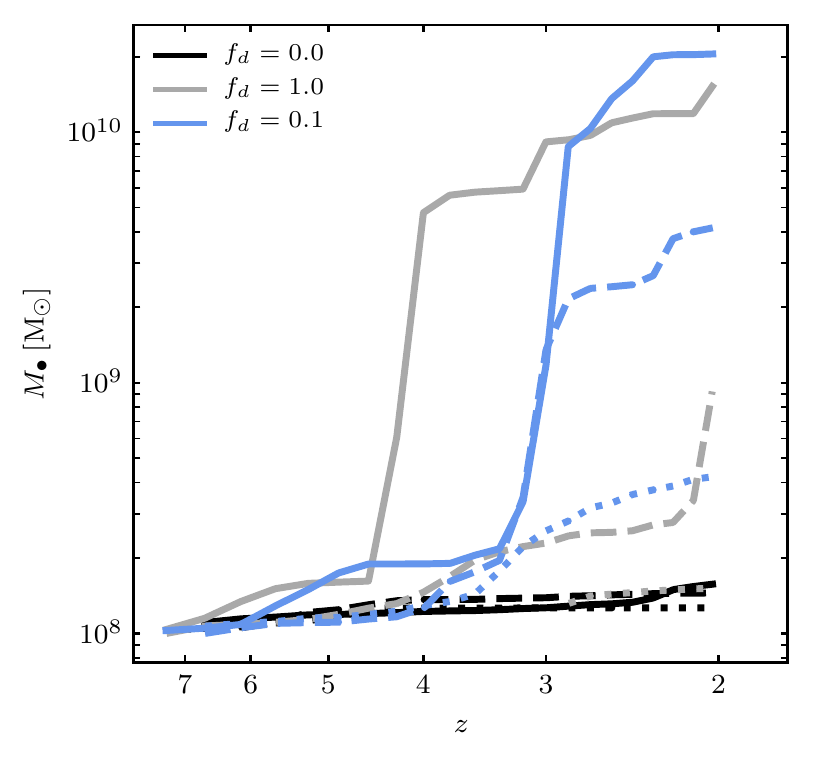}
    \caption{Redshift evolution of the mass of three most massive SMBHs in the simulation with no (black, $f_d=0$), a maximal (grey, $f_d=1$) or mild (blue, $f_d=0.1$) `tidal' descent. The evolution of most massive sink is shown with solid lines, the second and third most massive SMBHs are shown with dashed and dotted lines respectively. We can directly see the effect of the orbital decay model on the SMBH mass growth. The most massive SMBH in the simulation shows a greater mass growth for the maximal descent ($f_d=1$) simulation compared to the simulation using $f_d=0.1$. As the result, the simulation with $f_d=1$ profits from an early strong AGN activity which smother the mass growth of the other SMBHs. However, thanks to a smoother gas accretion in the simulation using a mild descent ($f_d=0.1$), SMBHs can reach higher masses at later times, and therefore induce a later but greater and steady AGN activity.}
    \label{SMBH_Descent_Mbh}
\end{figure}

\subsubsection{Validation in Simulations}
\paragraph*{Impact on black hole growth.} Starting from a simulation with $\rho_{\rm peak}=0.5$, $\rho_{\rm saddle}=2$ (the original \Rg setup), we test the effect of different values of $f_d$. In \Fig{SMBH_Descent_Mbh}, we show the mass growth of the three most massive SMBHs in the simulation as a function of $f_d$. As a consequence of oscillations in the host potential, we confirm that in the $f_d=0$ case (i.e. without orbital decay) the SMBHs (1) barely accrete any mass between their birth with a seed mass of $10^8\Msun$ at $z\simeq 7$ and $z\simeq 2$, as well as (2) cannot grow through mergers as those become unlikely as well. This is in stark contrast to the simulations with $f_d>0$ where we see dramatically boosted mass growth.
In the case of a strong decay with $f_d=1$, SMBHs are essentially pinned to the halo centers at most times. The most massive SMBH accretes gas at high redshift and experiences, below $z=5$, frequent mergers which leads to a very massive central SMBH by $z=2$. However, the rapid mass growth of this central SMBH at high redshifts is also responsible for a very efficient and early AGN feedback (which peaks at $z\sim4$), thus effectively strangulating the growth of the other black holes. 

To reach middle ground between the artificial pinning and the large swinging of black holes, we found $f_d=0.1$ to yield reasonable results, but more detailed investigations to tune this parameter might be helpful in the future.  In fact, observations of AGNs in dwarf galaxies show that BHs are not located at the centers of their host galaxies with an offset between tens of parsecs and a few kiloparsecs \citep[e.g.][]{Shen2019,Reines2020,Mezcua2020}. A detection of an isolated stellar-mass black hole located $\sim$1.6\,kpc away from the galactic center of the Milky Way has been recently reported by \citet{Sahu2022}. Recent simulations \citep{Pfister2019,Bellovary2021,Bellovary2019,Boldrini2020, Ma2021} show that BHs in dwarf galaxies are expected to be wandering around the central regions after the occurrence of mergers or due to tidal stripping or dynamical friction heating. We observe in the $f_d=0.1$ case a steadier mass growth of SMBHs which is mainly driven by accretion of gas down to $z\sim3$. As a result the ICM is heated more gradually by AGN activity, cold gas clumps can form and be later accreted onto SMBHs. As a consequence, we observe at $z\lesssim3$ a boosted gas accretion in the less massive black holes, compared to the simulation with $f_d=1$, by almost an order of magnitude.

\begin{figure}
    \includegraphics{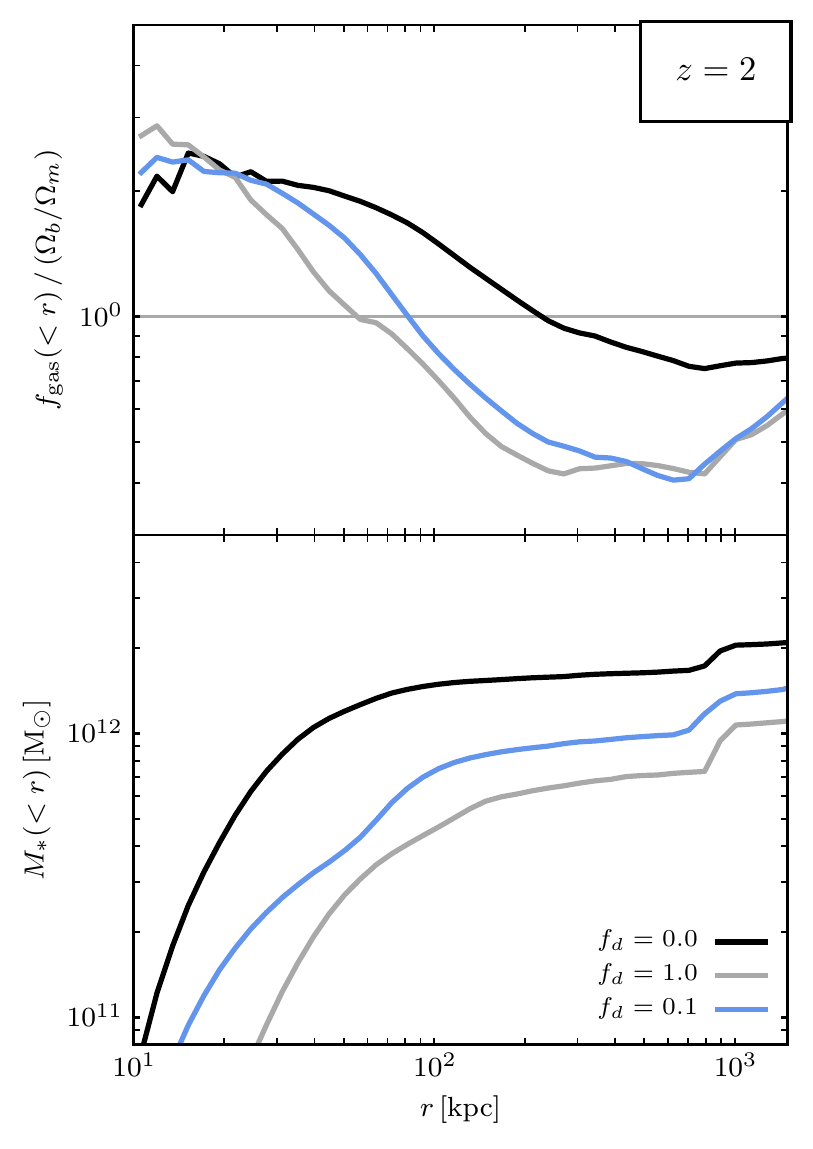}
    \caption{Gas depletion (top) and stellar mass (bottom) radial profile for the simulations with no (black, $f_d=0$), strong (grey, $f_d=1$) or mild (blue, $f_d=0.1$) tidal descent measured at $z=2$. We show the universal baryon fraction, $\Omega_{\rm b}/\Omega_{\rm m}$, with the horizontal grey line. We notice that simulations including the SMBH orbital decay (grey and blue) show lower amount of gas in the ICM ($\sim$40\,per cent). The greater is the SMBH decay, the stronger is the stellar mass reduction inside the proto-cluster ($40$\,per cent for $f_d=0.1$ and $60$\,per cent for $f_d=1.0$).}
    \label{SMBH_Descent_fg_Ms}
\end{figure}

\paragraph*{Impact on gas and stars. } 
Finally, in \Fig{SMBH_Descent_fg_Ms}, we show the gas depletion profile as well as the cumulative stellar mass in the proto cluster region at $z=2$. At this time, the proto-cluster has a virial radius of $\sim$500\,kpc. Clearly, in the $f_d=0$ case, the AGN has not heated the proto-ICM leading to a very high gas fraction at all radii. With enforced orbital decay, the AGNs become active and we observe as a consequence a stark reduction of the gas fraction. Thanks to the tidal descent of SMBHs, AGN feedback is able to deplete the gas from the central region and efficiently offset radiative losses in the forming proto-cluster. In the lower panel of \Fig{SMBH_Descent_fg_Ms}, we can see the resulting reduction of the stellar mass formed. We observe, inside the virial radius, a reduction by a factor of $40$ and $60$\,per cent for the simulations with  $f_d=0.1$ and $f_d=1$ respectively. This result is largely consistent with the recent work of \cite{Bahe2021} who find that the magnitude of the AGN feedback suppression depends on the `drift speed' towards the center, where a slower SMBH drift speed toward the halo center in their case also leads to systematically higher stellar masses.

This new sub-grid model is a first step towards a more physical solution to the `sinking problem' of SMBH in numerical simulations. We studied here, the dramatic effect it can have on the SMBH mass growth, hence AGN activity, which induces amplified gas depletion and a greater star formation quenching. The dimensionless $f_d$ parameter was tuned to reproduce the observed values of stellar masses.\\
We note here, that in our \simu simulation benefiting from this new sub-grid model, the \citet{Booth2009} boost (used in the previous \Rg simulations) was dropped. Indeed, the found SMBH accretion rates are already high enough once the sink particles are more stably confined to the gas rich centre of haloes.

\subsection{Delivering AGN feedback}
\label{sec:agn}

AGN feedback is believed to proceed in two distinct modes \citep{Best2012}. The quasar mode (or thermal) feedback occurs when the gas accretion is comparable to the Eddington limit. A large amount of radiation is emitted from the accretion region which is able to photoionise and heat the gas in the BH vicinity.
In contrast, the radio mode (or kinetic) feedback, preferentially triggered during low-accretion-rate episodes, drives powerful well-collimated radio-emmiting jets coinciding with cavities in the X-ray emission \citep{McNamara2007,Fabian2012}. In some cases both mechanisms can be found in the same object \citep[i.e., radio-loud quasar, see e.g. ][]{Banados2021}.  

\subsubsection{Implementation}
Thermal feedback is usually implemented in astrophysical codes through the injection of energy or momentum in the surrounding gas \citep[e.g.][]{Schaye2015,McCarthy2017,Tremmel2017}. Radio-mode feedback is often implemented as a second sub-resolution feedback channel once the accretion rate falls below a threshold value \citep[e.g.][]{Dubois2014,Weinberger2017,Henden2018}. In this work, we focus purely on thermal feedback and will come back to the impact of kinetic AGN feedback in future work.

Once an AGN event is triggered, in the thermal feedback model, the released energy is assumed to thermalise within the `sink sphere' (defined as a sphere of radius 4 high-resolution elements i.e. $4\Delta x$ around the SMBH particle) thus effectively leading to an increase in thermal energy in those cells. Even though these are operations at the resolution level, multiple ways to distribute this energy among those few cells are possible, with important consequences. We will focus on two distinct weighting schemes here.

\paragraph*{Mass-weighted (MW) injection.} Here the total AGN energy $E_{\rm AGN}$ is injected at every fine time step proportionally to the gas mass in a cell $i$ inside the sink sphere as
\begin{equation}
    E_{{\rm AGN},i} = E_{\rm AGN}\frac{\rho_i \Delta x_i^3}{\sum_i \rho_i \Delta x_i^3},
\end{equation}
In this case, energy is preferentially injected in denser regions (with shorter cooling times). The MW scheme predominantly heats the accretion region fuelling the central SMBH growth and thus reduces future accretion.

\paragraph*{Volume-weighted (VW) injection.} Here the AGN energy is injected at every fine time step proportionally to the volume of the cell $i$ inside the sink sphere as
\begin{equation}
    E_{{\rm AGN},i} = E_{\rm AGN}\frac{\Delta x_i^3}{\sum_i \Delta x_i^3},
\end{equation}
Compared to the MW scheme, here relatively more energy is given to lower density cells, which for a given energy, leads to a higher cell temperature. As a result stronger outflows through lower density regions can be driven, and the immediate accretion gas supply is less affected.

\begin{figure}
    \includegraphics{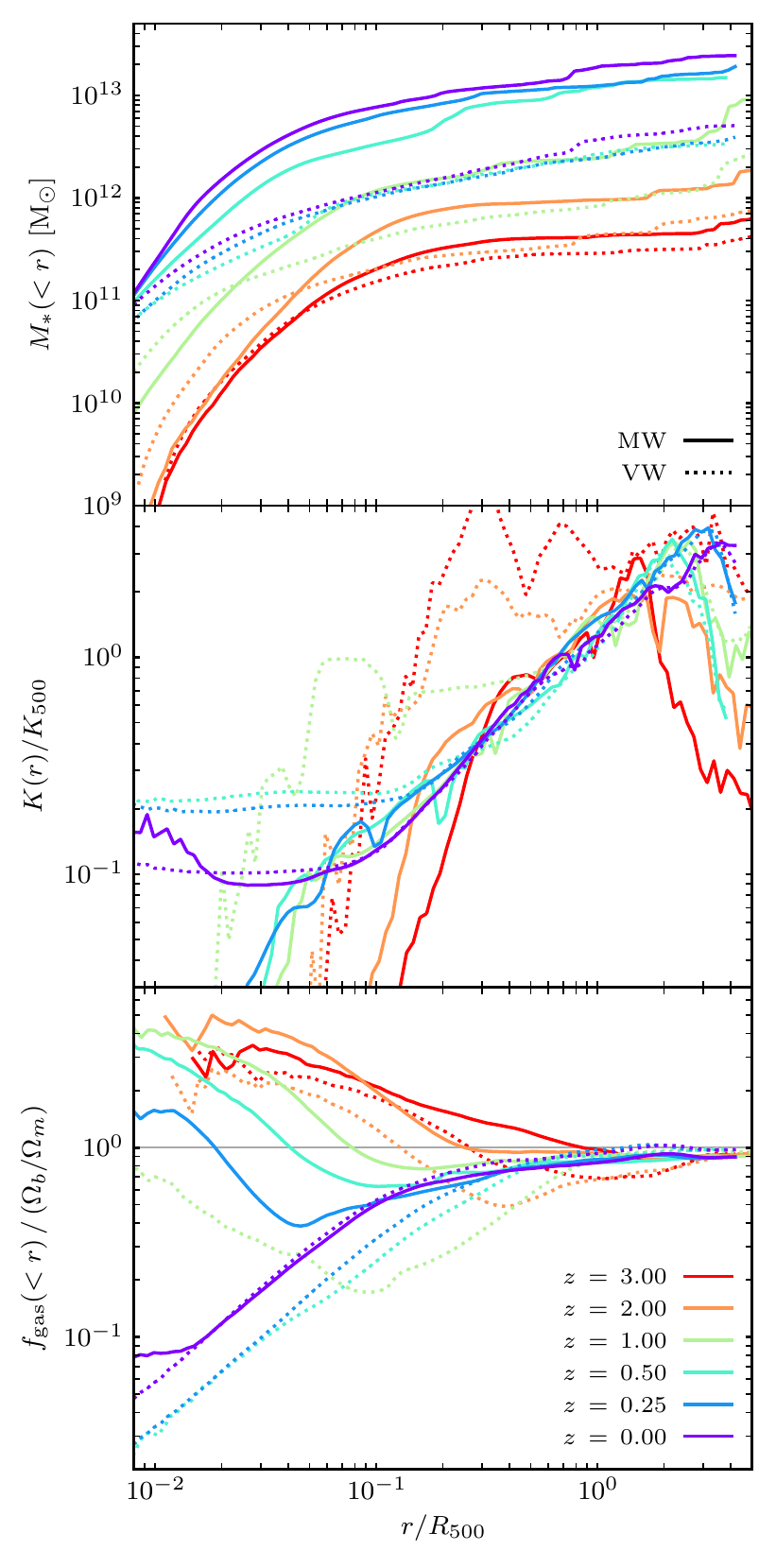}
    \caption{Total enclosed stellar mass (top), ICM entropy (middle) and enclosed gas fraction (bottom) radial profiles. Radii have been normalised to $\Rfiveh$ corresponding to the radius enclosing 500 times the critical density at the indicated redshift (we have $\Rvir \sim 1.7 - 2\Rfiveh$). Solid and dotted lines show the radial profiles of the MW and VW simulations respectively and line color indicates the redshift at which the radial profile is computed. The entropy profiles has been rescaled by the self-similar value $K_{500}$ of \citet{Nagai2007} to compare profiles at different redshifts more easily. The universal baryon fraction is shown in the bottom pannel by the horizontal grey line. We see the greater effect of the AGN with a volume weighted energy deposition in quenching star formation at all radii. The entropy profiles indicate that the  VW AGN is more efficient in heating the ICM up to large radii at $z>1$. It also allows an earlier transition to a NCC cluster by $z=0.5$, while the MW simulation still shows in the core low entropy and  high $f_{\rm gas}$ values.}
    \label{fig:AGN_weights}
\end{figure}

\subsubsection{Validation in simulations} 
\label{sec:MWVW}
In \Fig{fig:AGN_weights} we show the effect of the mass-weighted (MW) compared to the volume-weighted (VW) energy injection on both the thermodynamics of the intra-cluster gas and the stellar content of the cluster over cosmic time. In both simulations, the energy accumulation threshold $\Delta T$ is kept constant (cf. \Tab{tab:sims}). We find that the stellar content in the cluster has been reduced by a factor of $\sim 6-7$ at $z=0$ ($\Rvir\sim 2\Rfiveh$) for the simulation using the VW AGN feedback model compared to the MW model (top panel of \Fig{fig:AGN_weights}). 

Regarding the ICM, the volume-weighted entropy profiles of the MW or VW simulations differ at all redshifts and become similar only at $z=0$, except in the core ($\sim0.1\Rfiveh$).
Outside the core (i.e. $r>0.1\Rfiveh$), the entropy profiles of the MW simulation do not significantly change out to the virial radius (i.e. $1.6\Rfiveh$ and $1.9\Rfiveh$ at $z=3$ and $z=0$ respectively). At the same time, the VW simulation shows a higher ICM entropy at high redshifts, which drops by $z=0.5$ to values similar to the MW simulation. Similarly, the gas fraction drops below the universal $\Omega_b/\Omega_m$ value for the VW simulation at $z>0.5$ whereas the MW simulation shows a higher amount of gas. The higher entropy and lower gas fraction observed for the VW simulation shows the efficiency of VW AGN deposition at heating the ICM consistent with a strong star formation quenching.

From $z=0.5$, we observe an earlier flattening of the entropy profile in the core of the VW simulation compared to the MW simulation, but both settle with a similar core entropy of $\sim 100\keV\cm^2$ at $z=0$. In other words, the VW AGN feedback model can prevent gas cooling in the core from $z=1$ in contrast to the MW model.
The reason behind is that the MW model deposits the AGN feedback energy preferentially in the dense accretion region and therefore has difficulty to escape from the sink accretion sphere. Meanwhile, the gas continues to cool outside the accretion region and star formation proceeds at high rates. Thanks to the large reservoir of cold gas surrounding the central SMBH, the AGN activity remains energetic down to $z=0$ and can therefore gradually heat the cluster core until it reaches a core entropy comparable to the VW simulation.

\begin{figure*}
    \raggedright
    \includegraphics[width=\textwidth    ]{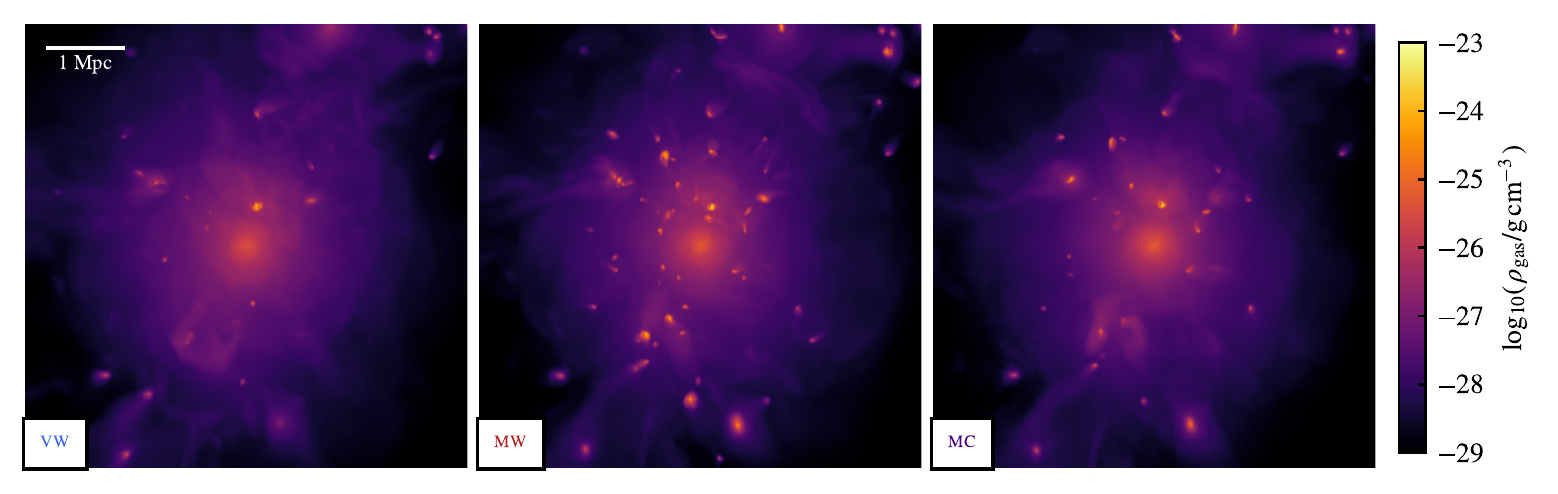}
    \includegraphics[width=0.90\textwidth]{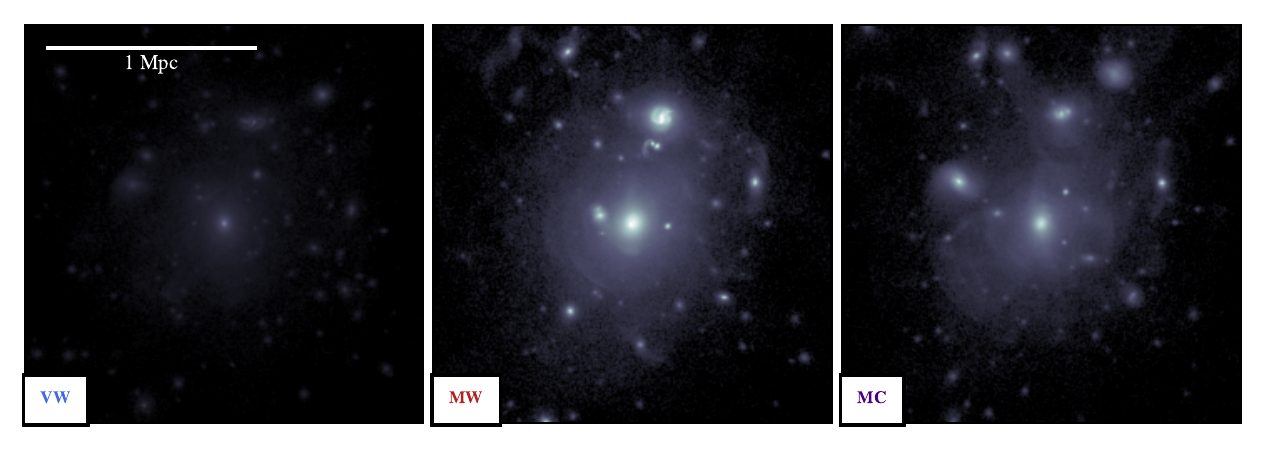}
    \caption{We show the maximal intensity maps of the gas (top panels) and stellar (bottom panels) density in a box of $5.9\Mpc$ (top) and $1.9\Mpc$ (bottom) on the side at $z=0$ respectively. We have from left to right, the VW, MW and MC simulations. We can see the effect of the VW AGN feedback model at removing the cold dense gas clumps compared to the MW model. We can also notice a similar effect, albeit lower, of the anisotropic thermal conduction in the MC simulation at preventing the pile up of gas in dense clumps. As a result, the stellar content in the VW and MC simulations is greatly reduced compared to the MW simulation.}
    \label{fig:models_maps}
\end{figure*}

Despite the relative similarity of the entropy profiles at $z=0$, the evolution of the AGN activity differs significantly. In the MW simulation, the inefficient AGN feedback at early times fails to regulate the star formation in the proto-cluster which leads to over-massive cluster galaxies, whereas in the VW simulation, we see a strong quenching of the star formation at earlier times as a result of the VW deposition injecting more energy in the more diffuse gas.
These differences can be seen in the gaseous and stellar maps at $z=0$ of the VW (left) and MW (center) simulations shown in \Fig{fig:models_maps}. Thanks to the heating at large radii enabled by the VW deposition model, the pile up of cold gas observed in the MW simulations has been prevented. Consequently, the stellar content in the cluster has been greatly reduced in the VW simulation as well as a lower number of cluster galaxies formed.\\

In this study, we do not vary the size of the AGN energy injection, However using a VW deposition scheme, \citet{Dubois2012} showed that the size of the AGN feedback deposition significantly impacts the evolution of the SFR, galaxy and SMBH masses. Indeed, a larger AGN injection region extends to less dense regions, hence far away from the galaxies, which are more easily affected by AGN feedback - which is less the case with a MW deposition.

\subsubsection{On the energy accumulation threshold} 
\label{sec:MWT2}

\citet{Lebrun2014} showed that the entire gas profile can be varied by tuning the energy accumulation threshold of the feedback model of \citet{Booth2009}. In contrast, \citet{Hahn2017} found that none of the AGN models tested on a CC cluster of the \Rg sample had a significant effect outside the core.\\ 
Similarly, we would like to test the robustness of the ICM properties to changes in the AGN energy injection threshold, $\Delta T$, over two orders of magnitude. We emphasise that this parameter does not control the total energy injected in an AGN blast, but only its proportions: a higher value of $\Delta T$ results in a less frequent but more energetic AGN blast and reciprocally, a lower $\Delta T$ value induces more frequent and less energetic AGN blasts.\\

\begin{figure}
    \includegraphics{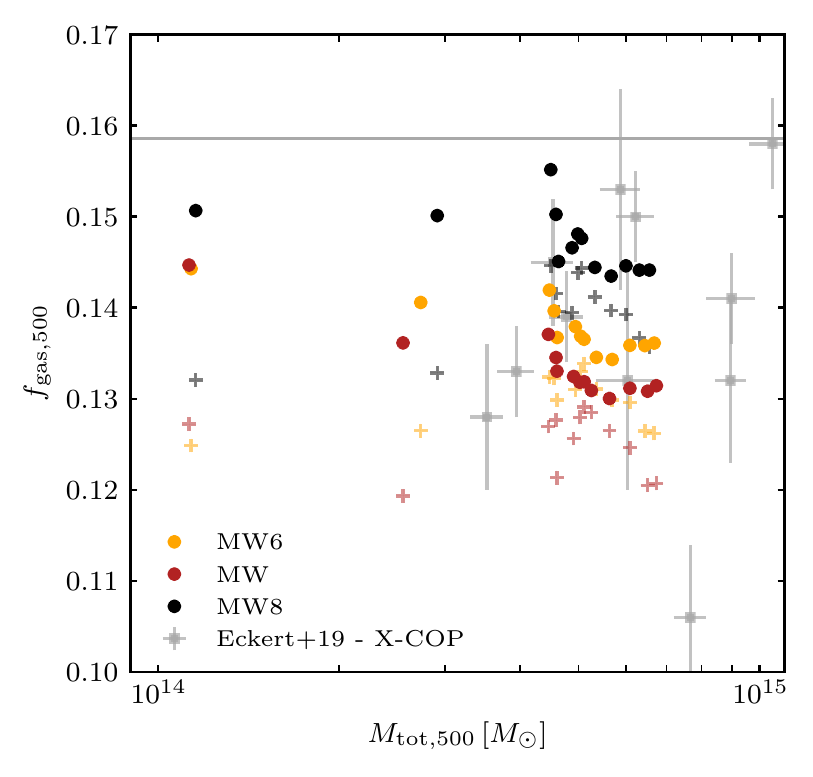}
    \caption{Evolution of the cluster gas fraction as a function of its total mass measured within $\Rfiveh$ for the simulation using a AGN energy injection threshold of $\Delta T=10^6$ (MW6, orange), $10^7$ (MW, dark red) and $10^8\,{\rm K}$ (MW8, black). The circles show the total gas fraction while the lighter-colored crosses show only the X-ray emitting gas fraction (i.e. the gas with $T>0.5\keV$). We compare our data to the hydrostatic gas fractions and total masses corrected for the non-thermal pressure of the X-COP sample \citep{Eckert2019} and show the universal baryon fraction $\Omega_{\rm b}/\Omega_{\rm m}=0.1586$ by the gray horizontal line. 
    Variations of the $\Delta T$ parameter by an order of magnitude, compared to the MW simulation, increase by 10\,per cent at most the gas fraction in the $\Mfiveh>4\times19^{14}\Msun$ range. 
    The trend in gas depletion is observed to be non-linear with respect to $\Delta T$.}
    \label{fig:T2min_fgas}
\end{figure}

We simulate the same halo with the thermal AGN model using a MW energy deposition and change the energy accumulation threshold $\Delta T$. The simulations presented in this work all use $\Delta T=10^7\,{\rm K}$, for this section we ran two additional simulations with $\Delta T=10^6\,{\rm K}$ (MW6) and $10^8\,{\rm K}$ (MW8) that we compare with the fiducial MW simulation presented in the previous \Sec{sec:MWVW}.
We show in \Fig{fig:T2min_fgas}, the evolution of the gas fraction as a function of the cluster mass measured in the $\Rfiveh$ aperture.

We see that that the simulations with higher (MW8) or lower (MW6) $\Delta T$ value shows a systematically higher gas fraction at all cluster masses compared to the fiducial MW simulation. Hence, we observe a non-linear dependence of the gas fraction on the energy accumulation threshold. This is at odds with the findings of \citet{Lebrun2014} who report a decreasing $f_{\rm gas}$ with increasing $\Delta T$ values.\\
With a higher energy accumulation threshold, the AGN feedback in the MW8 simulation show the highest amount of gas. Due to energetic AGN blasts, gas is efficiently depleted from the core region. However, as the AGN blasts are less frequent, the ICM cools efficiently and gas condenses towards the cluster core between consecutive AGN blasts. As a result, a cool core forms which can no longer be impacted by the AGN activity.
With a lower energy accumulation threshold, AGN feedback events are not energetic enough (albeit more frequent) to counterbalance the gas cooling outside the core in the MW6 simulation. Consequently, it results in a greater amount of gas within the $\Rfiveh$ region, compared to the MW simulation.

We see that a higher/lower $\Delta T$ does not systematically lead to smaller/larger gas fractions in such a simulation. Moreover, it does not significantly affect the overall amount of gas in the cluster. Changes in the energy accumulation threshold only induce a variation of 10\,per cent at most of the cluster gas content. This is considerably less compared to the overall 30\,per cent gas fraction reduction observed by \citet{Lebrun2014} when increasing by 0.5 dex the $\Delta T$ value.
This study is consistent with the findings of \citet{Hahn2017} and shows that the $\Delta T$ parameter indeed has little impact of the gas content of GCs in our simulations. 
However, compared to \citet{Hahn2017}, the differences that we observe originate from the inclusion of our SMBH orbital decay model (presented in \Sec{sec:SMBHdescent}) which keeps SMBHs close to their host halo centre, resulting in a greater effect on the gas outside the cluster's core. \\

\subsubsection{Summary on the SMBH modelling} 
In this section, we discussed the response of the stellar and gaseous cluster components to small changes of the SMBH and AGN feedback modelling. The seeding of SMBHs efficiently regulates the star formation in the ICM: less massive SMBH seeds trigger more spatially frequent SMBH formation hence more efficient AGN gas heating and SF quenching.\\
Enabled by our new model using the tidal field information, the orbital decay of SMBH towards the potential minimum can be robustly controlled. SMBH stable orbits enable a greater gas accretion. The resulting enhancement of the AGN activity quenches the SF in the ICM and deplete gas to large cluster radii.\\
Regarding AGN feedback, the energy injection scheme appreciably impacts the ICM gas heating and controls its thermal evolution. When the AGN feedback energy is delivered proportionally to the local gas density, it remains confined to the core region but gradually progresses late to more intermediate radii. With a homogeneous AGN energy injection, more energy is deposited in diffuse gas region and thus can escape the cluster core and heat a large radii without preventing the accretion of cold gas fuelling the central SMBH activity. In that case, the ICM is almost entirely impacted by an early strong heating which prevents the build up of cold gas and SF later on.
On the other hand, the amount of gas is relatively robust to changes in the energy accumulation threshold (i.e. the AGN duty cycle and energetics) of the purely thermal AGN model.


\section{Anisotropic thermal conduction}
\label{sec:atc}

In the simulations of \citet{Hahn2017}, the gaseous content of the \Rg haloes was found to suffer from the over-cooling problem with a too gas-rich ICM. None of their AGN feedback models were able to impact the gas outside the cluster core to bring it towards more realistic values. This suggested that AGN feedback was likely not the sole solution to regulating galaxy cluster thermodynamics. Thanks to improvements in our models, we have seen in the previous section that AGN feedback is now able to impact the gas on large scales but its impact remains extremely dependent to the  choice of parameters.\\
Anisotropic thermal conduction, in conjunction with AGN heating and radiative cooling, likely plays an important role in setting the gas properties of clusters \citep{Kannan2017,Barnes2019}. 
However, in the presence of thermal conduction, the heat buoyancy instability (HBI) \citep{Quataert2008,Parrish2009} can reorient the magnetic field lines to tangential configurations leading to the suppression of conductive heat fluxes \citep{Parrish2009,Bogdanovic2009}. Simulations of \citet{Ruszkowski2011} showed that turbulence can counteract the HBI and re-randomise the magnetic field. The recent work of \citet{Beckmann2022b} showed in idealised massive galaxy cluster simulations that spin-driven AGN feedback cannot counteract alone the HBI in the cluster center and suggested that volume-filling turbulence is needed to restore significant thermal conduction. However, in isolated CC cluster simulations, \citet{Yang2016} found that magnetic tension can suppress a significant portion of the HBI-unstable modes which completely inhibit or significantly impair the HBI for realistic field strengths on scales smaller than $\sim 50-70\kpc$.
Therefore, if thermal conduction is not suppressed by the HBI, it can transport heat to redistribute it in the ICM. It could help alleviating the SMBH/AGN model parameter dependency found in the previous sections and help to reach ICM regulation.\\

The MW simulation shows greater temperature gradients compared to the VW, hence we expect that thermal conduction has a larger effect in that case.
We investigate to what extent anisotropic thermal conduction (ATC) is able to offset radiative losses in the ICM. In idealised adiabatic simulations, we have observed that anisotropic thermal conduction can act as an efficient cooling or heating source depending on the sign of the ICM temperature gradient (see \App{app:atc}) where heat is transported from or to the cluster outskirts in order to flatten out temperature inhomogeneities.
In the previous section, we have seen that a mass-weighted AGN feedback model deposits its energy predominantly in the gas-rich regions. It causes the AGN feedback energy to stay confined close to the central SMBH with difficulty to escape the accretion region. We would like to determine whether ATC can transport the centrally injected AGN energy on large distances to regulate the high cooling losses and star formation rates. \citet{Ruszkowski2011} found that ATC was able to noticeably reduce the effective radiative cooling driven gas accretion in idealised cool core cluster simulations.
In that context, we explore the effect of ATC on the MW simulation with the shortest cooling times in the core. We label the simulation with ATC and a mass-weighted AGN deposition model as MC. 
The effective conductivity is given in \Eq{eq:conductivity} for which we recall the canonical \citet{Spitzer1962} value,
\begin{equation}
    \kappa_{\rm Sp}=n_{\rm e}{\rm k_{\rm B}}D_c,
\end{equation}
where $D_c$ is the thermal diffusivity
\begin{equation}
    D_c = 8\times 10^{31}\left(\frac{k_{\rm B} T_{\rm e}}{10\keV}\right)^{5/2}\left(\frac{n_{\rm e}}{5\times 10^{-3}\,{\rm cm^{-3}}}\right)^{-1}\,{\rm cm^2\,s^{-1}}.\\
\end{equation}

\begin{figure}
    \includegraphics{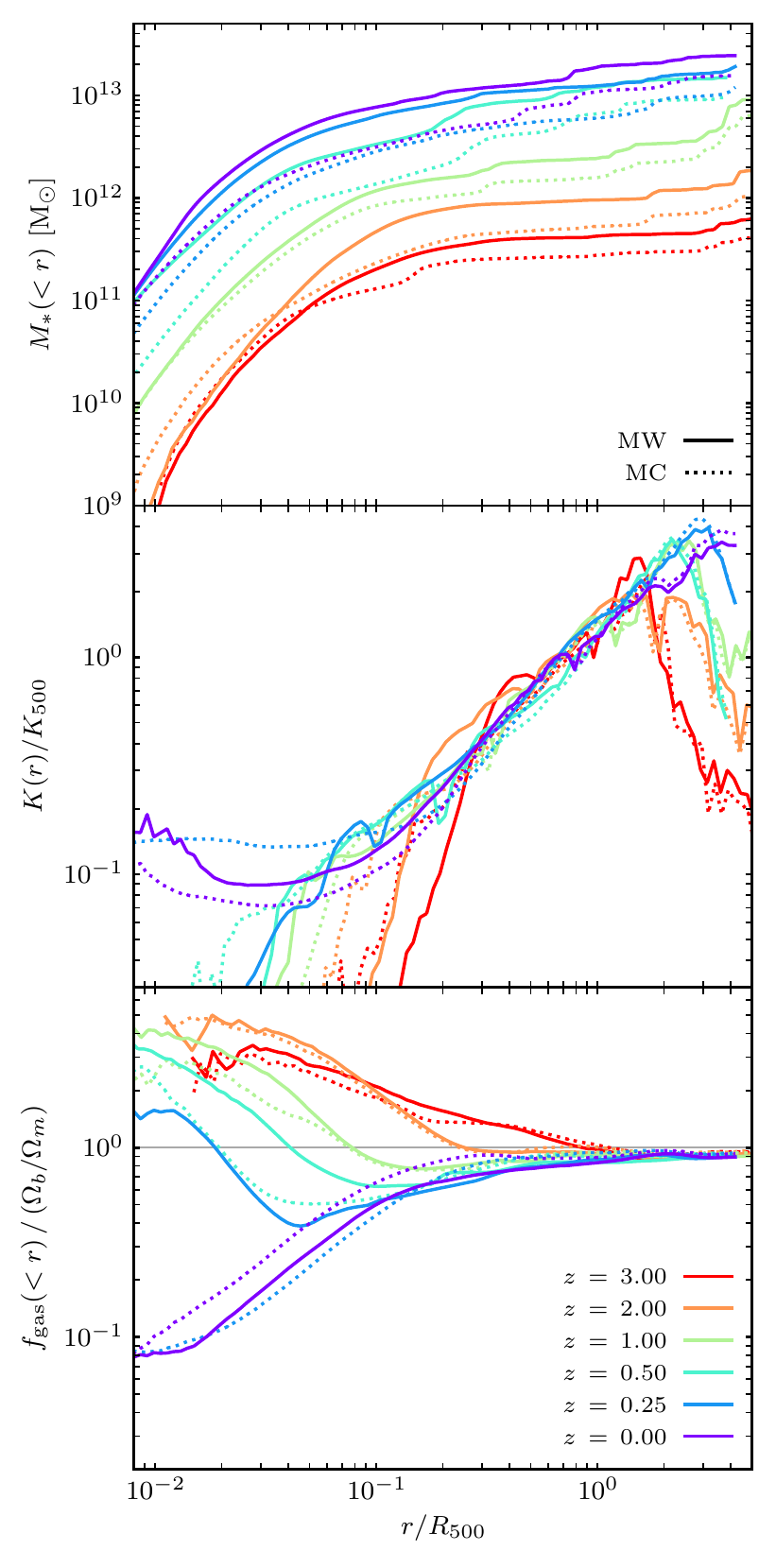}
    \caption{Similarly to \Fig{fig:AGN_weights}, we show the radial profiles for the simulations with and without anisotropic thermal conduction, labelled MC and MW respectively. Anisotropic thermal conduction allows to reduce the stellar content in the intra-cluster medium by a factor of $\sim$2 already by $z=3$ due to the transport of heat in the ICM that prevents the formation of cold gas clumps. However, it also leads to the reduction of the AGN activity which causes a build-up of cold gas at later times, as we can see from the gas depletion profiles at $z=0$.}
    \label{fig:Conduction_profiles}
\end{figure}

We show in the top panel of \Fig{fig:Conduction_profiles} the effect of ATC on star formation and the distribution of gas within the ICM. We also show in the right panels of \Fig{fig:models_maps}, the impact of ATC on the distribution of stars and gas compared to the simulation without ATC (middle panels) at $z=0$. With ATC, we observe the lower amount of gas clumpiness in a more diffuse ICM which hosts less massive galaxies.
As seen in the profiles, the amount of stars in the MC simulation has been reduced by $\sim$40\,per cent already by $z=3$, before the peak of the AGN activity. By smoothing out temperature gradients in the ICM, ATC prevents the formation of cold gas clumps where stars should form. Therefore, the lower amount of stars in the cluster is not due to an enhanced AGN activity but due to a suppression of cold gas clump formation hence star formation. The entropy profiles at $z=3$ shown in the middle panel of \Fig{fig:Conduction_profiles}, reveal a higher core entropy in the MC simulation as ATC transports heat to the central region from the reservoir of hot gas at larger radii. 
In this case, thermal conduction contributes in fact to the suppression of cold gas accretion onto the central SMBH. In consequence, the AGN activity declines and leads to a build-up of cold gas at later times, as we can see from the gas fraction profiles at $z=0$, where the MC simulation shows a higher amount of gas at all radii with a core contraction.\\

Similarly to what we find, the simulations of \citet{Altamura2023} showed a higher cluster core entropy when their artificial conduction model is switched on.
\citet{Kannan2017} found that ATC isotropises the injected AGN energy and enhances its coupling with the ICM. They found that the SFR is hence reduced by an order of magnitude while the overall amount of AGN feedback energy deposited in the ICM is lower. They also show that the earlier quenching comes with an earlier transition to a NCC cluster.
From the gas depletion profiles at $z=0.25$ shown in the bottom panel of \Fig{fig:Conduction_profiles}, we also witness an earlier transition to a NCC cluster where the MC simulation shows a lower amount of gas (and a higher entropy) in the core compared to the MW simulation which does not include conduction. Additionally, we also found lower SFRs in the galaxies within the halo by roughly a factor of two which is however lower than the order of magnitude reduction found by \citet{Kannan2017}.
In spite of these similarities, we do not observe the reported strong quenching induced by a greater AGN heating efficiency. We find that the simulation with conduction shows a lower amount of injected AGN feedback energy by almost a factor of 2. 
In our simulations, conduction reduces the AGN activity by lowering the amount of cold gas available in the ICM which should fuel the SMBH accretion. 
To summarize, heat transport smoothes  temperature inhomogeneities early-on which decreases star formation in the ICM. As a result, less cold gas clumps are available in the ICM for SMBH accretion which results in weakening of the AGN activity. 
Consequently, this decline in AGN feedback heating leads to the contraction of the ICM at low redshifts.


\section{Galaxy properties and ICM profiles}
\label{sec:galandicm}

\subsection{The stellar mass-to-halo mass relation}
\label{sec:smhmrelations}

\begin{figure*}
    \centering
    \includegraphics{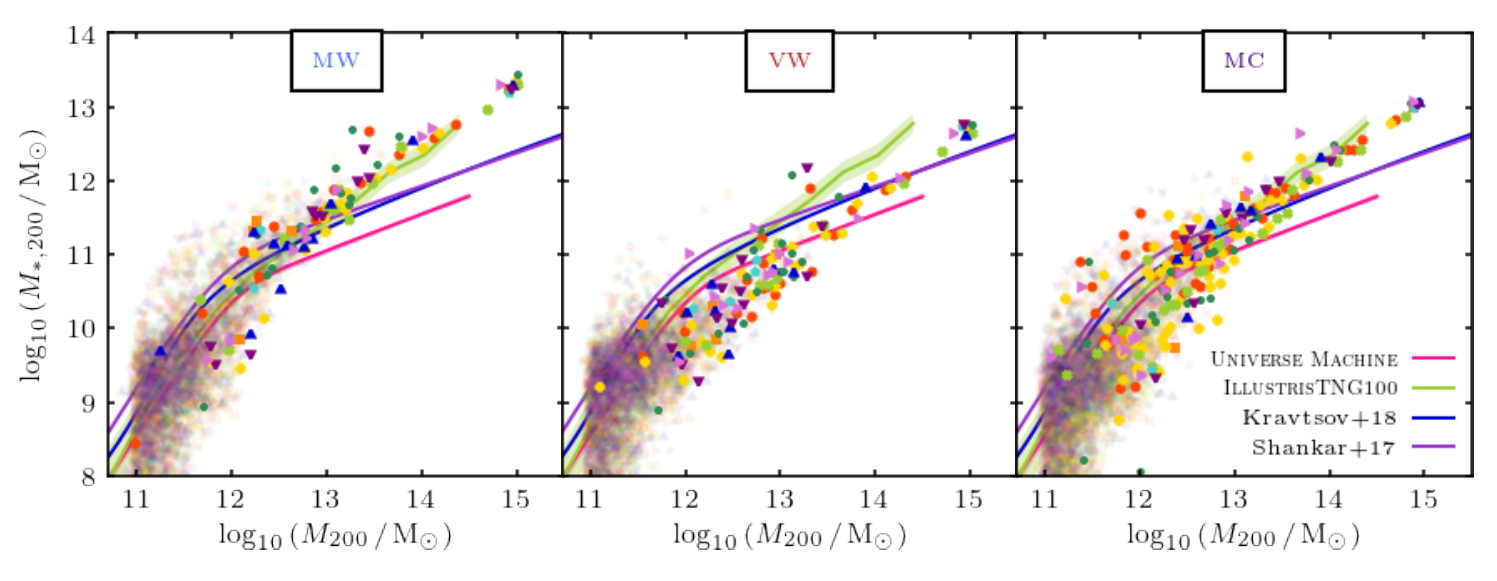}
    \caption{Comparison of the $M_*-M$ relations for the MW (right), VW (middle) and MC (right) simulations for all haloes in the nine \simu clusters at $z=0$. Each color represents one of the nine simulations and we show the central galaxies as well as the satellite populations in transparency. We show the total stellar mass as a function of the total mass within $R_{200}$, the radius enclosing 200 times the critical density. Similarly, we also show the total stellar mass inside $R_{200}$ for the haloes of the {\sc IllustrisTNG100} simulation with the $1\sigma$ scatter ribbon.
    Note that at group and cluster scales (i.e. $M_{200}\geqslant10^{13}\Msun$), the relations of \citet{Shankar2017}, \citet{Kravtsov2018} and {\sc Universe Machine} \citep{Behroozi2019} indicate for lower stellar masses as they only take into account the stellar mass of the central galaxies, while the {\sc IllustrisTNG100} relation and our data provide the total stellar mass in haloes i.e. the central galaxies and the intra-cluster light.
    }
    \label{fig:galaxies}
\end{figure*}

We next study the properties of the galaxies in and around our simulated clusters. Since our simulations have a high-resolution region of twice the virial radius around each cluster at $z=0$, we are probing a wide range of halo mass which enable a statistical comparison.
A rightful test for the realism of our simulations is to compare the stellar mass of our galaxies (both centrals and satellites) as a function of their halo mass with results obtained using the abundance matching technique \citep{Shankar2017}, the {\sc Universe Machine} with the semi-empirical model of \citet{Behroozi2019}, with cluster X-ray mass measurements \citep{Kravtsov2018} and with the {\sc IllustrisTNG100} simulation \citep{Nelson2018,Marinacci2018,Springel2018,Pillepich2018,Naiman2018}.\\
In \Fig{fig:galaxies} we plot, for all haloes found in the nine \simu clusters at $z=0$, the total stellar mass $M_{*,200}$ within $R_{200}$, the radius enclosing 200 times the critical density of the Universe, as a function of the total halo mass $M_{200}$. The halo were identified using the heavily modified version of {\sc Rockstar-Galaxies}, a special version of the original version of {\sc Rockstar} \citep{Behroozi2013}, designed to work with AMR tree and to compute a wide range of observables during the sub-halo finding (for more detail see \citealt{Hahn2017}) .\\

We show the different effect of the energy deposition scheme of our thermal AGN feedback model in the left (MW) and middle (VW) panels as well as the addition of thermal conduction (MC, right panel). We observe a strong reduction of the star formation in the VW simulations by almost an order of magnitude (see \Sec{sec:MWVW}). As the stellar masses are systematically lower for the VW simulations compared to the  {\sc IllustrisTNG100} simulation or the relations of \citet{Shankar2017}, \citet{Kravtsov2018} and \citet{Behroozi2019}, the stellar masses of the MW simulation are larger at $M_{200}>10^{13}\Msun$.
On the other hand, the stellar masses of the MC simulations (left panel), which use anisotropic thermal conduction in contrast to the MW simulation, show stellar masses in good agreement with the various studies. We see the effect of thermal conduction at regulating the star formation in haloes (cf. \Sec{sec:atc}). However, despite efficient gas deletion (cf. right panel of \Fig{fig:scaling_fM}), the VW simulations cannot reproduce realistic galaxy properties since masses are systematically too low.

\begin{figure*}
    \centering
    \includegraphics{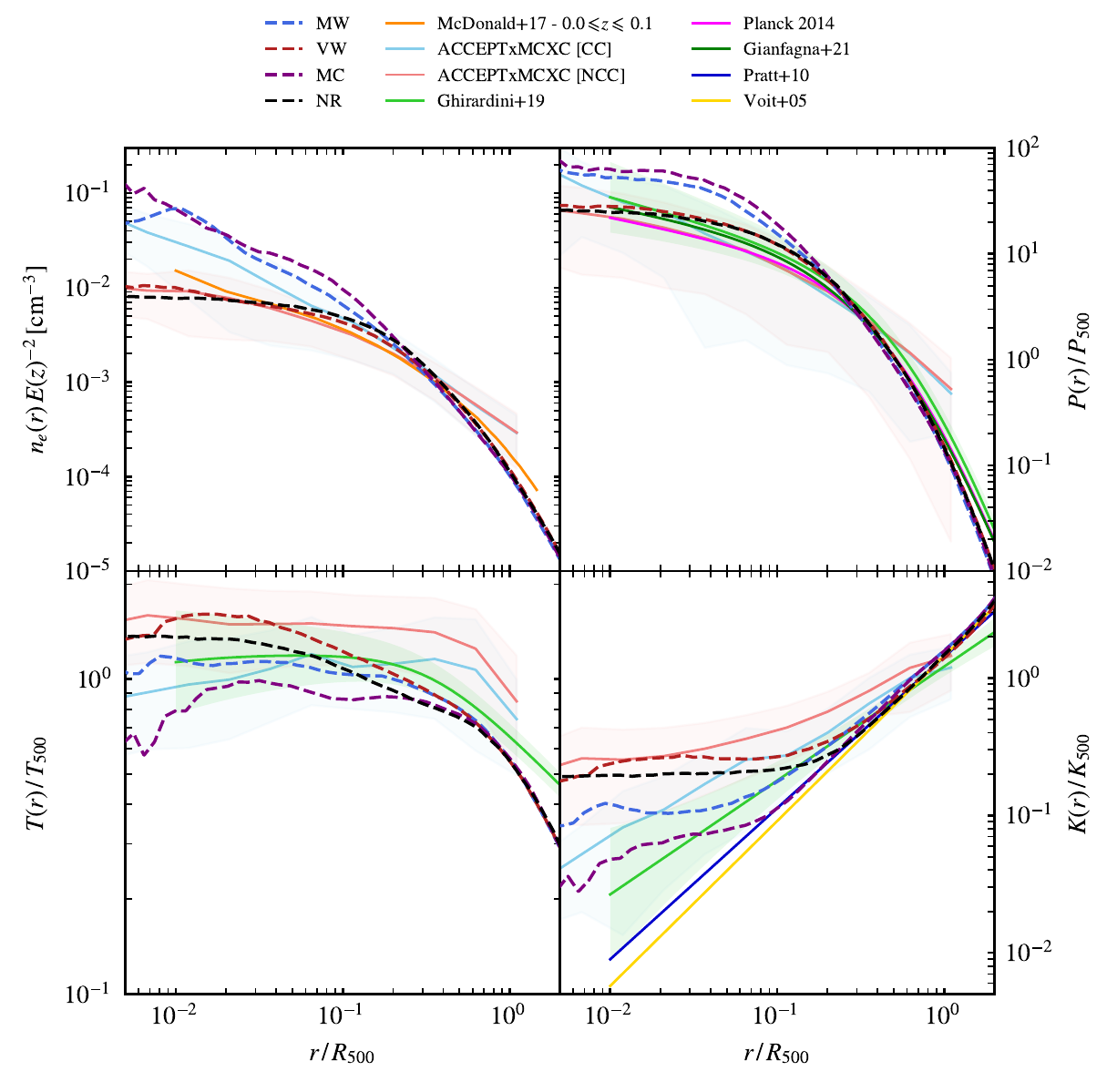}
    \caption{Mean ICM radial profiles of the electronic number density (top left), dimensionless pressure (top right), temperature (bottom left) and entropy (bottom right) of our VW, MW, MC and NR simulations for $z\leqslant 0.5$  in comparison with the matched sample of ACCEPT \citep{Cavagnolo2009} and MCXC \citep{Piffaretti2011} and the sample of \citet{McDonald2017}. We show the best fit thermodynamic profiles of \citet{Ghirardini2019}, the pressure profiles of \citet{Planck2013} and \citet{Gianfagna2021} as well as the outer entropy slopes of \citet{Pratt2009} and \citet{Voit2005}}
    \label{fig:ICMprofiles}
\end{figure*}

\subsection{Structure of the ICM}
\label{sec:ICMprofiles}

We perform a comparison of electron number density, pressure, temperature and entropy in \Fig{fig:ICMprofiles} with numerical simulations, SZ and X-ray observations. We consider stacked profiles over cosmic time of all our clusters for each type of simulations (NR, MW, VW and MC, see details in \Tab{tab:sims}) separately in order to quantify the mean profiles. We selected only snapshots in the $0\leqslant z \leqslant 0.5$ redshift range to be consistent with the studies to which we compare our data.\\

In the top left panel of \Fig{fig:ICMprofiles}, we compare the electron number density profiles with the low-$z$ sub-sample of \citet{McDonald2017} of (X-ray-selected) galaxy clusters at $z=0-0.1$. Additionally, we compare our simulations to the publicly available  data from the ACCEPT Chandra archival project described in \citet{Cavagnolo2009}. From the 242 ACCEPT galaxy cluster profiles and using the right ascensions, declinations and redshifts, we match 141 objects to the MCXC sample \citep{Piffaretti2011} which gives access to  X-ray radius and mass estimates. \citet{Cavagnolo2009} found a bi-modality in the central entropy excess ($K_0$) distribution with two distinct population separated at $K_0 \sim 30-50\,{\rm keV cm^2}$. We therefore classify the ACCEPTxMCXC clusters as cool-core (CC) and non-cool-core (NCC) if the core entropy excess $K_0$ is respectively below or above $50\,{\rm keV cm^2}$.

The simulations using the MW AGN model (MW and MC) show a denser core than the CC population of the ACCEPTxMCXC sample. In contrast, the VW simulations agree almost perfectly with the NCC ACCEPTxMCXC population within the 1$\sigma$ scatter shown by the ribbon. The VW simulations approach the mean radial profile of \citet{McDonald2017}, but have a flatter core electronic density. The non-radiative simulation shows the flattest core mean density profile and is consistent with the NCC ACCEPTxMCXC clusters.
On the other hand, outside the core and especially at $r>0.7\Rfiveh$, our simulations consistently show  a steeper decrease of the electron density with radius.

Our VW and NR simulations are consistent with both the CC and the NCC ACCEPTxMCXC pressure profiles within scatter as shown in the top right panel of \Fig{fig:ICMprofiles}. The VW and NR simulations are in good agreement with the mean pressure profiles of \citet{Planck2013}, the X-COP sample \citep{Ghirardini2019} and the MUSIC simulated clusters \citep{Gianfagna2021} out to large cluster radii. On the other hand, the MW and MC simulations show a factor 4 higher core pressure, but, meet at $r\geqslant0.2\Rfiveh$ the VW and MC profiles.

In the ICM temperature profiles shown in the bottom left panel, both ACCEPTxMCXC clusters and \citet{Ghirardini2019} show large uncertainties. The CC and NCC populations of the ACCEPTxMCXC sample show flat temperature profiles with high ICM temperatures out to $\Rfiveh$. \citet{Ghirardini2019} also show, to a lesser extend, higher temperatures outside the core compared to our simulations.
In the core region, we can see that the VW and NR simulations approach the NCC mean profiles of the ACCEPTxMCXC sample and the MW and MC simulations tend towards the NCC mean profile.

In the bottom right panel, the entropy slope of all our simulations at large radii is consistent with the simulations of \citet{Voit2005} and the observations of the REXCESS sample \citep{Pratt2009}, but is found to be steeper than the work of \citet{Ghirardini2019}. The ACCEPTxMCXC entropy profiles shows a shallower slope with a higher normalisation (consistent with the high ACCEPTxMCXC ICM temperatures). 
The VW simulations agree well with the NCC ACCEPTxMCXC population within scatter while the MW simulations better match the CC sub-sample and the relation of \citet{Ghirardini2019}. The NR simulations are somehow in between but the MC shows low core entropy still compatible the CC ACCEPTxMCXC clusters.\\

Overall, compared to the ACCEPTxMCMC CC and NCC mean profiles, we can see that the simulations implementing a VW AGN feedback model tend to produce GCs with NCCs. On the other hand, the simulations implementing a MW AGN feedback model tend to produce more CC clusters which is even accentuated by the addition of thermal conduction. Our ICM radial profiles demonstrate that our simulations are in relatively good agreement with both observations and state-of-the-art simulations. We note, however, that the MC and MW mean density profiles show a slight excess of gas in the core compared to observations.


\section{Evolution along cluster scaling relations}
\label{sec:sr}

Well-calibrated scaling relations between observed (X-ray, SZ or optical) quantities and the total mass of GCs are not only important to understand the physical processes that give rise to these relations, but are a crucial ingredient for cosmology \citep{Giodini2013}. 
Hydrodynamical simulations can model the complex processes of structure formation with the inclusion of baryonic physics in a cosmological context. 
Having access to the true cluster mass, such simulations can be used to explore possible biases in the mass estimation methods and can help to obtain a definitive measure of the true cluster mass scale to enhance cosmological parameter analysis using cluster counts \citep[see][for a review]{Pratt2019}.
However, it is first important to estimate the degree to which numerical models impact and potentially bias cluster scaling relations before directly confronting them with observational results.
We studied independently in \Sec{sec:agn} and \Sec{sec:atc} the impact of the AGN deposition schemes and the addition of ATC, respectively, on the stellar and gaseous content of a single \simu halo. However, the effect of such numerical schemes on the global properties of the whole \simu sample still needs to be assessed.
In this section, we extend the analysis to the full \simu sample and quantify the changes in the cluster scaling relations with the variation of sub-grid baryonic models.\\

We summarise the simulations details in \Tab{tab:sims}, where we label the various simulations NR, VW, MW and MC for convenience.
In short, all simulations share the same resolution and numerical strategy, with the same modelling for gas cooling, star formation, stellar feedback, black hole seeding and growth (except for the adiabatic run, NR). The only differences are in the AGN energy injection scheme and whether ATC is included.

\subsection{Synthetic X-ray observables}
\label{sec:Xobs}

We chose a different methodology from \citet{Hahn2017} to compute the X-ray observables from the simulation. Instead of using simple weighting schemes, we produce a synthetic X-ray spectrum from which the temperature (and gas density) of the ICM can be estimated as closely as possible to the observer's methodology. For each cell in the range $0.15 \leqslant r/\Rfiveh\leqslant 1$,\footnote{We exclude the core ($r<0.15\Rfiveh$) from the analysis to avoid being biased by the presence of any cool-core or central AGN activity. See in \App{app:ci} for more details.} we read the gas density, temperature and metallicity from the simulation and compute the emissivity $\epsilon(T,Z)$ (atomic lines and continuum) using tabulated emission models from the Astrophysical Plasma Emission Code (APEC, version 3.0.9, \citealp{Smith2001}). In each spectral bin, we compute the photon emission rates\footnote{We omit any redshift or column density dependence} $\phi_i$ of all the cells $i$,
\begin{equation}
    \phi_i = \epsilon(T_i,Z_i) \,n_{{\rm e},i}\, n_{{\rm H},i}\, \Delta x_i^3.
\end{equation}
We sum the individual spectra of all gas cells in the core-excluded region inside $\Rfiveh$ to produce a mock X-ray spectrum (more details on the computation of X-ray observables are given in \App{app:Tx}).
We then fit the obtained spectrum, using MCMC via the {\tt emcee} python library \citep{emcee}, with a single temperature APEC model generated with {\tt PyAtomDB} \citep{Foster2020}. 
The X-ray luminosity is obtained by integrating over the spectra measured from the simulation (not from the best-fit model) in the soft X-ray ($L_{\rm X,500}$) and bolometric band ($L_{\rm X,bol,500}$) i.e. 0.5--$2\keV$ and 0.0--$100.0\keV$ respectively, with no instrument spectral response.

\subsection{X-ray scaling relations}

We use the publicly availabe Bayesian regression scheme {\tt LIRA} of \citet{Sereno2016a,Sereno2016b} for our cluster scaling relation analysis. 
We consider a power-law function of the form $Y\propto10^\alpha X^\beta E(z)^\gamma$ that describes the average scaling relation of a given cluster observable $Y$ with another cluster observable $X$. 
For the fitting procedure, we focus on logarithms of the cluster observables of $Y$ and $X$ which are normalised by their respective pivot values $Y_0$ and $X_0$ :
\begin{equation}
    \log_{10}\left(\frac{Y}{Y_0}\right) = \alpha + \beta\log_{10}\left(\frac{X}{X_0}\right) + \gamma E(z) \pm \sigma_{Y|X},\label{eq:fitform}
\end{equation}
where $\alpha$ is the normalisation, $\beta$ is the slope, $\sigma_{Y|X}$ is the intrinsic scatter of $Y$ at fixed $X$ and $E(z)=H(z)/H_0$ is the expansion function, which describes the evolution of the Hubble parameter with redshift for a given cosmology. We fix its evolution with redshift to the self-similar expectation, i.e. $\gamma=\gamma_{\rm ss}$, and we chose pivot values to be the average sample values which we list in \Tab{table:fittedparams}. We additionally fit the intrinsic scatter of $Y$ at fixed $X$.
As we are using `true' observables computed directly from the simulation, we do not assume any selection effects or prior distributions on the regression parameters.

\begin{table}
\begin{center}
\begin{tabular}{ l | l l | l l }
 \hline
 $Y,X$ & $\beta_{\rm ss}$ & $\gamma_{\rm ss}$ & $Y_0$ & $X_0$ \\ 
 \hline
 $T_{\rm X},M$             & 2/3 & 2/3   & $5.0\keV$ & $5.0\times10^{14}\Msun$ \\
 $L_{\rm X},M$             & 1   & 2   & $4.0\times10^{44}\,{\rm erg/s}$ & $5.0\times10^{14}\Msun$ \\  
 $L_{\rm X,bol},M$         & 4/3 & 7/3 & $1.0\times10^{45}\,{\rm erg/s}$ & $5.0\times10^{14}\Msun$ \\  
 $L_{\rm X},T_{\rm X}$     & 3/2 & 1   & $4.0\times10^{44}\,{\rm erg/s}$ & $5.0\keV$ \\
 $L_{\rm X,bol},T_{\rm X}$ & 2   & 1   & $1.0\times10^{45}\,{\rm erg/s}$ & $5.0\keV$ \\
 $M_{\rm gas,X},M$         & 1   & 0   & $5.0\keV$ & $5.0\times10^{14}\Msun$ \\
 $Y_{\rm SZ},M$            & 5/3 & 2/3 & $40\kpc^2$ & $5.0\times10^{14}\Msun$ \\
 $Y_{\rm X},M$             & 5/3 & 2/3 & $3.0\times10^{14}\Msun{\rm keV}$ & $5.0\times10^{14}\Msun$ \\
 \hline
\end{tabular}
\end{center}
\caption{Scaling relations in the form $Y\propto X{^\beta_{\rm ss}}\,E(z)^{\gamma_{\rm ss}}$ expected from the self-similar theory. We note $Y$ the integrated Comptonization Y parameter and $M$ the total cluster mass, $T_{\rm X}$, $L_{\rm X}$ and $L_{\rm X,bol}$, the X-ray temperature and the soft band and bolometric luminosity. The last two columns list the pivot values used for the fitting the (non-self-similar) scaling relations (\Eq{eq:fitform}).}
\label{table:fittedparams}
\end{table}

\subsubsection{The $f_{\rm gas}-M_{\rm tot}$ relation}
\label{sec:fgas}

\begin{figure*}
    \includegraphics[width=\textwidth]{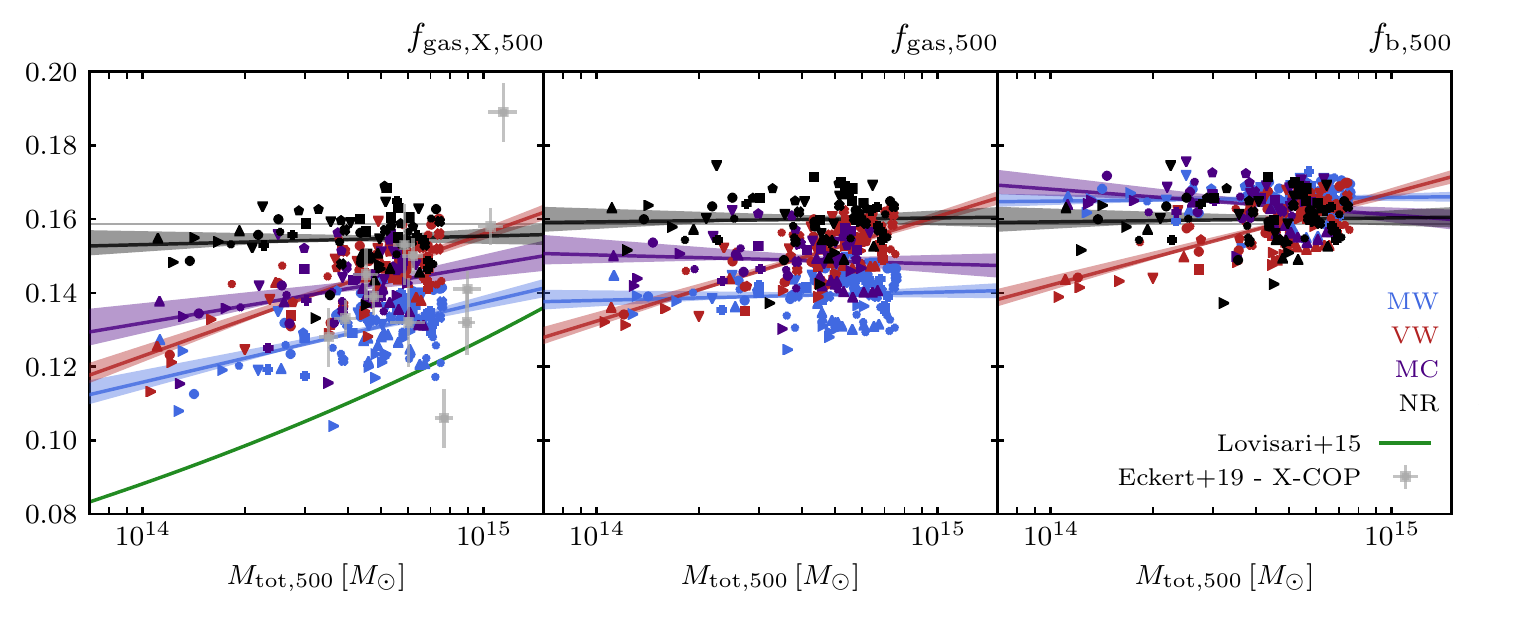}
    \caption{Similarly to \Fig{fig:T2min_fgas}, we show here the fraction of the X-ray emitting gas (left), gas fraction (middle) and baryonic fraction (right) as a function of the mass, all measured within $\Rfiveh$. The horizontal grey line show our cosmic baryon fraction of 0.1586. We distinguish the simulation types (MW, VW, MC or NR) by using different colors (blue, red, purple and black respectively) while individual \simu haloes have different symbols. The colored solid lines and shadding show the fitted relations reported in \Tab{tab:slopesfgas} with the 1$\sigma$ statistical error. The X-ray emitting gas fraction is an increasing function of the total mass for full-physics simulations except in the non-radiative case (NR), which does not implement physical processes that deplete gas from cluster central regions. In the middle panel, when we consider all gas and not only the hot gas, the $f_{\rm gas,500}$ values show a constant evolution with mass, i.e. time, except in the case of the VW simulations. The early and strong AGN heating happening at low cluster masses efficiently depletes gas from the $\Rfiveh$ region and quenches star formation. However, in the VW simulation, due to strong cooling losses at high masses (later times), gas condenses to the cluster centre and the baryonic fraction rises to values comparable to the other simulations (NR, MW, MC). On the other hand, the origin of the offset between the gas fraction of the simulation without (MW) and with ATC (MC) comes from the ability of the thermal conduction to prevent the formation of stars in the ICM, leading to a gas rich ICM. }
    \label{fig:scaling_fM}
\end{figure*}

We start our study of the cluster scaling relations with the ratio of the gas mass to the total cluster mass. This quantity is a crucial ingredient for cosmology because in combination with external information on the baryon density parameter $\Omega_{\rm b}$, it has provided some of the earliest and most robust constraints on the cosmic matter density $\Omega_{\rm m}$ and dark energy \citep[][e.g. for recent measurements]{Ettori2009,Allen2011,Mantz2022}. 
Moreover, a constant gas fraction is a key assumption in the self-similar model of \citet{Kaiser1986} from which the cluster scaling relations ensue but observations indicates for a mass-dependence  \citep{Pratt2009,Lovisari2015,Eckert2016}.

In this section, we discuss how this quantity evolves with mass when different baryonic physical models are considered. Interestingly, we will see in the following sections that when discussing the properties of other cluster scaling relations, their outcomes can already be predicted by the mean of the gas fraction evolution. A detailed characterisation of this dependence will be investigated in future work.\\

We show in the left panel of \Fig{fig:scaling_fM} the X-ray emitting gas fractions measured inside $\Rfiveh$, $f_{\rm gas,X,500}$, of all \simu haloes for all MW (blue), VW (red), MC (purple) and NR (black) simulations as a function of their total mass enclosing 500 times the critical density. This X-ray emitting gas fraction is the ratio of the mass of the hot gas, i.e. with $T_{\rm gas}>0.5 \keV$, to the total mass inside $\Mfiveh$.
We compare our data to the hydrostatic gas fractions and total masses corrected for the non-thermal pressure of the X-COP sample \citep[values are taken from Table 2 of][]{Eckert2019} and with the relation of \citet{Lovisari2015}. At the high-mass end, our simulations are in good agreement with the results of \citet{Eckert2019} but systematically show higher gas fraction than the result of \citet{Lovisari2015}, which use hydrostatic mass estimates.

\begin{table}
    \caption{Fitted normalisation $\alpha$ and slope $\beta$ parameters using {\tt LIRA} for haloes with $z\lesssim1.5$ along their standard deviations. In the bottom part of the table, we give the slopes found by observational studies.}
    \label{tab:slopesfgas}
    \small
    \centering
    \begin{tabular}{ c c c }
\hline
$f_{\rm gas,X,500}$--$\Mfiveh$ & $\alpha$ & $\beta$\Bstrut\\
\hline
MW & $0.873 \pm 0.004$ & $0.145 \pm 0.025$\Tstrut\\
VW & $0.973 \pm 0.004$ & $0.221 \pm 0.021$ \\
MC & $0.951 \pm 0.009$ & $0.103 \pm 0.039$ \\
NR & $1.030 \pm 0.006$ & $0.015 \pm 0.027$ \\[0.15cm]
\multicolumn{2}{r}{\citet{Sun2009}}      & $0.135\pm0.030$ \\
\multicolumn{2}{r}{\citet{Lovisari2015}} & $0.16\pm0.04$   \\
\multicolumn{2}{r}{\citet{Ettori2015}}   & $0.198\pm0.025$ \\
\multicolumn{2}{r}{\citet{Eckert2016}}   & $0.21\pm0.11$\\[0.15cm]
\hline
$f_{\rm gas,500}$--$\Mfiveh$ & $\alpha$ & $\beta$\Bstrut\\
\hline
MW & $0.930 \pm 0.004$ & $0.015 \pm 0.021$ \\
VW & $1.010 \pm 0.003$ & $0.189 \pm 0.018$ \\
MC & $0.991 \pm 0.007$ & $-0.016 \pm 0.031$\\ 
NR & $1.070 \pm 0.006$ & $0.007 \pm 0.026$\\[0.15cm]
\hline
$f_{\rm b,500}$--$\Mfiveh$ & $\alpha$ & $\beta$\Bstrut\\
\hline
MW & $1.100 \pm 0.002$ & $0.007 \pm 0.014$ \\
VW & $1.060 \pm 0.003$ & $0.167 \pm 0.018$ \\
MC & $1.090 \pm 0.006$ & $-0.047 \pm 0.026$ \\
NR & $1.070 \pm 0.006$ & $0.007 \pm 0.026$\\[0.15cm]
\hline
\end{tabular}
\end{table}

Each of the simulation types occupies a different place in the $f_{\rm gas,X,500}-\Mfiveh$ plane.
The NR simulations systematically show the highest $f_{\rm gas,X,500}$ values, as non-gravitational processes that could be responsible for any gas depletion are not included. On the other hand, the radiative runs reveal systematically lower $f_{\rm gas,X,500}$ for lower mass haloes.
The VW runs show the steepest increase with mass to reach comparable values to the NR haloes, but their values are still lower compared to the NR $f_{\rm gas,X,500}$ values. The MW and MC simulations also show positive slopes (although shallower than for VW) with increasing mass to reach different (X-ray emitting) gas fractions values at the highest halo masses, with the MW ones being the lowest.
We list the slopes and normalisations we found in \Tab{tab:slopesfgas} along scaling relations derived in observational studies. The slopes of our MW and VW simulations are in agreement with the slopes  of \citet{Sun2009,Lovisari2015,Ettori2015,Eckert2016}. On the other hand, the MC simulations using thermal conduction show shallower slope but is still consistent with the analysis of \citet{Lovisari2015}. As expected, a zero slope is found for the NR simulation that do not include radiative processes.\\

In the middle panel of \Fig{fig:scaling_fM}, we now plot the gas fraction $f_{\rm gas,500}$ without any cut in the gas temperature. This quantity does not reflect what X-ray observations measure but can help to understand the origin of the different slopes found in the $f_{\rm gas,X,500}-\Mfiveh$ relations. \\
We can see that the NR, MW and MC simulations now show a fairly constant fraction of gas within $\Rfiveh$. MW haloes have $\sim$10\,per cent lower values compared to their non-radiative counterparts, which is not the case if we look at the baryonic fraction $f_{\rm b,500}$ in the right panel of \Fig{fig:scaling_fM}. This suggests that the amount of gas `missing' in the MW simulations, compared to the NR simulations, has been converted into stars as $f_{\rm b}=f_{\rm gas}+f_{\rm stars}$. To a lesser extent, we see the same behaviour for the MC simulations, which indicates that ATC suppresses star formation at the expense of a denser ICM.

However, haloes in the VW simulations, which benefit from early and strong AGN activity, show the steepest increase in both $f_{\rm gas,X,500}$ and $f_{\rm gas,500}$ with mass (i.e. cosmic time). This demonstrates the ability of the VW AGN feedback model to efficiently deplete the gas from the $\Rfiveh$ region in lower mass haloes, where the hot gas can escape more easily in a shallower potential well.
Hence, at earlier times, the ICM temperatures rose to high values that significantly held back both the infall of gas (due to a high central gas pressure) and the formation of cold gas clumps, that can fuel both star formation and SMBH gas accretion. This explains why at low masses (i.e. earlier times), VW haloes also show low baryonic fractions. However, at later times, while this relatively hot gas is slowly radiatively cooling, it condenses toward the cluster centre to meet $f_{\rm b,500}$ values similar to the other simulations (see the right panel of \Fig{fig:scaling_fM}). This indicates that VW haloes host cooling flows at late times consequent to early and efficient AGN activity.

Our radiative simulations suggest a non constant evolution for the X-ray emitting gas fraction $f_{\rm gas,X,500}$ which increase with mass. 
In agreement with the self-similar model, we find a roughly constant evolution of the total gas fractions with mass (i.e.  cosmic time), with the exception of the VW simulations which show efficient gas depletion due to energetic AGN feedback events that deplete gas from the central regions of lower mass haloes. 

At our resolution, no baryons are expelled beyond $\Rfiveh$ in the MW and MC simulations in contrast to the VW simulations.
Although the VW model has an effect on the baryon content of our haloes, the galaxies properties are off (as we can see in \Fig{fig:galaxies}) while galaxies of the MW and MC reproduce more realistic properties.

\subsubsection{The $T_{\rm X}-M_{\rm tot}$ relation}

\begin{figure}
    \includegraphics[width=0.47\textwidth]{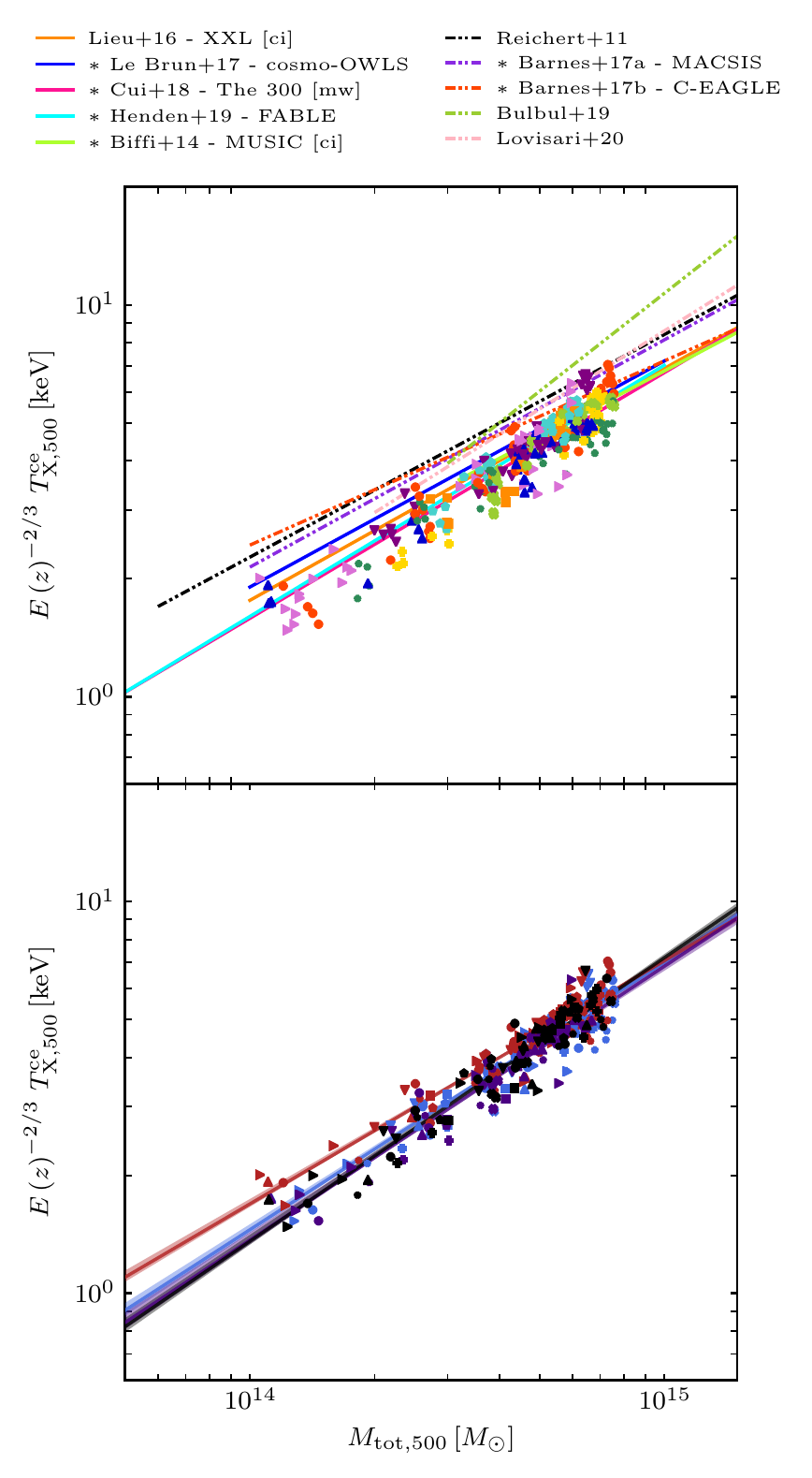}
    \caption{Core excluded X-ray temperature as a function of the total mass within $\Rfiveh$. In the top panel we show the evolution of all our \simu haloes along published scaling relations. We compare our data to the studies using hydrostatic mass estimates or true/weak-lensing masses using dash-dotted and solid lines, respectively. The scaling relations are plotted for the same mass considered in each of these works. In the top legend, we specify in brackets if the measurements include cluster cores (ci) and we indicate simulation works with asterisks. In the top panel, we distinguish each \simu halo with a unique symbol and color, while in the bottom panel, the color stands for the type of simulations with black, blue, red and purple used for the NR, MW, VW and MC simulations, respectively. We plot in the bottom panel the best fit scaling relations obtained for our haloes and give the values of the slopes and intercepts found inside the brackets located in the legend.}
    \label{fig:scaling_TM}
\end{figure}

In \Fig{fig:scaling_TM}, we show a comparison of the X-ray temperatures of the \simu clusters as a function of their mass with various scaling relations from the literature, both simulations and observational studies, plotted in their respective studied mass ranges\footnote{We only show the scaling relation at $z=0$ for the redshift-dependent scaling relations of \citet{Giles2016,Lieu2016,Mantz2016,LeBrun2017,Bulbul2019,Henden2019,Lovisari2020}}. As observational mass estimates may be biased with respect to the true three-dimensional spherical masses in simulations such as ours, we differentiate them in \Fig{fig:scaling_TM} by using dash-dotted lines for studies that use hydrostatic mass estimates. Indeed, we can see that, at a given temperature, hydrostatic mass based studies systematically indicate a lower total mass.

In the top panel of \Fig{fig:scaling_TM}, without making any distinction between the simulation types, we plot the evolution of each \simu halo along published scaling relations as they grow in mass. In the high mass end, i.e. for $\Mfiveh \geqslant 5\times 10^{14}\Msun$, our results are in good agreement with studies using `unbiased' mass measurements i.e. weak lensing mass estimates for observational studies \citep{Lieu2016} or true total masses measured from simulations \citep{Biffi2014,Lieu2016,LeBrun2017,Cui2018,Henden2019}. Accounting for a hydrostatic mass bias of $\sim$20\,per cent brings our data into agreement with the observational studies using hydrostatic mass estimates \citep{Reichert2011,Bulbul2019,Lovisari2020} or simulations estimating mass with a mock X-ray analysis \citep{Barnes2017a,Barnes2017b}. However, a more precise characterisation of the hydrostatic mass bias measured in our simulations will be the subject of a future paper.\\

\paragraph*{Comparison with observations}
Compared to the literature, our simulations indicate slightly steeper slopes than most of the studies with the exception of the observations by \citet{Bulbul2019}. While being consistent in the high mass range ($\Mfiveh\geqslant5\times 10^{14}\Msun$), \citet{Lieu2016} indicates for higher X-ray temperatures in the lower mass range  compared to our results, which might be induced by the inclusion of the cluster core in the X-ray temperature measurements (see \App{app:ci}).

\paragraph*{Comparison with simulations}
Similarly, while being in agreement with \citet{Cui2018} and \citet{Henden2019}, we report systematically slightly steeper slopes compared to the other simulation works but our data agree well within scatter with these studies in the high mass range. The shallower slope of \citet{Biffi2014} could be induced by the inclusion of the core in the temperature estimation (see \App{app:ci}). 
However, we are aware that our spectral fits for the temperature estimation in the low mass range can be slightly biased low (as discussed in \App{app:Tx}), which could explain the steeper slopes we find.
As discussed more extensively in \App{app:Tx}, we show that the mass-weighted temperature estimates are a factor of $\sim$2 lower than the ones resulting from our spectral fit. However, assuming such a factor of 2 lower temperatures shifts the scaling relation of \citet{Cui2018} to even lower temperatures.
\\

\begin{table}
    \caption{Similarly to \Tab{tab:slopesfgas}, we show the fitted normalisation ($\alpha$) and slope ($\beta$) for the $T_{X,500}^{\mathrm{ce}}$--$\Mfiveh$ scaling relation for haloes with $z\lesssim1.5$. In the second and last part of the table, we show the slopes found by the analyses based on numerical simulations and observations respectively, for which we compare our data to in the upper panels of \Figs{fig:scaling_TM}. We highlight both in \Figs{fig:scaling_TM} and this table, the simulation works with asterisks.}
    \label{tab:slopesTX}
    \small
    \centering
    \begin{tabular}{ c c c }
\hline
$T_{X,500}^{\mathrm{ce}}$--$\Mfiveh$ & $\alpha$ & $\beta$\Bstrut\\
\hline
MW & $-0.059 \pm 0.003$ & $0.683 \pm 0.016$\Tstrut\\
VW & $-0.037 \pm 0.003$ & $0.621 \pm 0.014$       \\
MC & $-0.075 \pm 0.005$ & $0.699 \pm 0.021$       \\
NR & $-0.060 \pm 0.004$ & $0.724 \pm 0.018$       \\[0.15cm]
\multicolumn{2}{r}{$\ast$~\citet{Barnes2017a}} & $0.58\pm0.01$ \\
\multicolumn{2}{r}{$\ast$~\citet{Barnes2017b}} & $0.47\pm0.07$ \\
\multicolumn{2}{r}{$\ast$~\citet{Biffi2014}} & $0.56\pm0.03$   \\
\multicolumn{2}{r}{$\ast$~\citet{Cui2018}} & $0.627\pm0.007$   \\
\multicolumn{2}{r}{$\ast$~\citet{Henden2019}} & $0.64\pm0.02$  \\
\multicolumn{2}{r}{$\ast$~\citet{LeBrun2017}} & $0.577\pm0.006$\\[0.15cm]
\multicolumn{2}{r}{\citet{Bulbul2019}} & $0.83\pm0.10$         \\
\multicolumn{2}{r}{\citet{Lieu2016}} & $0.56\pm0.12$           \\
\multicolumn{2}{r}{\citet{Lovisari2020}} & $0.66\pm0.06$       \\
\multicolumn{2}{r}{\citet{Reichert2011}} & $0.57\pm0.03$\Bstrut\\[0.15cm]
\hline
\end{tabular}
\end{table}

We quantitatively compare our scaling relation slopes with the above-mentioned studies in \Tab{tab:slopesTX}.
We observe that the inferred slopes and normalisations are rather insensitive to the physical models used for our simulations as the temperature reflect the depth of the cluster's potential well. The core-excluded X-ray temperatures are similar for simulations with or without ATC (MC and MW respectively). Therefore, thermal conduction does not play an important role in offsetting the X-ray temperatures outside the core in our simulations. 
While the VW simulations indicate a shallower slope with a higher normalisation, they converge to the same core-excluded X-ray temperature values in the high mass range. We see again the efficiency of the VW AGN heating in raising the ICM temperature to higher values, especially in lower mass haloes where the potential well is shallowest.
Most importantly, we see that all simulations converge to the same temperatures with similar scatter for masses above $5\times10^{14}\Msun$. On average, the slopes agree with a $\sim$2\,per cent steeper value than the self-similar expectation and no significant effect of the AGN models or ATC on our $T_{\rm X}-M_{\rm tot}$ scaling relation is seen.\\
Surprisingly, the non-radiative simulations are able to reproduce the same core-excised temperatures as the full-physics simulations. Besides the fact that the temperature is less affected by feedback processes as it reflects more the cluster potential well, this results also indicates that non-gravitational processes mostly affect the core. In the radial range $0.15 \leqslant R/\Rfiveh \leqslant 1$, radiative cooling, thermal conduction, AGN and SF feedback do not play a major role in offsetting the ICM core-excluded X-ray temperatures. We note that only the VW AGN feedback, being the most effective, is able to heat the gas at these radii in the lower mass regime.

From \Tab{tab:slopesTX}, we see that different slopes and normalisations for the $T_{\rm X}-M_{\rm tot}$ scaling relation are found in the literature.
These normalisation differences can be attributed to the method used to infer cluster masses, which might be biased compared to the true mass (e.g. due to the hydrostatic bias or biased weak lensing estimates).
The observational studies of \citet{Sun2009} and \citet{Lovisari2020} showed that the slopes remain consistent for low mass groups to massive galaxy clusters. This consistency implies that non-gravitational processes are not affecting the $T_{\rm X}-M$ scaling relation in a different manner in distinct mass (or temperature) ranges.
\citet{Bulbul2019} actually found the steepest slope in their observations. They explain this apparent tension by the fact that they simultaneously fit the mass and redshift trend of the scaling relation, in contrast to the assumed self-similar redshift evolution in other studies. \citet{Lovisari2020} claimed that it could also be explained if their SPT-SZ masses suffer from a mass-dependent bias (similar to the Planck mass estimates).

We observe slightly steeper slopes compared to the simulation works of \citet{Biffi2014,Barnes2017a,Barnes2017b} and \citet{LeBrun2017} but our results agree with the studies of \citet{Cui2018} and \citet{Henden2019}. The discrepancy could originate from the higher temperature found in lower mass haloes in those works. It can be attributed to the method used to estimates X-ray temperatures (see discussion in \App{app:Tx} and \ref{app:ci}) but also from the efficiency of the feedback model to deplete and heat the gas in lower mass halo.

\subsubsection{The $L_{\rm X}-M_{\rm tot}$ relation}

\begin{figure*}
    \includegraphics[width=0.47\textwidth]{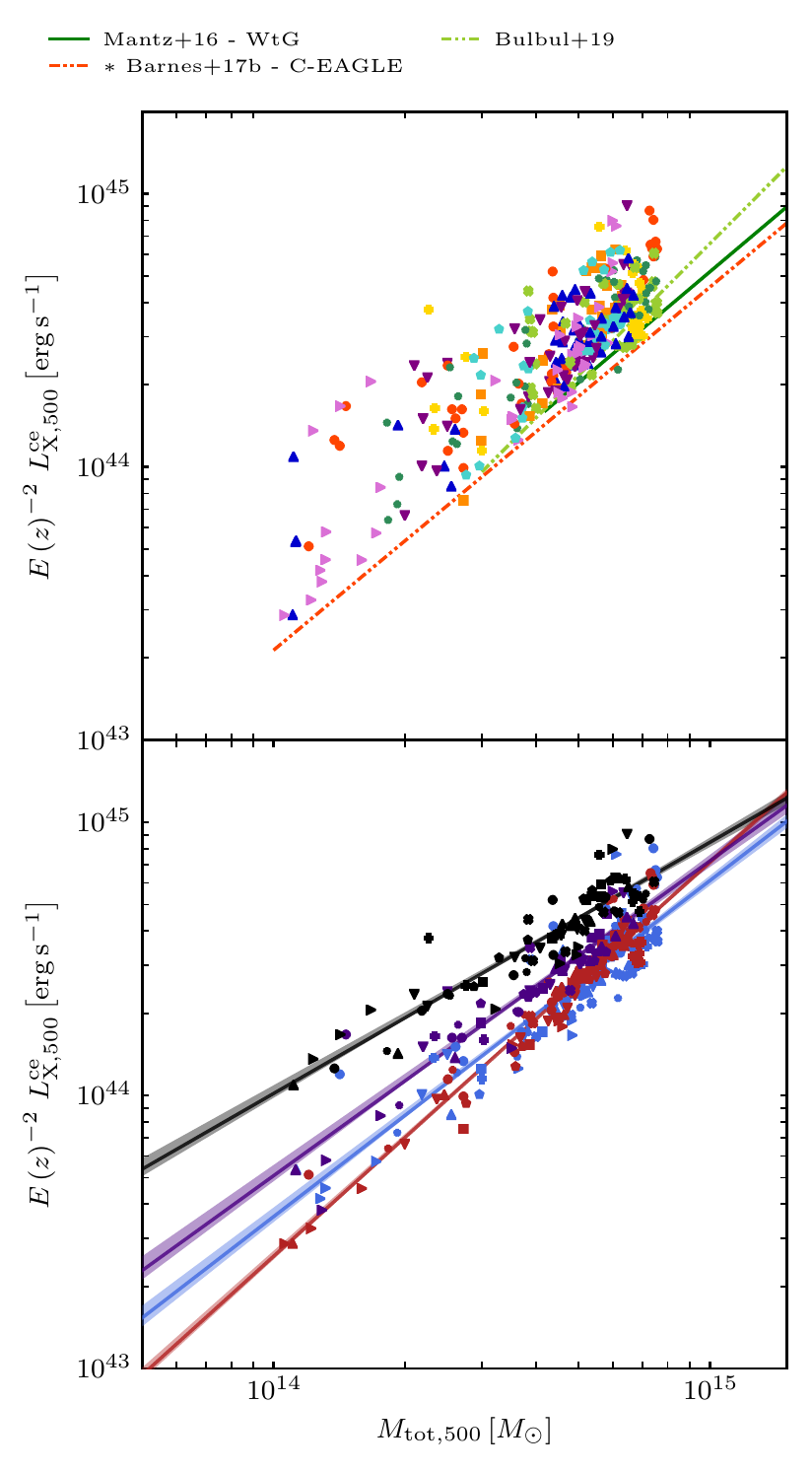}
    \includegraphics[width=0.47\textwidth]{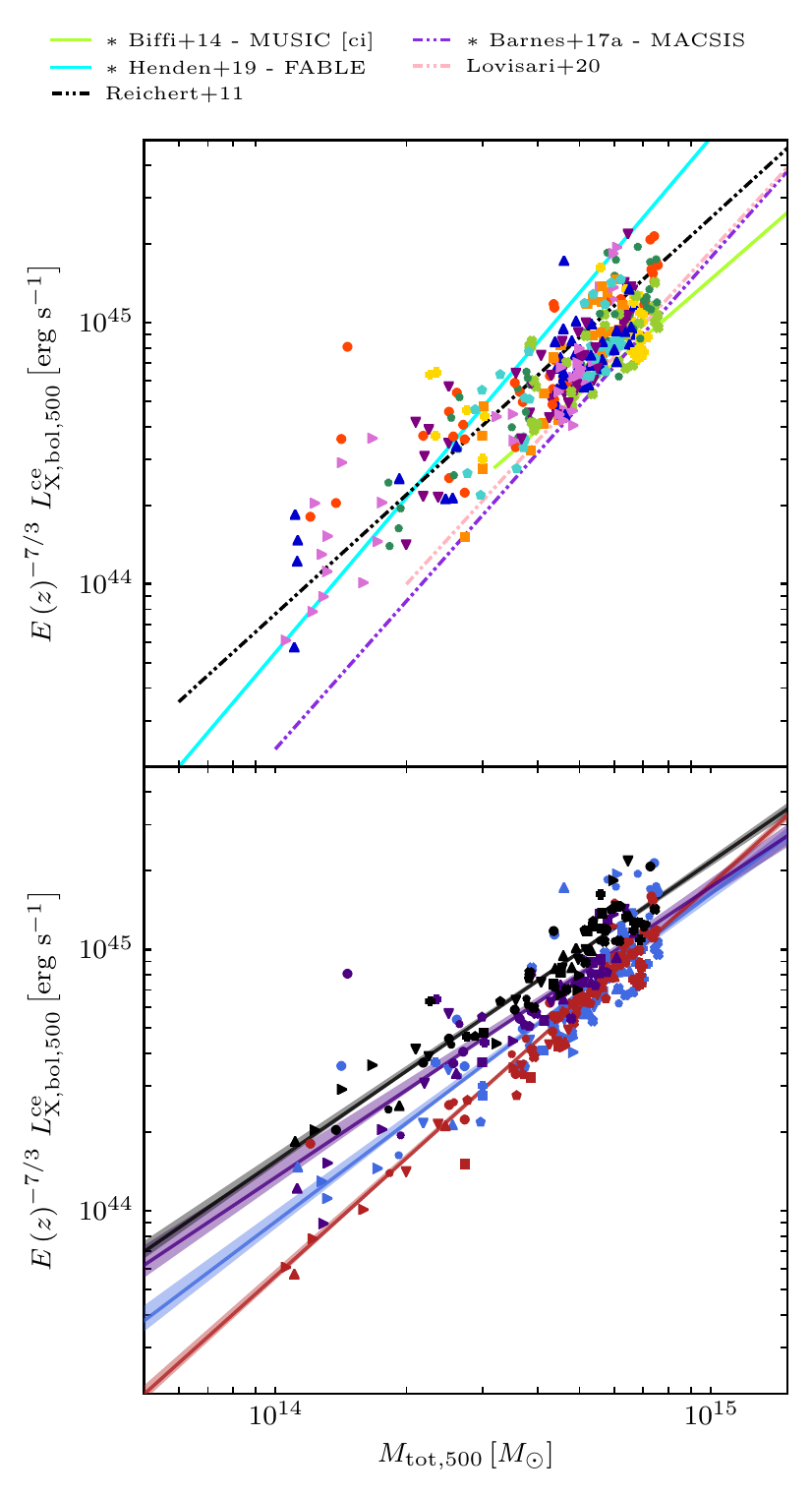}
    \caption{As \Fig{fig:scaling_TM}, but for the core-excluded X-ray soft band (left) and bolometric (right) luminosities measured in the radial range $0.15 \leqslant R/\Rfiveh \leqslant 1$ as a function of halo mass inside $\Rfiveh$. We notice the large differences in slope and intercept between the published scaling relations.}
    \label{fig:scaling_LM}
\end{figure*}

The X-ray luminosity mass scaling relation is important as it can relate one of the `cheapest' X-ray observables to the total cluster mass. 
To fully exploit the data from large galaxy cluster samples provided by X-ray surveys such as e-ROSITA \citep{Liu2021}, which collects too few photons to infer any spectra or construct any mass profiles, it is of great use to have a well calibrated $L_{\rm X}-M_{\rm tot}$ scaling relation and an accurate determination of its scatter. However, the X-ray luminosity measurement depends on the energy band and the aperture from which it is derived as well as the flux extraction method. As a consequence, among all the X-ray scaling relations, it is the one that shows the largest scatter. Moreover, due to its density squared dependence, the X-ray luminosity can be easily biased by the presence of gas-rich substructures, a cool core and non-gravitational processes \citep{Reichert2011} which motivates the exclusion of the core from analyses.\\

In figure \ref{fig:scaling_LM}, we show the core excluded soft-band ($0.5-2\keV$) and bolometric ($0.01-100\keV$) X-ray luminosities as a function of the cluster mass for all our haloes.
The X-ray luminosity is rather sensitive to non-gravitational processes and to the ICM clumpiness, and this can explain why such a diversity of slopes and normalisations is observed. As we can see in \Tab{tab:slopesLXM}, published studies show on average a pronounced deviation, which is on average 50 and 30\,per cent greater than the expected self-similar scaling for soft-band and bolometric luminosities respectively. In the bottom panels of \Fig{fig:scaling_LM}, we make a distinction between simulations that incorporate galaxy formation physics (MW, VW, MC) and those without (NR). 
The X-ray luminosity is more sensitive to the physical models used in the simulations, hence on radiative processes, compared to the temperature that do not show such large discrepancy between the different simulations. Therefore, the calibration of scaling relations using the X-ray luminosity is more complex than relations using the X-ray temperature or the X-ray analogue of the Sunyaev-Zeldovich $Y$ parameter (see later in \Sec{sec:Ysz}). 

\begin{table}
    \caption{Similarly to \Tab{tab:slopesTX} for the scaling of  $L_{\rm X,500}^{\mathrm{ce}}$ and $L_{\rm X,bol,500}^{\mathrm{ce}}$ with $\Mfiveh$. The scaling relations listed here are shown in \Fig{fig:scaling_LM}. In constrast to observations, we denote numerical works with an asterisk.}
    \label{tab:slopesLXM}
    \resizebox{0.48\textwidth}{!}{
    \centering
    \begin{tabular}{ c  c c | c c }
\cline{2-5}
  & \multicolumn{2}{c|}{$L_{\rm X,500}^{\mathrm{ce}}$--$\Mfiveh$} & \multicolumn{2}{c}{$L_{\rm X,bol,500}^{\mathrm{ce}}$--$\Mfiveh$}\Tstrut\\
  & $\alpha$           & $\beta$                                  & $\alpha$           & $\beta$\Bstrut\\
\hline 
MW & $-0.186 \pm 0.007$ & $1.230 \pm 0.038$                       & $-0.165 \pm 0.009$ & $1.255 \pm 0.049$\Tstrut\\
VW & $-0.182 \pm 0.004$ & $1.440 \pm 0.023$                       & $-0.203 \pm 0.005$ & $1.497 \pm 0.028$\\
MC & $-0.090 \pm 0.009$ & $1.150 \pm 0.043$                       & $-0.095 \pm 0.014$ & $1.112 \pm 0.062$\\
NR & $0.048 \pm 0.007$  & $0.920 \pm 0.033$                       & $-0.010 \pm 0.007$ & $1.146 \pm 0.033$\Bstrut\\[0.15cm]
\hline 
  & $\ast$~\citet{Barnes2017b} & $1.33\pm0.13$                   & $\ast$~\citet{Biffi2014}   & $1.45\pm0.05$\Tstrut\\
  &                            &                                 & $\ast$~\citet{Barnes2017a} & $1.88\pm0.05$\\
  & \citet{Mantz2016}  & $1.65\pm0.14$                           & $\ast$~\citet{Henden2019}  & $1.97\pm0.10$\\
  & \citet{Bulbul2019} & $1.60\pm0.17$                           &                            &        \\
  &                            &                                 & \citet{Reichert2011} & $1.52\pm0.04$\\
  &                            &                                 & \citet{Lovisari2020} & $1.82\pm0.25$\Bstrut\\
\cline{2-5}
\end{tabular}}
\end{table}

\paragraph*{Comparison with observations}
For the soft-band luminosity, we systematically find shallower slopes compared to the relations of \citet{Mantz2016} and \citet{Bulbul2019}. With the exception of the NR simulations which show a 25\,per cent higher value, our slopes for the bolometric luminosity do not significantly change from the ones derived using the soft-band luminosities.
With the exception of the VW simulations which agree within scatter with the slope of \citet{Reichert2011}, we find even greater discrepancy with observation for the bolometric luminosity. 
When setting the redshift evolution of the scaling relation to the self-similar value (which is the choice we have made for this work), \citet{Lovisari2020} find a slope of $1.45\pm0.10$ which agrees only the VW simulations.\\
If we account for a mass bias of 20 percent (which is also shown to be a good fit for the $T_{\rm X}-M$ scaling relation), we can bring our data in agreement with the study \citet{Reichert2011} which use hydrostatic mass estimates but widen the gap with the scaling relation of \citet{Bulbul2019} and \citet{Lovisari2020} which show lower luminosities (higher mass) at fixed mass (X-ray luminosity). We observe a higher normalisation than \citet{Mantz2015}.

\paragraph*{Comparison with simulations}
Our radiative simulations have slopes that are consistent with the simulations of \citet{Barnes2017b} for the soft-band luminosity but only the VW simulations  agree with the slope of \citet{Biffi2014} for the bolometric luminosity. \citet{Barnes2017a} and \citet{Henden2019} find 50\,per cent steeper slopes compared to our radiative simulations.
The {\sc Macsis} simulations \citep{Barnes2017a} indicate a 50 per cent steeper slope on average compared to our radiative simulations. Some of this discrepancy can be attributed the ability of their AGN feedback model to efficiently heat gas in lower mass haloes, lowering their X-ray luminosity.
Regarding normalisation, assuming a hydrostatic mass bias of 20 percent increases the offset with the {\sc C-Eagle} and the {\sc Macsis} simulations. This discrepancy could originate from the difference in the energy injection of the thermal AGN feedback model (which use different $\Delta T$ values and the number of heated neighbour particles).
The {\sc Music} simulations \citep{Biffi2014} show, at fixed total mass, lower X-ray luminosities compared to our data. \\

By looking at the differences in slope and normalisation between our simulation types, we observe that the X-ray luminosity follows the same trend with mass as the X-ray emitting gas fraction (see the left panel in \Fig{fig:scaling_fM}) with the steepest slope being for the VW simulations, then in descending order, MW, MC and finally NR. This illustrates once more the relation between the X-ray luminosity and the halo gas content, which is driven by the different physical models (AGN and ATC) that the simulations use.
The shallower slope of the NR compared to both the self-similar scaling and the other simulations originates from the absence of galaxy formation physics or radiative gas cooling, which could boost the X-ray luminosity. The higher NR normalisation can be explained by the higher gas content in haloes as no star formation (which should turn cold and dense gas into stars) or feedback processes (which could deplete the gas in haloes) occur.\\

\subsubsection{The $L_{\rm X}-T_{\rm X}$ relation}

\begin{figure*}
    \includegraphics[width=0.47\textwidth]{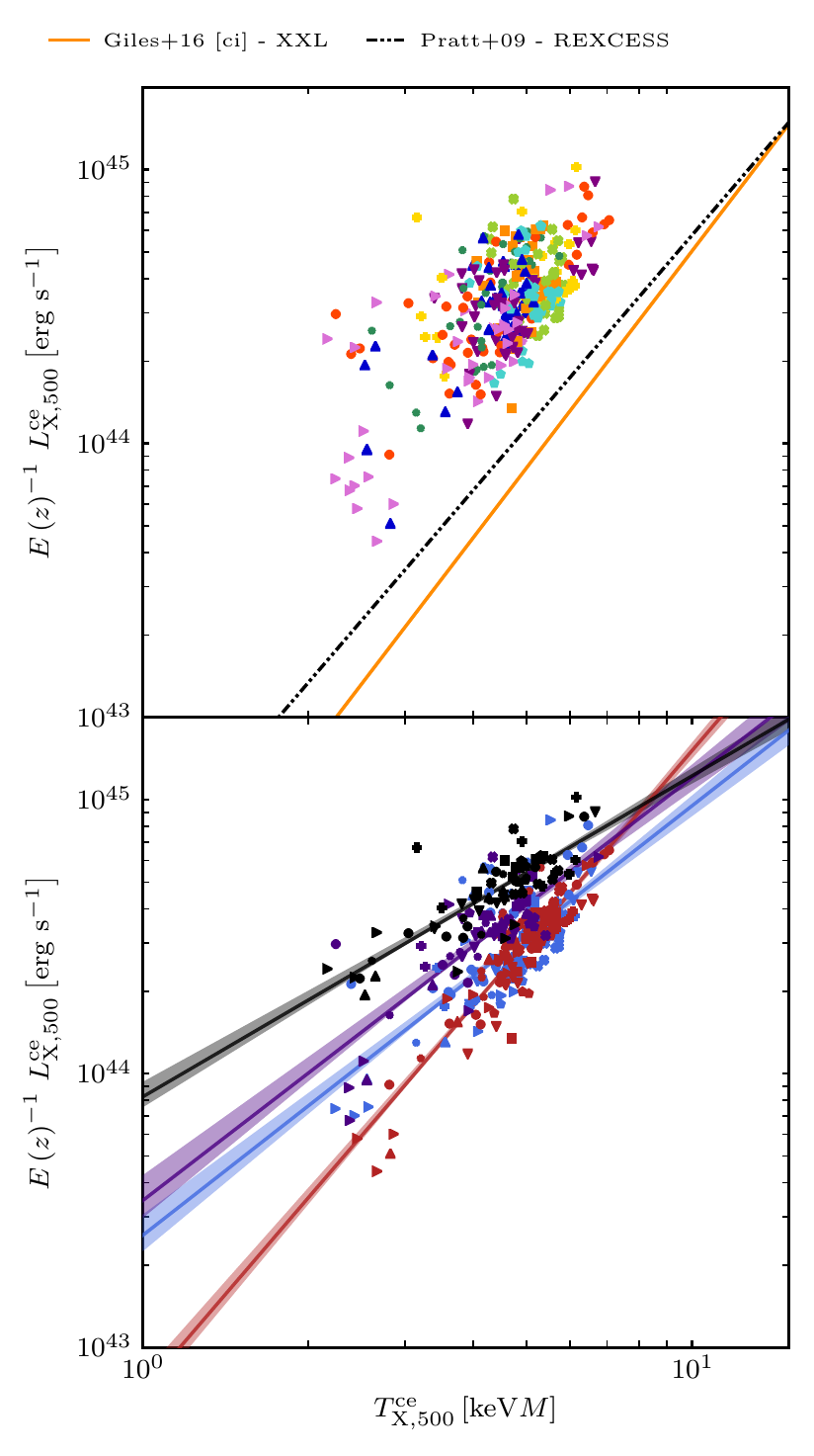}
    \includegraphics[width=0.47\textwidth]{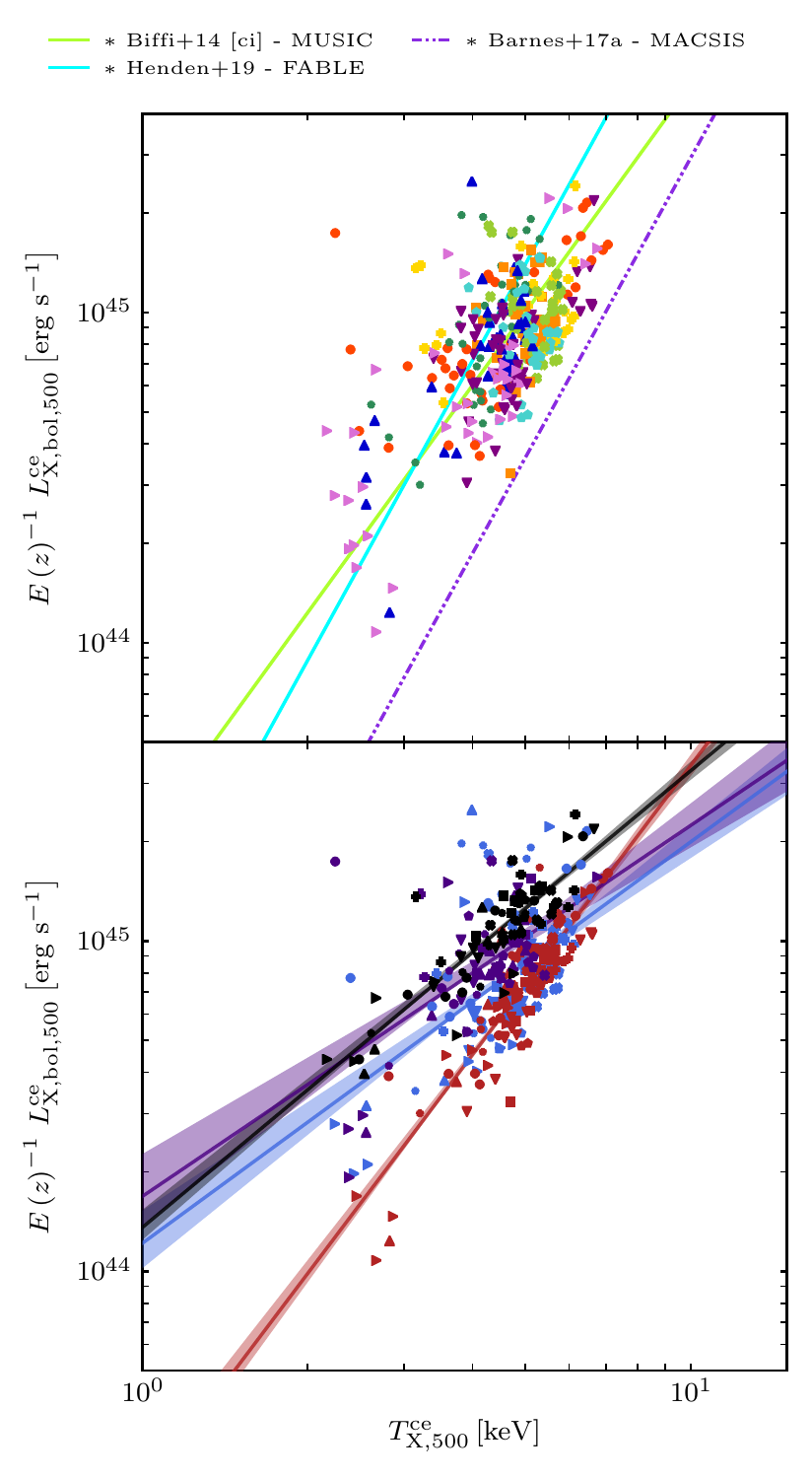}
    \caption{A pure X-ray scaling relation -- the core excluded X-ray soft band (left) and bolometric (right) luminosities as a function of the core excluded X-ray temperature. We keep the same figure properties as in \Fig{fig:scaling_TM}.}
    \label{fig:scaling_LT}
\end{figure*}

We now look at the X-ray scaling relations in \Fig{fig:scaling_LT}, which relates the core-excluded X-ray temperatures and luminosities, and collect their parameters (slope and intersect) in \Tab{tab:slopesLXT}. As for the $L_{\rm X}-M$ scaling relations, we observe a large offset between recent numerical and observational works both in normalisation and slope.

\begin{table}
    \caption{Similarly to \Tab{tab:slopesLXM} for the scaling of  $L_{\rm X,500}^{\mathrm{ce}}$ and $L_{\rm X,bol,500}^{\mathrm{ce}}$ with $T_{\rm X,500}^{\mathrm{ce}}$.}
    \label{tab:slopesLXT}
    \resizebox{0.48\textwidth}{!}{
    \centering
    \begin{tabular}{ c  c c | c c }
\cline{2-5}
  &  \multicolumn{2}{c|}{$L_{\rm X,500}^{\mathrm{ce}}$--$T_{\rm X,500}^{\mathrm{ce}}$} & \multicolumn{2}{c}{$L_{\rm X,bol,500}^{\mathrm{ce}}$--$T_{\rm X,500}^{\mathrm{ce}}$}\Tstrut\\
  & $\alpha$           & $\beta$                                                     & $\alpha$           & $\beta$\Bstrut\\
\hline
MW & $-0.096 \pm 0.009$ & $1.569 \pm 0.096$                                           & $-0.066 \pm 0.011$ & $1.215 \pm 0.125$\Tstrut\\
VW & $-0.126 \pm 0.006$ & $2.339 \pm 0.076$                                           & $-0.132 \pm 0.007$ & $2.202 \pm 0.084$ \\
MC & $0.014 \pm 0.014$  & $1.545 \pm 0.112$                                           & $0.012 \pm 0.020$  & $1.123 \pm 0.155$ \\
NR & $0.132 \pm 0.008$  & $1.173 \pm 0.067$                                           & $0.099 \pm 0.008$  & $1.383 \pm 0.065$\Bstrut\\[0.15cm]
\hline
  & \citet{Giles2016}  & $2.63\pm0.15$                                               &  $\ast$~\citet{Biffi2014}  & $2.29\pm0.07$ \Tstrut\\
  & \citet{Pratt2009}  & $2.34\pm0.13$                                               &  $\ast$~\citet{Barnes2017a}& $3.01\pm0.04$ \\
  &                    &                                                             &  $\ast$~\citet{Henden2019} & $3.02\pm0.15$ \Bstrut\\
\cline{2-5}
\end{tabular}}
\end{table}

\paragraph*{Comparison with observations}
The relations of \citet{Giles2016} and \citet{Pratt2009}, from observations of the {\rm XXL} and {\rm REXCESS} clusters respectively, are rather shifted to higher temperatures at fixed soft-band X-ray luminosity. We note however that \citet{Giles2016} use core-included measurements, unlike \citet{Pratt2009}, which can significantly bias the X-ray luminosity (see \App{app:ci}). 
While the VW simulations agrees with the slopes of the \citet{Giles2016} and \citet{Pratt2009}, our haloes follows on average 40\,per cent shallower scaling relations as we can see in \Tab{tab:slopesLXT}.

\paragraph*{Comparison with simulations}
All haloes agree, within scatter, with the values found by the {\sc Music} and {\sc Fable} simulations while the {\sc Macsis} simulations indicate a slightly lower normalisation i.e. higher temperatures at fixed bolometric luminosity. When comparing quantitatively the slopes for the $L_{\rm X,bol}-M$ scaling relations with these simulations, we found shallower slopes by more than a factor 2 on average. Only our VW simulations agree with the slope of \citet{Biffi2014} which however consider core-included bolometric luminosities.\\ 

In more detail, we systematically find significantly shallower slopes compared to the literature as we can see in \Fig{fig:scaling_LT} and \Tab{tab:slopesLXT}. The VW simulations, however, show the steepest slopes compatible with the studies of \citet{Pratt2009} and \citet{Biffi2014} for the soft-band and bolometric luminosities respectively. 
The steeper evolution of the X-ray luminosity with temperature of the VW simulations is the result of both 
the more effective AGN feedback at lower halo masses and the gas enrichment at high halo masses (see \Fig{fig:scaling_fM}), which significantly boosts the X-ray luminosities. The effect the different radiative models explored in this work is the most visibly seen for the $L_{\rm X}-T_{\rm X}$ relations. Although this scaling relation might be the most straightforward to derive from observations, its calibration remains challenging as it demonstrates the most significant sensitivity on the physical models used in simulations.\\

\begin{figure}
    \includegraphics[width=0.47\textwidth]{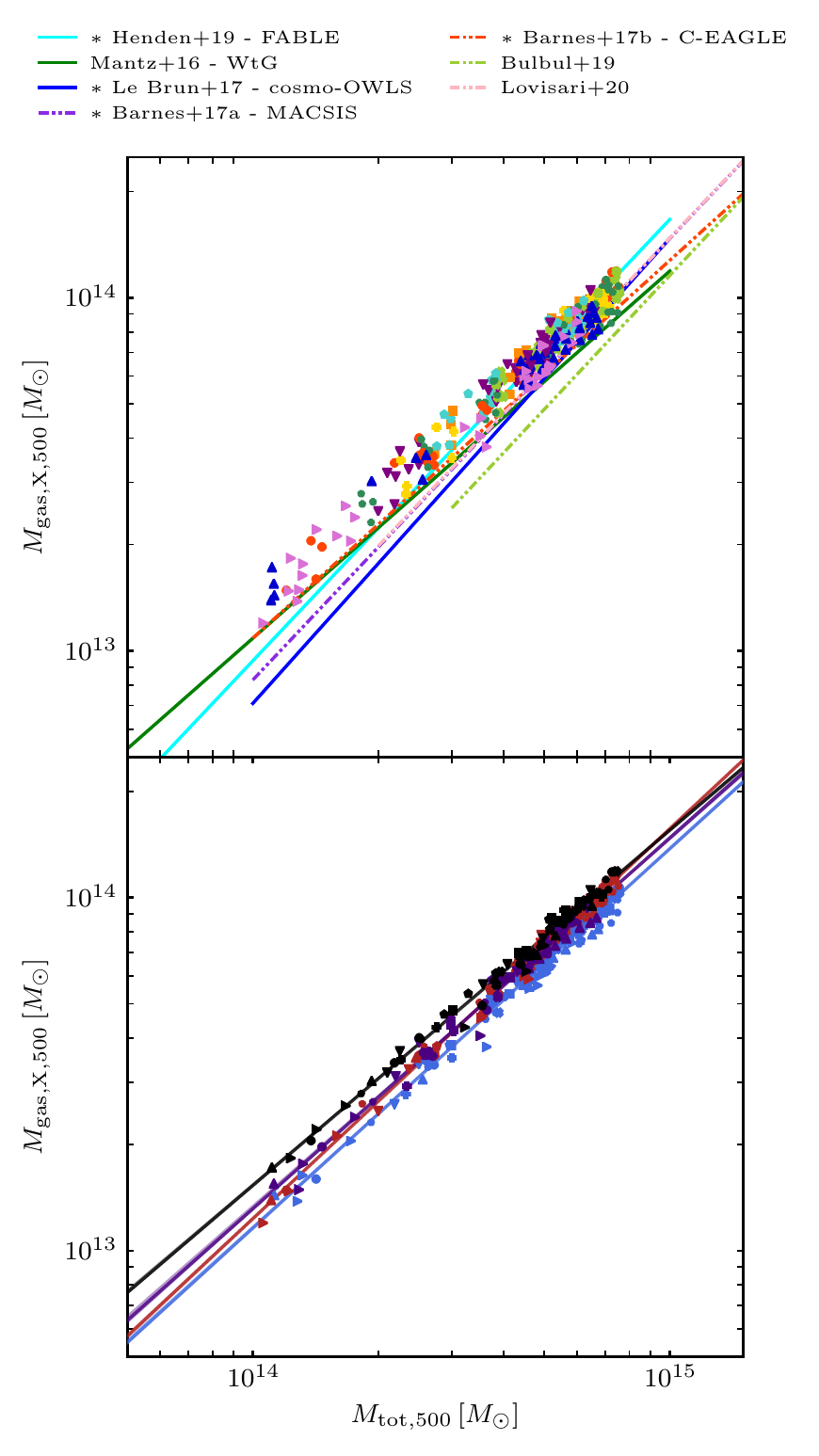}
    \caption{As \Fig{fig:scaling_TM}, but for the evolution of the (X-ray emitting) gas mass to the total cluster mass inside $\Rfiveh$ compared to both numerical and observational studies.}
    \label{fig:scaling_MM}
\end{figure}

\subsubsection{The $M_{\rm gas,X}-M_{\rm tot}$ relation}

We show in \Fig{fig:scaling_MM} the evolution of the X-ray emitting gas mass (i.e. the gas with $T>0.5\keV$) with the cluster total mass enclosed within $\Rfiveh$.
We see that the ICM mass correlates well with the total cluster mass with a relatively small scatter. 
The study of observed relaxed and disturbed clusters by \citet{Lovisari2020} showed that this relation is quite insensitive to the dynamical state of the clusters, however with a higher scatter for their disturbed sample.
In the upper panel of \Fig{fig:scaling_MM}, we can see that our haloes show a relatively higher gas mass for haloes with $\Mfiveh<5\times10^{14}\Msun$. As we can see in \Tab{tab:slopesY}, our simulations indicate slopes consistent with the observations of \citet{Mantz2016} (1.04) and the {\sc C-Eagle} simulations \citep[1.07, ][]{Barnes2017b}, albeit shallower compared to the observations of \citet{Bulbul2019} and \citet{Lovisari2020} but also the {\sc cosmo-OWLS} \citep{LeBrun2017} {\sc Macsis} \citep{Barnes2017a} and {\sc Fable} \citep{Henden2019} simulations.

The dependence of the gas fractions on the physical models of our simulations obviously translates in this scaling relation. Therefore, similarly to the findings of \Sec{sec:fgas}, we see a $\sim$10\,per cent steeper evolution of the gas mass with the total cluster mass for the VW simulation while the NR, MW and MC show relatively similar slopes. 

\subsection{Sunyaev--Zeldovich scaling relations}
\label{sec:Ysz}
The Sunyaev-Zel'dovich (SZ) effect \citep{Zeldovich1969,Sunyaev1970} - which is the distortion of the cosmic microwave background (CMB) spectrum by the inverse Compton scattering of the low-energy CMB photons with free electrons in the ICM - provides an unique view of the ICM baryons. 
By probing the line-of-sight integral of the ICM thermal pressure support, it yields an ideal proxy for the gas mass in a galaxy cluster and therefore the total mass.\\

We compute $Y_{\rm SZ,500}$, the integrated Comptonization parameter $Y$ within $\Rfiveh$, directly from the simulation using the cell gas temperature, $T_i$, and electronic density, $n_{{\rm e},i}=\rho_{{\rm gas},i}/(\mu_{\rm e} {\rm m_p})$, as
\begin{equation}
    Y_{\rm SZ,500} = \frac{\sigma_T}{\rm m_e c^2} \sum^{r_i\leqslant \Rfiveh}_i {\rm k_{\rm B}} T_i n_{{\rm e},i} {\rm d}V_i,
\end{equation} 
where $\sigma_T$, ${\rm m_e}$, ${\rm c}$ and ${\rm d}V_i$  are respectively the Compton cross section, the electron mass, the speed of light and the volume of the considered gas cell.\\
The $Y_{\rm SZ,500}$ quantity does not show any particular scatter as the ICM pressure profiles in clusters tend to be universal within $\Rfiveh$ \citep{Arnaud2010}. It is less sensitive to the gas density than X-ray observables due to its linear dependence, and hence core exclusion is not necessary. Therefore measure $Y_{\rm SZ,500}$ in the $r\leqslant\Rfiveh$ range.\\

\begin{figure}
    \includegraphics[width=0.47\textwidth]{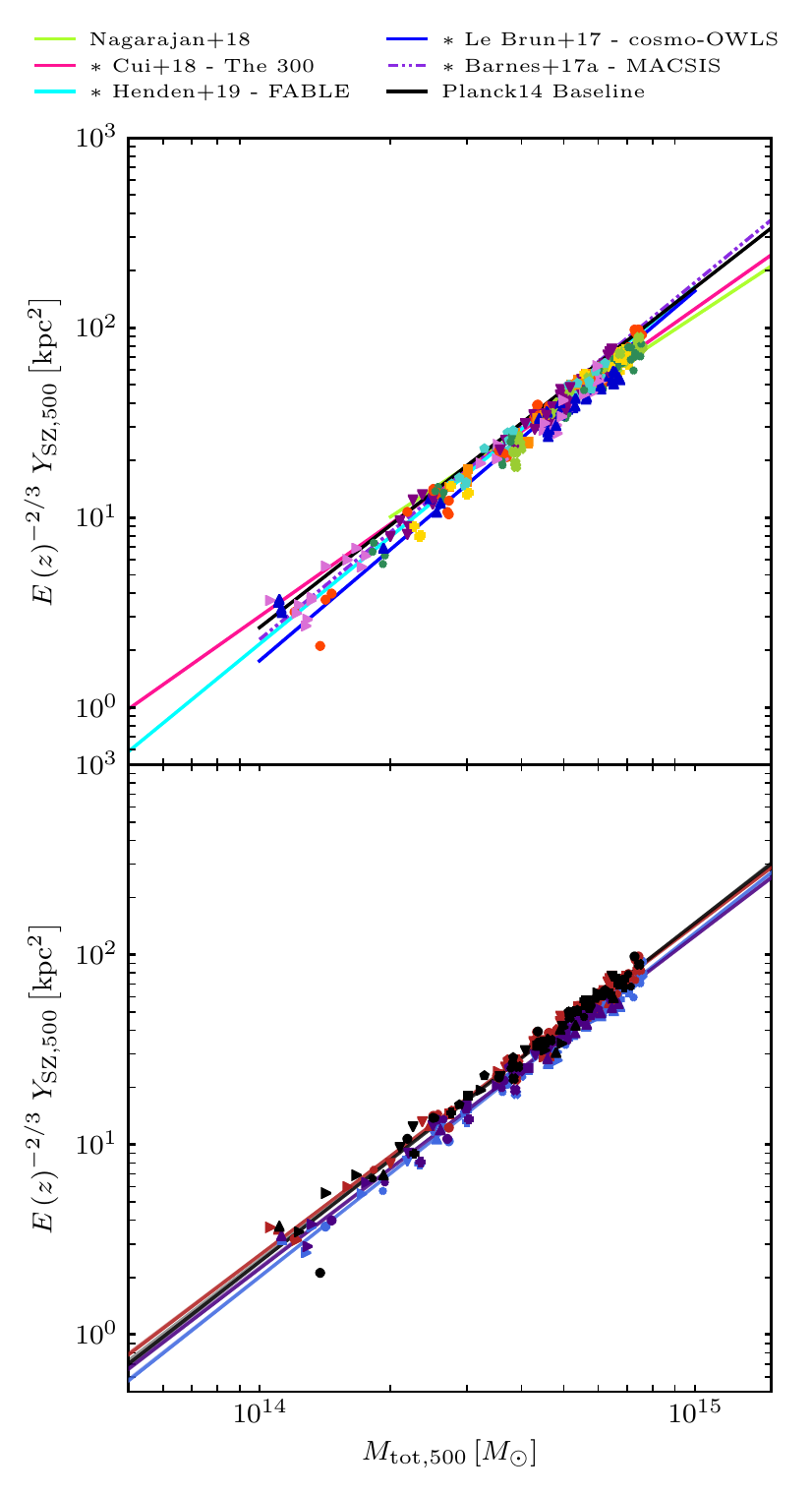}
    \caption{Evolution of the integrated Compton $Y_{\rm SZ,500}$ parameter of the Sunyaev-Zel'dovich effect as a function of the total halo mass computed within $\Rfiveh$. We keep the same properties as of \Fig{fig:scaling_TM}. We see in the upper panel the tight evolution of the \simu haloes along published scaling relations irrespective of the physical models used in our simulations. In the bottom panel, we see that the models used in our simulations induce a very slight change in the slope of our scaling relations.}
    \label{fig:scaling_YszM}
\end{figure}

In \Fig{fig:scaling_YszM} we show the $Y_{\rm SZ,500}-M_{\rm tot,500}$ scaling relations. We see that the $Y_{\rm SZ,500}$ parameter is tightly connected to the cluster mass, where we observe the lowest scatter compared to the X-ray scaling relations. Indeed, this parameter probes the mass-weighted temperature, which is much less sensitive to the gas clumpiness (as opposed to the emission measure weighted temperature of X-ray quantities).

Our \simu haloes agree well with all previously published scaling relations from both numerical simulations \citep{Cui2018,Henden2019,Barnes2017a,LeBrun2017} and observational works \citep{Planck2014,Nagarajan2019}. Due to the very low scatter in the $Y_{\rm SZ}-M$ scaling relation, its use seems well suited for studies that aim to constrain cosmological parameters. 
In a more quantitative comparison, we can see in \Tab{tab:slopesY} that \citet{Planck2014,Nagarajan2019,Cui2018,Nagarajan2019} have slope values in agreement with our simulations (within errors), but the latter shows a 17\,per cent lower slope value on average. On the other hand, the simulations of \citet{LeBrun2017} and \citet{Henden2019} indicate slightly steeper slopes. 
For observational studies, this difference can be understood by $Y_{\rm SZ,500}$ being a mass-dependent observable and therefore less constrained at lower masses, which can explain the difference in the slopes of \citet{Nagarajan2019} and the \citet{Planck2014} baseline that probe different cluster mass ranges.

Between our simulations, the slopes of the radiative simulations are in very good agreement and the non-radiative simulations. We find similar slopes for all simulations with a difference being at most 4\,per cent between the MW and VW simulations. 
Although the difference in the AGN feedback model between the MW and VW simulations yields disparate X-ray scaling relations, the difference here weakens for the $Y_{\rm SZ}-M$ scaling relation.
We can understand this slight difference as the VW simulations produce higher ICM temperatures (and hence higher pressures) at lower halo masses due to a more efficient AGN gas heating at early times. 
The slightly shallower slope observed for simulations with conduction (MC) compared to the simulations without (MW) is a consequence of the higher fraction of cooling gas at high halo masses which results in a lower pressure support, hence lower $Y_{\rm SZ,500}$ values at higher halo mass (cf. \Sec{sec:fgas}).
Therefore, we see here that the physical models used in our simulations do not play a particular role in offsetting the $Y_{\rm SZ,500}-M_{\rm tot}$ scaling relation.

Again, the slight normalisation and slope changes can be understood in the same way as the conclusions of \Sec{sec:fgas}: the simulations with ATC show a shallower evolution with mass than simulations without, as the ICM is more quiescent with a suppressed star formation, and the higher normalisation of VW simulations can be ascribed to their higher ICM pressure support caused by their AGN feedback model.\\
\begin{figure}
    \includegraphics[width=0.47\textwidth]{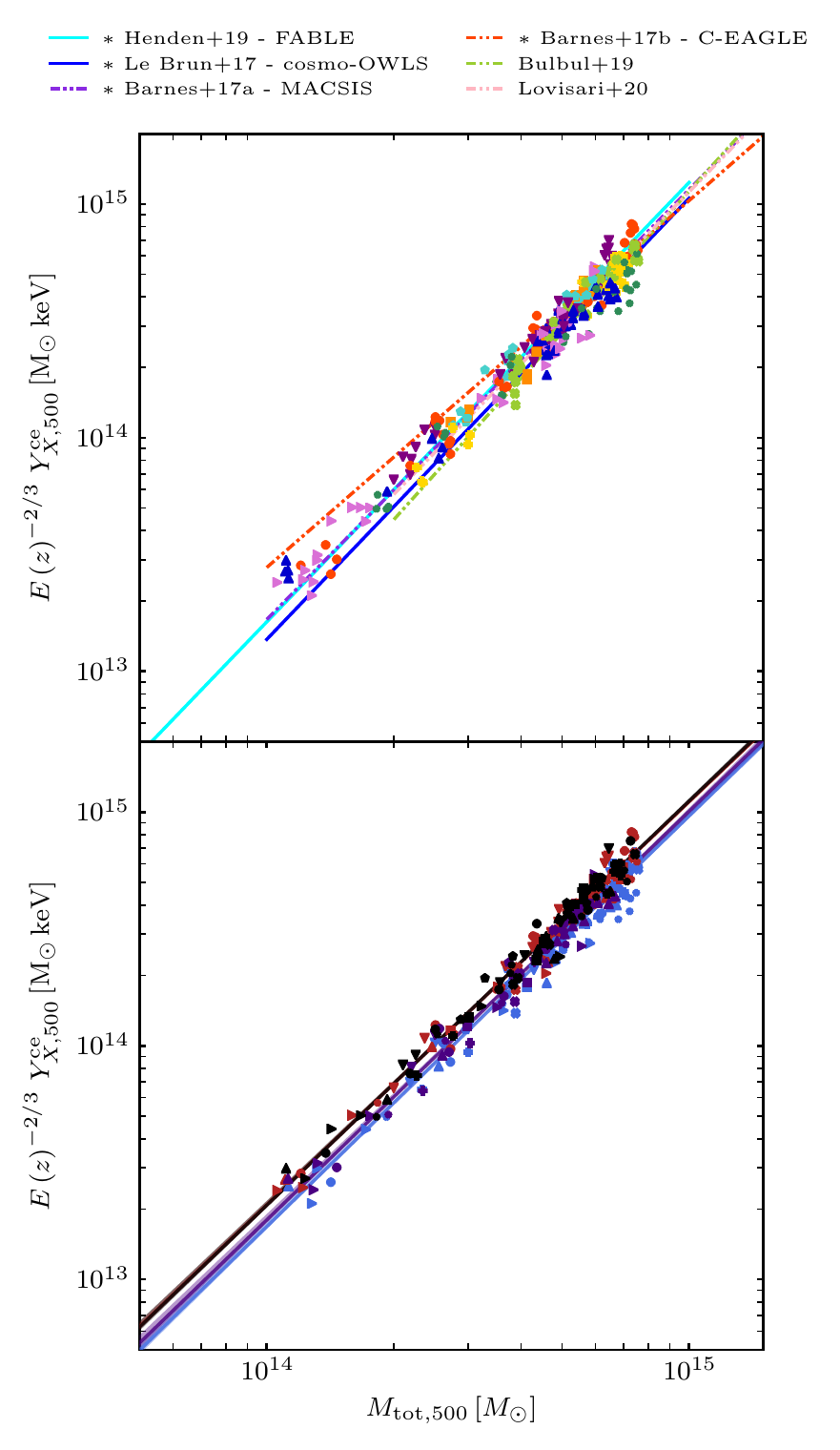}
    \caption{Similarly to \Fig{fig:scaling_YszM}, we show the X-ray analogue of the core-excluded integrated SZ signal, i.e. $Y_{\rm X}=M_{\rm gas,X}\times T_{\rm X}^{\rm ce}$, as a function of the total halo mass. Compared to $Y_{\rm SZ}-M$, we observe a greater scatter, expected from the sensitivity of X-ray observables. However, it shows the lowest scatter compared to the other X-ray scaling relations ($T_{\rm X}-M$, $L_{\rm X}-M$ or $L_{\rm X}-T_{\rm X}$). Our data is in fair agreement with published studies even at lower halo masses where the scatter is the greatest. Similarly to \Fig{fig:scaling_YszM}, the difference in the used physical models only induce a slight difference in the inferred slopes.}
    \label{fig:scaling_YxM}
\end{figure}

We also measure the X-ray analogue of $Y_{\rm SZ,500}$ by taking the product of the mass of the X-ray emitting gas ($M_{\rm gas,X,500}$) and the X-ray temperature from our spectral fit ($T_{\rm X,500}^{\rm ce}$).
In \Fig{fig:scaling_YxM} we show our data along other $Y_{\rm X}-M$ scaling relations and we observe a increased scatter than from the SZ scaling relation, as this $Y_{\rm X}$ parameter is more sensitive to the internal structure of the ICM\footnote{For instance, a completely smooth ICM will give equality between $Y_{\rm X,500}$ and $Y_{\rm SZ,500}$, as in this case we have $\langle n^2\rangle = \langle n\rangle^2$ (which shows the respective dependence of the $Y_{\rm X,500}$ and $Y_{\rm SZ,500}$ on the gas number density).}.
Our slope values remain rather constant and very similar to the slopes found for the $Y_{\rm SZ,500}-M_{\rm tot,500}$ relation (see \Tab{tab:slopesY}) and overall, our data is consistent with the results of the studies shown in \Fig{fig:scaling_YxM}, but our slopes indicate a $\sim$6\,per cent lower value on average. 
In the lower mass range, where differences between published scaling relations become more obvious, our haloes follow the relations of the numerical studies of \citet{LeBrun2017,Barnes2017a} and \citet{Henden2019} while the former indicates somewhat lower $Y_{\rm X,500}$ values. On the other hand, the C-EAGLE simulations \citep{Barnes2017b} indicate a 10\,per cent shallower slope with higher $Y_{\rm X,500}$ values, especially at low halo masses.\\

To summarize, we have seen in this section that the evolution of our haloes along cluster scaling relations are not significantly affected by changes in the AGN feedback model or the inclusion of ATC. Yet, in this study we focused on massive systems whereas the effect of feedback processes might be more pronounced in the group regime. Despite their profound impact on the cluster gaseous and stellar components, the global properties of our haloes and their evolution with mass are not significantly affected by the baryonic processes in the ICM (with the exception of the X-ray luminosity, which is relatively sensitive to the ICM thermodynamical state).  \\
This finding is good news for cluster cosmology, which relies on scaling relations to derive cluster total masses. Numerical simulations are proven to be a reliable and suitable tool for such calibrations. 

\begin{table*}
    \caption{Same as \Tab{tab:slopesTX} for the X-ray emitting gas mass ($M_{\rm gas,X,500}$) integrated $Y_{\rm SZ}$ parameter and its X-ray analogue ($Y_{\rm X}$).}
    \label{tab:slopesY}
    \begin{tabular}{ c c c | c c | c c}
  \cline{2-7}
  & \multicolumn{2}{c|}{$M_{\rm gas,X,500}$--$\Mfiveh$} & \multicolumn{2}{c|}{$Y_{\rm SZ,500}$--$\Mfiveh$} & \multicolumn{2}{c}{$Y_{\rm X,500}$--$\Mfiveh$}\Tstrut\\
  & $\alpha$           & $\beta$                        & $\alpha$           & $\beta$                     & $\alpha$            & $\beta$\Bstrut\\
  \hline
MW & $-0.030 \pm 0.002$& $1.075 \pm 0.009$              & $-0.034 \pm 0.003$ & $1.812 \pm 0.014$           & $-0.022 \pm 0.004$ & $1.757 \pm 0.022$\Tstrut\\
VW & $0.018 \pm 0.001$ & $1.105 \pm 0.007$              & $0.029  \pm 0.003$ & $1.737 \pm 0.014$           & $0.047  \pm 0.003$ & $1.725 \pm 0.017$ \\ 
MC & $0.008 \pm 0.003$ & $1.050 \pm 0.013$              & $-0.031 \pm 0.004$ & $1.752 \pm 0.019$           & $-0.000 \pm 0.006$ & $1.748 \pm 0.027$ \\ 
NR & $0.043 \pm 0.002$ & $1.006 \pm 0.008$              & $0.027  \pm 0.005$ & $1.784 \pm 0.023$           & $0.050  \pm 0.004$ & $1.730 \pm 0.021$\Bstrut\\
\hline
   & $\ast$~\citet{Barnes2017a} & $1.25\pm0.03$         & $\ast$~\citet{Barnes2017b} & $1.69\pm0.07$       & $\ast$~\citet{Barnes2017a} & $1.84\pm0.05$\Tstrut\\
   & $\ast$~\citet{Barnes2017b} & $1.07\pm0.05$         & $\ast$~\citet{Cui2018}     & $1.62\pm0.31$       & $\ast$~\citet{Barnes2017b} & $1.57\pm0.07$ \\
   & $\ast$\citet{LeBrun2017}   & $1.32\pm0.01$         & $\ast$~\citet{Henden2019}  & $1.88\pm0.05$       & $\ast$~\citet{Henden2019}  & $1.88\pm0.05$  \\
   & $\ast$~\citet{Henden2019}  & $1.25\pm0.04$         & $\ast$\citet{LeBrun2017}   & $1.948\pm0.018$     & $\ast$\citet{LeBrun2017}   & $1.948\pm0.018$ \\[0.15cm]
   & \cite{Mantz2016}           & $1.04\pm0.05$         & \citet{Nagarajan2019}      & $1.51\pm0.31$       & \citet{Bulbul2019}         & $2.01\pm0.20$ \\
   & \cite{Bulbul2019}          & $1.26\pm0.10$         & \citet{Planck2014}         & $1.79\pm0.065$      & \citet{Lovisari2020}       & $1.85\pm0.10$\\
   & \cite{Lovisari2020}        & $1.25\pm0.05$         &                            &                     &                            & \Bstrut\\
\cline{2-7}
\end{tabular}
\end{table*}


\section{Summary and conclusions}
\label{sec:conclusion}

We presented the \simu suite, a series of zoom-in magneto-hydrodynamical simulations of massive galaxy clusters ($M_{\rm vir}\sim10^{15}\Msun$) with a physical resolution of $2.8\kpc$. The simulations include radiative gas cooling, star formation, feedback from supernovae (SN) and active galactic nuclei (AGN) as well as anisotropic thermal conduction (ATC). This work was motivated by the \Rg suite \citep{Wu2015,Hahn2017,Martizzi2016}, which suggested shortcomings in the thermal AGN model and the need for additional sources of energy injection. Hence, in this paper we revisit thoroughly the seeding of super massive black holes (SMBHs), introduce a new model for their dynamical evolution, and consider different AGN energy deposition schemes as well as the anisotropic transport of heat within the intra-cluster medium (ICM). 
We study the impact of each of the models on the cluster stellar component and examine how they shape the intra-cluster gas. We next investigate the evolution of our simulated clusters over cosmic time with a range of cosmological observables that serve as mass proxies when the above-mentioned sub-grid models are changed. We focus in this analysis on the total cluster mass versus X-ray temperature, luminosity, gas mass and the integrated Comptonization parameter.
The main findings of our analysis are as follows :

\begin{itemize}
    \item The star formation in the proto-cluster can be efficiently controlled by the seeding of the SMBHs in the ICM. With a low number of SMBH seeds, the AGN heating cannot prevent the gas from over-cooling in the proto-cluster. Seeding less massive but more numerous SMBHs enables an efficient, more frequent AGN heating in time and space. Our simulations indicate that this might be one of the most crucial parameters to regulate feedback. \\
    \item We develop a new model for the SMBH dynamics, which is made publicly available for the {\sc Ramses} code. It consists of decaying the SMBHs towards the local potential minimum along the steepest gradient with a magnitude that depends on the tidal forces experienced by the SMBH during its evolution. This new model makes the accretion of gas onto SMBHs easily tunable by keeping the SMBHs relatively close to the potential minimum. Consequently, the amount of AGN feedback energy injected into the ICM can be controlled and it is shown to have a significant effect on reducing the gas content in the proto-cluster as well as quenching star formation. The abundance (see point above) and the locations of the SMBHs therefore appear critical to regulate AGN feedback.\\
    \item By changing  only the AGN energy injection scheme (i.e. volume- or mass-weighted energy deposition), we observe a dramatic change in both the distribution of gas and in star formation.\\
    Mass-weighted deposition preferentially injects the AGN feedback energy into the dense accretion regions, which then has difficulty escaping and is thermalised by the cold gas reservoir surrounding the central SMBH. As a result, a build-up of cold gas occurs in the ICM and a high star formation rate follows. At late times, this larger cold gas reservoir fuels AGN activity that can increase the gas entropy in the core to produce a non cool-core (NCC) state at $z=0$. \\
    On the other hand, volume-weighted deposition injects more energy in more diffuse regions, which allows the AGN feedback energy to escape the accretion region early to heat the gas over large distances. Star formation is dramatically quenched, the gas in the core is efficiently depleted and the transition to an NCC proceeds from $z=1$. When volume-weighted deposition is used, the more diffuse ICM leads to the cessation of AGN activity at lower redshifts because of a low cold gas supply. Thus, with the decline of the AGN activity, the ICM gradually cools to settle into a similar ICM thermodynamical state as the simulation implementing mass-weighted AGN energy deposition at $z=0$. \\
    In spite of this relative similarity between the two simulations at $z=0$, the star formation and AGN activity histories greatly differ, as do the galaxy masses and the ICM clumpiness.\\
    \item Anisotropic thermal conduction appears to reduce star formation in the ICM by almost a factor of 2 by flattening out temperature gradients in the ICM. ATC leads to an earlier transition to an NCC cluster thanks to the transport of heat within the ICM. However, in our simulations, we do not observe enhanced AGN activity but the opposite: ATC delays the AGN activity by preventing the formation of cold gas in the ICM that would otherwise fuel the SMBH cold gas accretion.\\
    \item The cluster gas fractions are not appreciably altered by changes in the energy accumulation threshold of the thermal AGN feedback. The observed slight gas depletion does not linearly scale with this threshold. It suggests that purely thermal feedback cannot shape the ICM on large scales.\\
    \item The evolution of our simulated cluster observables over cosmic time is in relatively good agreement with both observational and numerical studies. Among the X-ray scaling relations, the $T_{\rm X}-M$ relation is rather insensitive, especially in the high halo mass regime ($ \Mfiveh\geqslant 5\times 10^{14}\Msun$), to the use of the different galaxy formation models (radiative gas cooling, ATC, mass- or volume-weighted AGN energy deposition) as they show no noticeable difference with the scalings derived for adiabatic simulations. The same conclusions hold for the mean Sunyaev-Zeldovich flux scaling relations (and its X-ray analogue) with much lower scatter. This suggests that galaxy formation physics does not play a particular role in significantly offsetting the global cluster observables, especially at the high mass end.
\end{itemize}

To aid the astrophysical and cosmological interpretation of current and future galaxy cluster surveys, we have increased the complexity of the \simu high-resolution cosmological simulations by including anisotropic thermal conduction and various SMBH/AGN models, as well as generating more sophisticated X-ray observables compared to the \Rg simulations. Despite the apparent sensitivity of the ICM and cluster galaxies to the numerical models used, the evolution of the simulated cluster observables over cosmic time is remarkably insensitive to changes in our astrophysical models. A notable exception of the X-ray luminosity, which is sensitive to clumping. The power of the \simu simulations is to have a sample of clusters sharing a similar mass at $z=0$ but with diverse assembly histories, shapes and richness. We are therefore in a position to study the scatter around the mean scaling relations and relate it to the astrophysical processes that shape each cluster. This will be investigated in future research.\\

In this work, by looking at the $T_{\rm X}-M_{\rm tot}$ scaling relation, we observed that accounting for a $\sim$20\,per cent mass bias can make our data consistent with studies based on hydrostatic mass estimates. A detailed study of the energy budget in our simulated galaxy cluster is needed to quantify the level of non-gravitational energy and non-thermal pressure support. Moreover, thanks to the implementation of the thermal conduction of \citet{Dubois2016} which includes a separate treatment of electrons and ions, we are able to resolve differences in their distribution and energetics. We will investigate these aspects in future work.


\section*{Acknowledgements}
We are grateful to Pawel Biernacki for helpful discussions about the modelling of super-massive black holes in {\sc Ramses}, to Yohan Dubois regarding the thermal conduction scheme, as well as Christian Garrel for his help with the {\tt LIRA} code. We are indebted to Lorenzo Lovisari and Yohan Dubois for thorough feedback on an early version of the manuscript, and we thank Ricarda Beckmann, Sandrine Codis, Fr\'ed\'eric Bournaud, Aoife Boyle, and Sunayana Bhargava for useful discussion and comments.

AP and OH acknowledges funding from the European Research Council (ERC) under the European Union's Horizon 2020 research and innovation programme (grant agreement No. 679145, project `COSMO-SIMS'). 
This work was granted access to the HPC resources of TGCC under the allocation A0040410487 made by GENCI.\\
{\small This work made use of the following open source software: Matplotlib \citep{Hunter2007}, NumPy \citep{Harris2020}, SciPy \citep{Virtanen2020}, emcee \citep{emcee}, PyAtomDB \citep{Foster2020}, APEC \citep{Smith2001}, Astropy \citep{Price2018}.}


\section*{Data Availability}
The simulation data and post-processed data can be made available per reasonable request to the authors on an individual basis.


\bibliographystyle{mnras}
\bibliography{main.bib}


\begin{appendix}

\section{Anisotropic thermal conduction as a heating or cooling source}
\label{app:atc}

To understand the effect of anisotropic thermal conduction (ATC) on the ICM before adding physics related to galaxy formation. We run our simulation at a lower resolution corresponding to an effective resolution of $4096^3$ cells for the full simulation box, yielding a DM particle mass  of $8.22\times10^8 \hMpc$.\\

\begin{figure}
    \centering
    \includegraphics{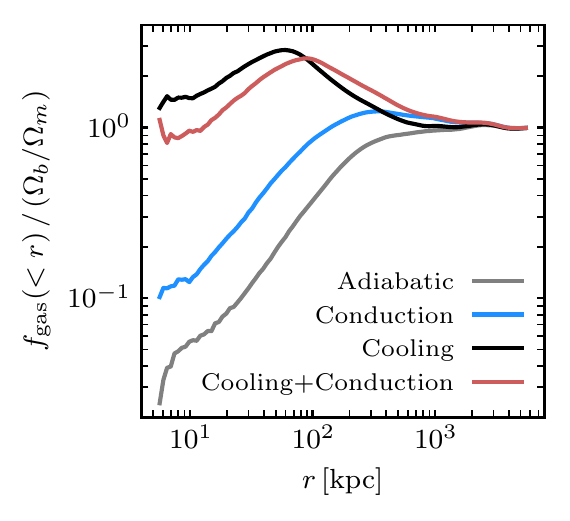}
    \caption{Gas depletion profiles for our 4 different cluster simulations. The blue and grey line shows the non-radiative simulation with and without anisotropic thermal conduction respectively. Red and black lines show respectively simulations when radiative gas cooling is coupled or not to thermal conduction. When ATC is added, see the contraction of the gas rich core in the non-radiative simulation (blue to be compared with grey), while the core expands when in simulation including gas cooling (red compared to black). Depending on the temperature gradient in the core ($\leqslant200\kpc$), anisotropic thermal conduction can act as a cooling or heating source and endeavours to smooth out substructures in the ICM.}
    \label{fig:condfgas}
\end{figure}

We show in \Fig{fig:condfgas}, the gas depletion profiles for an adiabatic simulation (grey line) and where only ATC was added (blue). Similarly, we compare a simulation where the gas is allowed to radiatively cool (black) to the same simulation when ATC is switched on (red).

In the adiabatic case, as the cluster forms, the gas in the center is adiabatically heated and the higher pressure support prevents the gravitation collapse of gas from the cluster outskirts.\\
However when thermal conduction is added, we observe that while the cluster forms and the more gas collapse towards the center, heat generated during this gas compression is now transported outwards were the gas temperature is lower. As the result more low-entropy gas fall inwards and the cluster centre gets denser. With thermal conduction, the heat generated by the gas compression during the cluster evolution is transported to the colder outskirts which lowers the pressure support in the cluster center. Consequently, gas flows inwards, the cluster core contracts and a higher gas fraction can be observed at all radii in \Fig{fig:condfgas}.
In that case, thermal diffusion is actually behaving like a gas cooling source driving a cooler and denser cluster core.\\

We now allow the gas to radiatively cool. Because this process is quadratically sensitive to the gas density, as the gas in the center is getting denser, it also cools at faster rates, which leads to a runaway instability: the cooling catastrophe. To prevent the overcooling of the gas, we set an artificial limit by imposing a pressure floor below which no gas can cool but can only increase its density along the adiabat $P_{\rm gas}\propto\rho_{\rm gas}^\gamma$.
In this configuration, we observe the formation of a much lower entropy and higher density core compared to the adiabatic case which translates into a high fraction in the central $100\kpc$.
By comparing the gas depletion profiles of the radiative simulations in \Fig{fig:condfgas}, we can see by comparing the black and red line that thermal conduction can deplete gas from the core to the outskirts. In the cooling-only simulation (black), we see the fraction of gas peaking at $60\kpc$, defining the extent of the core, which drops to reach the universal baryon faction at $700\kpc$. With the addition of thermal conduction (red), the core extends now to $100\kpc$ with a lower amount of gas. The gas fraction decreases less steeply towards the $\Omega_b/\Omega_m$ value at a greater distance of $2\Mpc$. In that configuration, thermal conduction drives the transport of heat from the outskirts toward the overcooling core: it now acts as a heating source.\\

Interestingly, we see that thermal conduction can act as a cooling source by transporting heat from the core to larger radii, or, as a heating source by transporting heat inwards. This behaviour is dictated by the sign of the temperature gradient in the inner cluster region. More generally, thermal conduction is effective at flattening out temperature substructures in the ICM.

\section{Estimating the X-ray temperature}
\label{app:Tx}

Having access to the true density and pressure of the gas inside a simulation cell, we can estimate the true gas temperature from our simulations. However, the comparison between real and simulated data is complicated by different problems like sky background, projection effects and instrumental noise. Additionally there is a possible mismatch between the spectroscopic temperature estimated from X-ray observations and the temperatures usually defined in numerical studies.
The former is a mean projected temperature obtained by fitting a single (or multitemperature) thermal model to the observed photon spectrum while the later fully exploits the three-dimensional thermal information of gas cells/particles. 
The average gas temperature in simulated cluster can be obtained by a weighted sum of all cell/particle temperatures $T_i$ as
\begin{equation}
    T_w = \frac{\sum_i w_i T_i}{w_i},
\end{equation}
where $w_{i,{\rm vw}}={\rm d} V_i$ in case of a volume-weighted temperature (${T_{\rm vw}}$), with ${\rm d} V_i$ is the AMR cell volume determined by the local resolution of the simulation, or with $w_{i,{\rm mw}}=n_i{\rm d} V_i$ for a mass-weighted temperature (${T_{\rm mw}}$) with $n_i$ the gas cell density. This mass-weighted temperature gives a more `physical' average as it emphasizes dense regions that participate more to the X-ray emission.\\

However, X-ray astronomy suffers from well-known biases due as its intensity depends quadratically on the density since both Bremsstrahlung and the collisional excitation responsible for the metal line emissions results from two-body processes. For this reason, X-ray observations are especially biased by the dense regions such as the cluster core or the presence of gas-rich substructures which motivate the definition of a `emission-weighted' temperature (${T_{\rm emw}}$) where the weighting function is proportional to the emissivity of each gas element $w_{i,{\rm emw}}=n_i^2\Lambda(T_i)$ with the cooling function $\Lambda(T)$ mainly being the so-called bolometric cooling function $\Lambda(T)\propto \sqrt{T_i}$ which implicitly assume that bremsstrahlung (free-free) emission is the dominant mechanism at high X-ray temperatures ($>3\keV$)\footnote{Nevertheless, at lower temperatures, metal lines participate significantly to the X-ray emission which becomes temperature and metallicity dependent.}. 
\citet{Mazzotta2004} found that the above-defined emission-weighted temperature over-estimates the projected spectroscopic X-ray temperatures of thermally complex clusters and propose another spectroscopic-like temperature $T_{\rm sl}$ with $w_{i,{\rm sl}}=n_i^2T_i^{3/4}$ which better approximates the spectroscopic temperature in Chandra and XMM-Newton observations. This weighting function, beside being biased toward the densest regions of the clusters such as in the emission-weighted case, will also be biased toward the coolest regions.\\

\begin{figure}
    \includegraphics{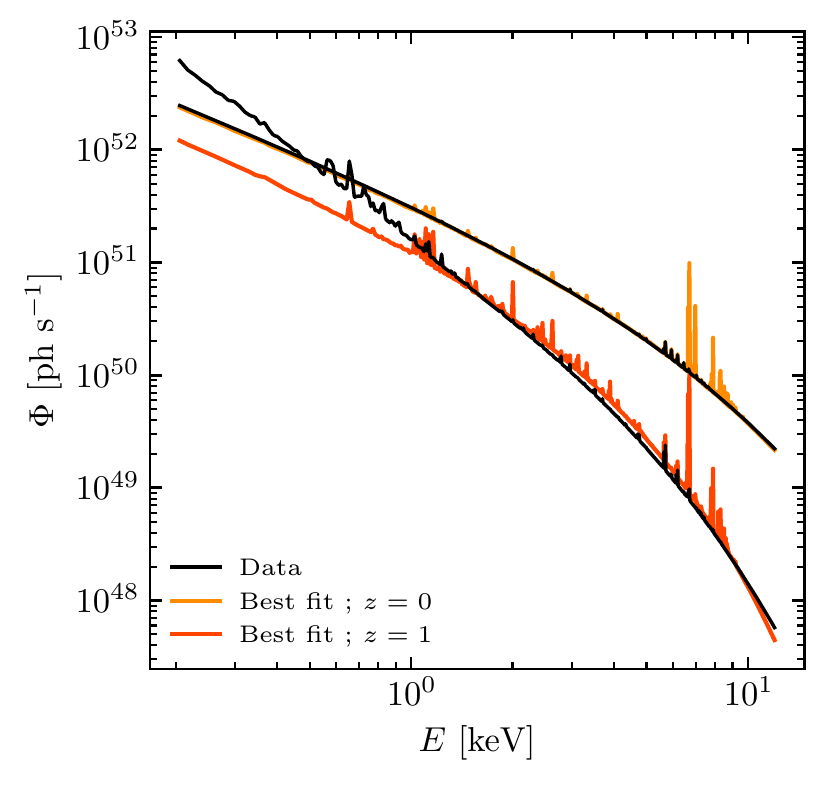}
    \includegraphics{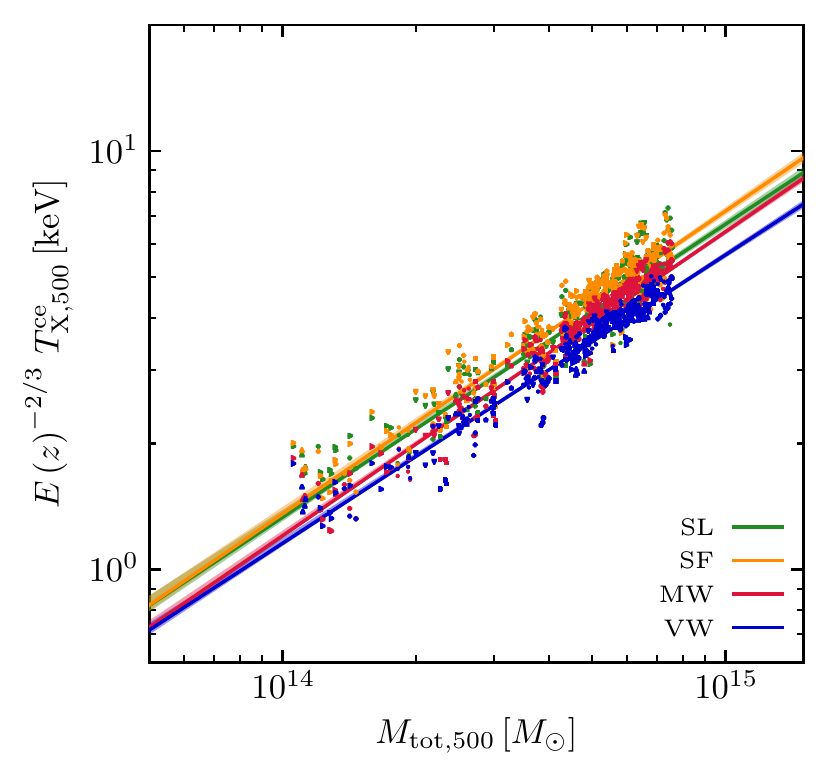}
    \caption{Top panel: ICM photon emission rate directly computed from the simulation in the core-excised $\Rfiveh$ sphere (black) along the best fit APEC model resulting from our MCMC sampling for the same halo at $z=0$ (orange) and $z=1$ (dark orange). At low redshifts, our fitting procedure performs well. We note that the gas metallicity responsible for the emission lines is not well constrained but is not relevant for our analysis as the lines does not significantly participate to the X-ray flux. For the $z=1$ spectrum, the high fraction of cold gas (i.e. $E\leqslant 1\,\keV$) lead to a slight overestimation of the density (i.e. higher normalisation) and an underestimation of the gas temperature (i.e. steeper spectrum).\\
    Bottom panel: Scaling of the core-excluded temperature with the total mass inside the $\Rfiveh$ using different temperature estimates. ${T_{\rm vw}}$  (blue) and ${T_{\rm mw}}$ (red) are a weighted average using respectively the cell volume and the cell density. ${T_{\rm sl}}$ (green) uses the cell emissivity and the \citet{Mazzotta2004} weights of hot ($E\geqslant0.5\keV$) gas cells only. We show the temperatures resulting from the spectral fits ${T_{\rm sf}}$  in orange. We can see that our ${T_{\rm sf}}$ estimates approach well the spectroscopic-like temperature.
    }
    \label{fig:spectralfit}
\end{figure}

To circumvent the shortcomings of simple weighting schemes, we instead produce an X-ray spectrum from which we derive the gas temperature and density using a single thermal as close as possible to the observers' methodology. 
The simulated ICM spectrum is obtained by computing the continuum and line emission of a gas cell with $T\geqslant0.5\keV$ and a metallicity $Z$, with the publicly available AtomDB atomic database\footnote{We use the abundances of \citet{Anders1989} as well as APEC equilibrium line and continuum fits files from the 3.0.9 version of AtomDB - \href{https://atomdb.org/}{https://atomdb.org/}}.
As such, from the continuum and line emissivity $\epsilon_i(T_i,Z_i)$, we compute the rate of emitted photon $\Phi_i$ of the cell $i$
\begin{equation}
    \phi_i = \epsilon_i(T_{{\rm e},i},Z_i)\sum_i n_{{\rm e},i} \,n_{{\rm H},i} {\rm d}V_i,
\end{equation}
which allow us to produce a mock X-ray spectrum by summing of the individual rest frame spectra of each gas cell.\\
In X-ray observation, the most common method to obtain the ICM temperature and density is to fit the observed spectra by a single temperature APEC model. Therefore, we follow a similar methodology and chose to perform fits using a Monte Carlo Markov Chain (MCMC) sampling method thanks to the {\tt emcee} python library \citep{emcee}. We fit our data to a single temperature spectra generated using the {\tt PyAtomDB} library.
We show the result of a such fits in the upper panel \Fig{fig:spectralfit} .

We note that the presence of cold gas at high redshifts can produce a low-energy bump in the spectrum which complicate the fitting procedure. As the result, it could induce the fit to converge faster to high values of the density (i.e. higher spectrum normalisation). In order to maximize the likelihood, the MCMC chain will later try converging to lower temperatures (i.e. steeper cutoff) to compensate for the overestimated gas density. Consequently, the spectral fit can slightly underestimate the temperature of haloes in the case of a high fraction of cold the gas, which is predominantly the case at high redshifts ($z\geqslant1$). \\
To overcome the overestimation of the gas density (and a underestimation of the temperature), we first fit the gas density 0.20--2.00~keV band first as the X-ray flux is not very sensitive on the temperature and metallicity in this band (as long as the metallicity is low, i.e. $Z\lesssim0.5$). We use the posterior distribution of this first MCMC sampling as the prior distribution of the gas density for a second MCMC while using flat priors for the gas temperature and metallicity. The fits in \Fig{fig:spectralfit} result from this two-step MCMC. However, the low energy bump cannot be constrained with a single temperature model and a double (or multi) temperature model would be more suited.\\

We show in the bottom panel of \Fig{fig:spectralfit}, the differences between the different gas temperature estimates (${T_{\rm vw}}$, ${T_{\rm mw}}$, ${T_{\rm sl}}$ and ${T_{\rm sf}}$). To compute the average within $\Rfiveh$ of ${T_{\rm sl}}$ and ${T_{\rm sf}}$, only the hot X-ray emitting gas ($E\geqslant 0.5\,\keV$) is considered while ${T_{\rm vw}}$ and ${T_{\rm mw}}$ use the information of all cells with no cut in minimum gas temperature. \\
We see that the ${T_{\rm sl}}$ and ${T_{\rm sf}}$ are similar and shows that the formula of \citet{Mazzotta2004} gives a good estimate of the spectroscopic temperature, especially in the lower mass range. However, these two X-ray' estimates are, on average,  10 and 20\,per cent higher than ${T_{\rm mw}}$ and ${T_{\rm vw}}$ respectively. This shows that accounting for the bias induced by the X-ray emitting gas offsets to higher temperatures the simple mass- or volume-weighted averages, which also do not have any cut in minimum gas temperature.\\
${T_{\rm mw}}$ is  10\,per cent higher compared to ${T_{\rm vw}}$ at higher halo masses but shows the steepest slope as we can see in \Fig{fig:spectralfit} (we have for haloes with $z\leqslant1.5$ slopes of $0.700 \pm 0.019$, $0.723 \pm 0.018$, $0.726 \pm 0.013$ and $0.689 \pm 0.014$ when using  ${T_{\rm sl}}$, ${T_{\rm sf}}$, ${T_{\rm mw}}$ and ${T_{\rm vw}}$ respectively).

\section{On the core inclusion}
\label{app:ci}

\begin{figure}
    \includegraphics[width=0.47\textwidth]{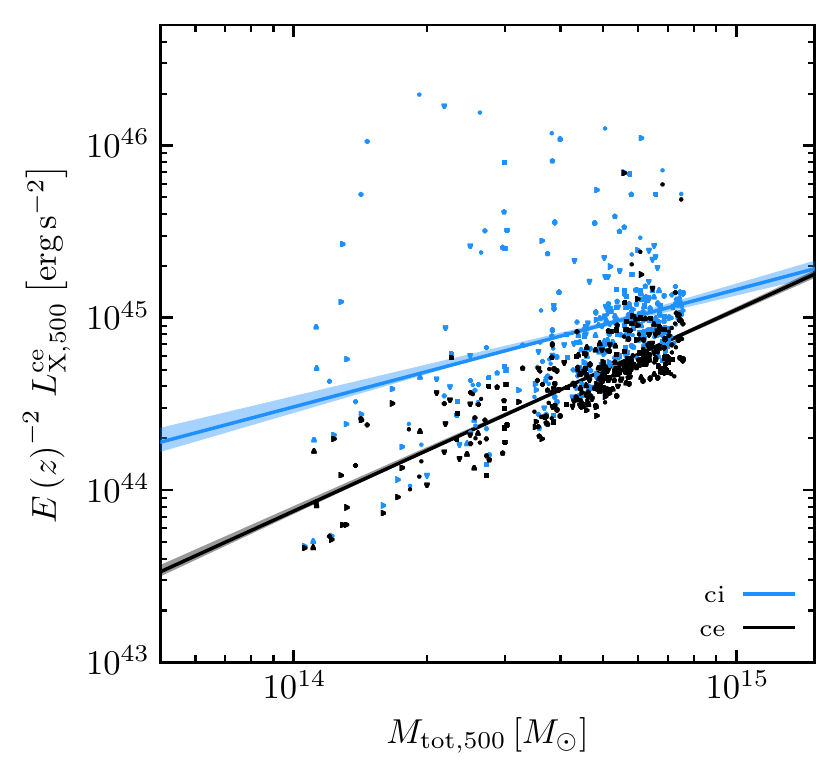}
    \caption{Mass versus X-ray luminosity for all haloes with $z\leqslant 1.5$ for the all simulations type combined (NR, MW, VW and MC, see details in \Tab{tab:sims}. The blue symbols show the core-included (ci) X-ray luminosity within $\Rfiveh$ and the black symbols shows the luminosity in the core-excluded region corresponding to the $[0.15-1]\,\Rfiveh$ aperture. The see that the exclusion of the core reduces dramatically the scatter and indicates for 40\,per cent the X-ray luminosities on average.}
    \label{fig:coreincl}
\end{figure}

The presence of AGN activity in the core of GCs can introduce a strong variability of the X-ray luminosity as the central density fluctuates and  unrealistically high luminosities can be obtained.  
We compare in \Fig{fig:coreincl} the core-included (ci) and core-excluded (ce) X-ray luminosities computed for two apertures: the entire cluster emission interior to $\Rfiveh$ and in the $[0.15-1]\,\Rfiveh$ aperture respectively.\\
The inclusion of the core typically boost the X-ray luminosity with a much greater scatter as it is greatly sensitive to the thermodynamic state of the cluster core (high gas density and possible AGN heating). On average, the exclusion of the core decreases by 40\,per cent the X-ray luminosity and shows a steeper slope of  ($1.169 \pm 0.033$, compared to $0.681 \pm 0.070$ for the ci).\\
It also significantly reduces the scatter and hence is more suited for galaxy cluster sample studies where clusters can have very different central states, consistent with the finding of \citet{Pratt2009} and \citet{Mantz2018}.

\begin{figure}
    \includegraphics{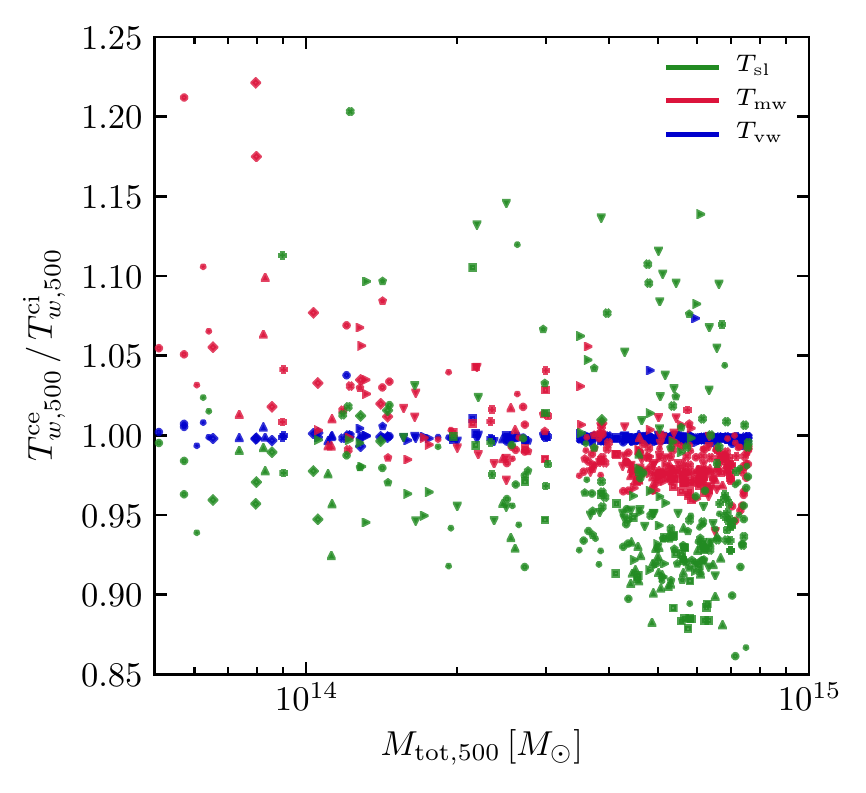}
    \caption{Ratio of the core-excluded to the core-included temperature versus mass for all haloes (NR, MW, VW and MC combined) with $z\leqslant 1.5$. We show the difference induced by the core exclusion on different temperature estimates: ${T_{\rm vw}}$  (blue) and ${T_{\rm mw}}$ (red) are a weighted average using respectively the cell volume and the cell density while ${T_{\rm sl}}$ (green) uses the cell emissivity and the \citet{Mazzotta2004} weights of hot ($E\geqslant0.5\keV$) gas cells only.}
    \label{fig:coreinclT}
\end{figure}

The inclusion of the core does not significantly impacts the $T-M$ scaling relation.\footnote{As the measurement of the temperature from a MCMC spectral fit is expensive, we only have measurement of ${T_{\rm sf}}$ in the core-excluded region. Therefore, we only discuss here the 3 other estimates (${T_{\rm vw}}$, ${T_{\rm mw}}$ and ${T_{\rm sl}}$) for which we have both ci and ce measurements. }
For instance, we find slopes of $0.717 \pm 0.020$ for the core-included and $0.700 \pm 0.019$ for the core-excluded ${T_{\rm sl}}-M$ scaling relation which are consistent.\footnote{Similarly, we find for the ${T_{\rm mw}}-M$ relation slopes of $0.745 \pm 0.014$ and $0.726 \pm 0.013$for the core-included and core-excluded estimates respectively and for the ${T_{\rm vw}}-M$ relation, $0.706 \pm 0.014$ and $0.689 \pm 0.014$ respectively.}
In \Fig{fig:coreinclT}, we see that the VW temperatures are widely insensitive to the core inclusion/exclusion because the volume inside $0.15\times\Rfiveh$ represents only a tiny fraction of the total $\Rfiveh$ sphere.

We can see that, in halo masses lower than $\sim2\times10^{14}\Msun$, the exclusion of the core tends to increase the MW temperatures as the measurement is biased by the presence of dense and cold gas in lower mass haloes (i.e. higher redshifts). On the other hand, for $\Mfiveh>3\times10^{14}\Msun$ the MW temperature is on average 2\,per cent lower when the core is excluded from the measurement . 
While showing a larger scatter, the ratio of the ce to ci ${T_{\rm sl}}$ can be as low as 0.85 with a median value of 0.94.\\

On average, we see that the ce temperature estimates are more biased low at higher halo masses which can help to explain why very slightly steeper slopes are found in the ci $T-M$ scaling relation.

\end{appendix}


\bsp	
\label{lastpage}
\end{document}